\documentclass[% 
twocolumn,
superscriptaddress,
nofootinbib,
amsmath,amssymb,
aps,
prd,
]{revtex4-1}

\usepackage{bm, graphics, graphicx, epsfig, soul, latexsym, hyperref, multirow}
\usepackage{comment}
\usepackage{subfigure}
\usepackage{graphicx,xcolor}
\usepackage{physics}
\usepackage{dcolumn}
\usepackage{bm,amsmath, amssymb,}
\usepackage{mathrsfs}
\usepackage{bigints}
\usepackage{float}
\usepackage{multirow}
\usepackage{footmisc}
\usepackage{placeins}
\usepackage{subfigure}
\usepackage[normalem]{ulem}
\usepackage{booktabs}
\usepackage{tabularx}
\usepackage{setspace}

\hypersetup{
    bookmarks=true,         % show bookmarks bar?
    unicode=false,          % non-Latin characters in Acrobat???s bookmarks
    pdftoolbar=true,        % show Acrobat???s toolbar?
    pdfmenubar=true,        % show Acrobat???s menu?
    pdffitwindow=false,     % window fit to page when opened
    pdfstartview={FitH},    % fits the width of the page to the window
	pdftitle={title},
    pdfauthor={authors},
    pdfsubject={Gravitational Waves},   % subject of the document
    pdfkeywords={gravitational waves, eccentric compact binaries, general relativity},
    pdfnewwindow=true,      % links in new window
    colorlinks=true,       % false: boxed links; true: colored links
    linkcolor=red,          % color of internal links
    citecolor=blue,        % color of links to bibliography
    filecolor=magenta,      % color of file links
    urlcolor=blue,           % color of external links
    linktocpage=true
}

\definecolor{darkgreen}{rgb}{0,0.5,0}
\definecolor{darkorange}{rgb}{1, 0.549,0}

\usepackage{amsmath}

\newcommand{\be}{\begin{equation}}
\newcommand{\ee}{\end{equation}}

\graphicspath{{./}}

\begin{document}

\title{Effect of double spin-precession and higher harmonics on spin-induced quadrupole moment measurements}
%\maketitle

\author{Divyajyoti}
\email{divyajyoti.physics@gmail.com}
\affiliation{Department of Physics, Indian Institute of Technology Madras, Chennai 600036, India}
\affiliation{Centre for Strings, Gravitation and Cosmology, Department of Physics, Indian Institute of Technology Madras, Chennai 600036, India}

\author{N. V. Krishnendu}
\email{krishnendu.nv@icts.res.in}
\affiliation{International Centre for Theoretical Sciences (ICTS), Survey No. 151, Shivakote, Hesaraghatta Hobli, Bengaluru, 560089, India}
\affiliation{Max Planck Institute for Gravitational Physics (Albert Einstein Institute), Callinstr. 38, 30167 Hannover, Germany}
\affiliation{Leibniz Universitat Hannover, D-30167 Hannover, Germany}

\author{Muhammed Saleem}\email{mcholayi@umn.edu}
\affiliation{School of Physics and Astronomy, University of Minnesota, Minneapolis, MN 55455, USA}

\author{Marta Colleoni}\email{marta.colleoni@uib.eu}
\affiliation{Departament de F\'isica, Universitat de les Illes Balears, IAC3 -- IEEC, Crta. Valldemossa km 7.5, E-07122 Palma, Spain}

\author{Aditya Vijaykumar}
\email{aditya@utoronto.ca}
\affiliation{Canadian Institute for Theoretical Astrophysics, University of Toronto, 60 St George St, Toronto, ON M5S 3H8, Canada}
\affiliation{International Centre for Theoretical Sciences (ICTS), Survey No. 151, Shivakote, Hesaraghatta Hobli, Bengaluru, 560089, India}

\author{K. G. Arun}
\email{kgarun@cmi.ac.in}
\affiliation{Chennai Mathematical Institute, Siruseri 603103, India}

\author{Chandra Kant Mishra}
\email{ckm@iitm.ac.in}
\affiliation{Department of Physics, Indian Institute of Technology Madras, Chennai 600036, India}
\affiliation{Centre for Strings, Gravitation and Cosmology, Department of Physics, Indian Institute of Technology Madras, Chennai 600036, India}

\date{\today}

\begin{abstract}
We investigate the prospect of performing a null test of binary black hole (BBH) nature using spin-induced quadrupole moment (SIQM) measurements. This is achieved by constraining a
deviation parameter ($\delta{\kappa}$) related to the parameter ($\kappa$) that quantifies the degree of deformation due to the spin of individual binary components on leading (quadrupolar) spin-induced moment.
%parameter ($\delta{\kappa}$) that characterises the deviations from general relativistic predictions for the parameter ($\kappa$) quantifying the degree of deformation due to the spin of individual binary component on leading (quadrupolar) spin-induced moment. 
%\saleemc{the previous sentence is too long. I propose: This is achieved by constraining the deviations from general relativistic predictions of a parameter $\kappa$, where $\kappa$ quantifies the . .  }\ckm{I have trimmed it down. I hope it reads better now.} 
Throughout the paper, we refer to $\kappa$ as the SIQM parameter and $\delta\kappa$ as the SIQM-deviation parameter. %\saleemc{Do we need this last sentence in abstract?}\ckm{Yes, this can be removed but may be after we have made related changes through the text.} \divya{I think it is fine to keep this sentence.} 
The test presented here extends the earlier SIQM-based null tests for BBH nature by employing waveform models that account for double spin-precession and higher modes. We find that waveform with double spin-precession gives better constraints for $\delta\kappa$, compared to waveform with single spin-precession. %\kga{is it deliberate that the names of the waveform families don't appear in the abstract?} \divya{I have no preference for/against this. If you think it'll be helpful, we can add the names.} 
We also revisit earlier constraints on the SIQM-deviation parameter for selected GW events observed through the first three observing runs (O1--O3) of LIGO-Virgo detectors. Additionally, the effects of higher-order modes on the test are also explored for a variety of mass-ratio and spin combinations by injecting simulated signals in zero-noise. Our analyses indicate that binaries with mass-ratio greater than 3 and significant spin precession may require waveforms that account for spin-precession and higher modes to perform the parameter estimation reliably. 
%\ckm{We should also add a sentence highlighting the impact of using a doubly-precessing waveform alone. At the moment it seems on its own it does not add value and must be supplemented with HMs.} \divya{I've added a sentence.} 
%We investigate the effect of spin-precession on the bounds of SIQM-deviation parameter when a waveform with doubly precessing spins is used for analysis, thus extending the previous SIQM-based binary black hole nature tests. 
%We also revisit the effects beyond the leading harmonic in the gravitational waveform on $\delta\kappa_s$ bounds for binaries with various mass and spin combinations. 
%The posterior distributions of $\delta\kappa_s$ for selected gravitational wave events observed through the first 3 observing runs of the LIGO-Virgo detectors employing a fully precessing dominant mode waveform model (\texttt{IMRPhenomXP}) as well as a companion higher order mode waveform model ({\tt IMRPhenomXPHM}) are also reported. 
%Through an injection study involving {\tt IMRPhenomXPHM} family of waveforms, we show that binaries with mass ratios greater than 3 and a significant spin precession signature will require waveforms with similar features for analyses. 
\end{abstract}
%TC:endignore\

\maketitle

\section{Introduction}
\label{sec:intro}

There have been over 90 statistically significant detections of binary coalescence events during the first three observing runs (O1-O3) of current interferometric gravitational wave (GW) detectors~\cite{Abbott:2016blz, TheLIGOScientific:2017qsa, LIGOScientific:2018mvr, LIGOScientific:2020aai, LIGOScientific:2020ibl}. These include the LIGO~\cite{LIGOScientific:2014pky, 2020PhRvD.102f2003B} and Virgo~\cite{VIRGO:2014yos, PhysRevLett.123.231108} detectors, as well as KAGRA~\cite{KAGRA:2020tym, Aso:2013eba} which has recently joined the network. Upon detection, an elaborate analysis is undertaken to deduce essential characteristics of the binary system, encompassing properties like masses, spins, orientation and location in the sky. This necessitates the utilization of accurate waveform models in conjunction with an efficient parameter inference algorithm to guarantee the accurate estimation of binary parameters \cite{LIGOScientific:2016vlm}. Each of these observations engender many significant follow-up analyses. One such endeavour involves determining the true nature of the compact objects in the binary system.
For instance, the nature of the secondary object in the binary coalescence event GW190814 \cite{LIGOScientific:2020zkf} is still a topic of discussion %\kga{Not sure we should invoke this example, as the referee will ask why you don't do the test for this event.} \divya{This is a compelling example and we can reply to referee that our test is only for spinning binaries. This event has $\rm{\chi_{eff}} \sim 0$.} 
as the secondary's mass is consistent with the lightest black holes (BHs) and heaviest neutron stars (NSs), along with other, more exotic composition stars~\cite{Clesse:2020ghq, Vattis:2020iuz, Huang:2020cab, Roupas:2020jyv, Biswas:2020xna, Tews:2020ylw}.

Various methods exist for investigating the true nature of the compact object in a binary system~\cite{Laarakkers:1997hb, PhysRevD.55.6081, Poisson:1997ha, Pacilio:2020jza, Gurlebeck:2015xpa, LeTiec:2020spy, Mendes:2016vdr, Uchikata:2016qku, Johnson-Mcdaniel:2018cdu, JimenezForteza:2018rwr, Abdelsalhin:2018reg, Datta:2019euh, Hartle:1973zz, Chatziioannou:2016kem, Datta:2020gem}. %\ckm{again a large number of citations}. \saleemc{I second Chandra's comment. The sentence can be expanded by naming a few of the tests and giving appropriate citations therein. Refer to NVK et al 2019 intro for example} \divya{I have removed some citations. The ones remaining seem relevant to me.} 
An analysis based on the spin-induced multipole moments is one among them. %\saleemc{I think this sentence till the end of the para should be at the biginning of this para with all previous SIQM works cited there. Then at the end of the para, say that there are other approaches/methods for testing the nature of the compact object and are applicable in various contexts etc.. with citations to a few named ones} \divya{I would like to keep it as such since changing the order is kind of disrupting the flow.} 
Spin-induced multipole moments arise due to the spins of individual compact objects in the binary \cite{Laarakkers:1997hb}. From observations, one can measure these spin-induced multipole moments and then use that information to distinguish black hole binaries from binaries composed of other compact objects. The method based on the leading order spin-induced multipole moment, the spin-induced quadrupole moment (SIQM) measurement, has been explored in detail~\cite{Krishnendu:2017shb, Krishnendu:2018nqa, Krishnendu:2019tjp, Krishnendu:2019ebd, Saleem:2021vph} and applied to the observed GW events from the first three observing runs of advanced LIGO-Virgo detectors~\cite{LIGOScientific:2020tif, LIGOScientific:2021sio}. Moreover, the possibility of measuring spin-induced quadrupole using future detectors, and simultaneous measurement of spin-induced quadrupole and octupole moment parameters~\cite{Krishnendu:2018nqa, Saini:2023gaw} have also been studied. A template bank for binaries of exotic compact object searches has recently been developed, accounting for the spin-induced quadrupole moment and tidal effects~\cite{Chia:2022rwc}. 
References~\cite{Krishnendu:2017shb, LIGOScientific:2020tif, LIGOScientific:2021sio} report the tests of binary black hole nature using spin-induced quadrupole moments using a phenomenological waveform model, \texttt{IMRPhenomPv2} \cite{Husa:2015iqa, Khan:2015jqa, Hannam:2013oca}, containing only the dominant modes ($\ell=2, |m|=2$) in the co-precessing frame %\kga{I thought Pv2 has 21 mode as well} \divya{Pv2 description on \href{https://lscsoft.docs.ligo.org/lalsuite/lalsimulation/group___l_a_l_sim_i_m_r_phenom__c.html}{\uline{LALSuite page}} says it is based on PhenomD. And \href{https://ui.adsabs.harvard.edu/abs/2023ascl.soft07019P/abstract}{\uline{PhenomD description}} only says the dominant mode. Please correct me if I'm wrong.} 
and an effective spin parameter. Recently, Ref.~\cite{LaHaye:2022yxa} came up with a fully precessing waveform implementation of SIQM test for low-mass binaries, focusing on binaries in which at least one object is in the lower mass gap ($< 3 M_\odot$) \cite{Lyu:2023zxv}. 
%\nvk{Lyu et al. also has results from GWTC-3, GW151226 and GW191216, but they find that the PhenomPv2 results are better for real events compared to PhenomXPHM, see Fig. 9 of arxiv version. So I suggest rewriting the above sentence to "Recently, Ref.~\cite{LaHaye:2022yxa} came up with a fully precessing waveform implementation of SIQM test with higher modes and reports measurement improvement in binaries consist of lower mass gap objects. Though we find overall consistency,  our findings disagree on the GWTC-3 results. "} \divya{Their study is using some XPHM+AI recovery. I would like to understand it better before boldly writing in the introduction that we disagree with their results. I want to make sure that it is an apples-to-apples comparison.} \nvk{I think it is a good idea to read through their work. I have gone through it. Since we are doing almost same thing as them, we will have to write about a comparison.} \kga{We should be specific about what is different from [48] and our work, as they seem along similar lines.}

\subsection{Current work}
\label{subsec:summary}

The effects of spin-precession and higher-order modes have been  extensively studied in the context of testing general relativity (TGR) using binary BHs
(see, for instance, Refs.~\cite{Bustillo:2016gid, Krishnendu:2021cyi, Krishnendu:2021fga, Puecher:2022sfm, Mehta:2022pcn, Islam:2021pbd, Breschi:2019wki}). 
By injecting the most up-to-date phenomenological waveform models with full spin-precession (\texttt{IMRPhenomXP}) and higher modes (\texttt{IMRPhenomXHM}, \texttt{IMRPhenomXPHM} \cite{Pratten:2020ceb, Pratten:2020fqn, Garcia-Quiros:2020qpx}) for binary black hole signals of varying masses and spins, we investigate the effects of spin-precession and higher modes on $\delta\kappa$ measurements. Specifically, our injections include binaries with mass ratios ($q=m_1/m_2$, where $m_1$ and $m_2$ are the detector-frame component masses and $m_1 > m_2$) in the range $q\in[1,5]$ and in-and-out-of-plane spin effects for a fixed mass binary ($M = 30 M_\odot$). We observe the difference in the bounds of $\delta\kappa$ between \texttt{IMRPhenomXP} (doubly spin-precessing) and \texttt{IMRPhenomPv2} (singly spin-precessing waveform model) in three scenarios: varying mass ratio, varying effective aligned spin parameter ($\chi_\text{eff}$), and varying effective spin-precession parameter ($\chi_\text{p}$). We note that in all cases \texttt{IMRPhenomXP} outperforms \texttt{IMRPhenomPv2}. Next, we employ the higher mode waveform models \texttt{IMRPhenomXHM} and \texttt{IMRPhenomXPHM} to study the effect of HMs on $\delta\kappa$. We observe that for higher mass ratios, the higher mode waveform models perform better compared to the dominant mode model \texttt{IMRPhenomXP}. Following the TGR analyses on the second and third GW transient catalogs (GWTC-2 \cite{LIGOScientific:2020tif} and GWTC-3 \cite{LIGOScientific:2021sio}), we measure $\delta\kappa$ of the binary systems using  \texttt{IMRPhenomXP} and \texttt{IMRPhenomXPHM}. We find that together with the effect of spin-precession, the inclusion of higher modes plays a critical role when analysing binaries with mass-asymmetries similar to that in the event GW190412 \cite{LIGOScientific:2020stg} (mass ratio $q\approx3.7$). For GW190412, the bounds on $\delta\kappa$ obtained with \texttt{IMRPhenomXPHM} are constrained enough to rule out the boson star binaries, subject to the assumptions in the current work. %\saleemc{Can we state it unconditionally? For example, we have ignored tidal effects and have not studied the systematics of ignoring them. So, it's not proven that a boson star binary can be captured with a BBH waveform if the tides are ignored. Since this is a strong statement about the physical nature of a real event, I think we should say subject to our assumptions.} \divya{I agree. I have modified the sentence a bit.} 
We also report the revised bounds on $\delta\kappa$ from selected GW events observed through the first three observing runs of the LIGO-Virgo detectors. This paper is organized as follows. In Sec~\ref{setup}, we detail the waveform model and parameter estimation method. Our results from simulated GW events are shown in Sec.~\ref{res:simulation}, and real GW observations are reported in Sec.~\ref{res:realevent}. We conclude with Sec.~\ref{sec:summary}.

\section{Analysis Setup}
\label{setup}

In this section, we review the details of waveform models used and the basics of Bayesian parameter estimation and hypothesis testing. Typically, the evolution of an inspiralling compact binary can roughly be divided into three stages: an early inspiral, late inspiral \& merger, and the final ringdown. During the early inspiral stage, the separation between the compact objects in the binary is large, and hence, their evolution can be modeled as a perturbation series in the velocity parameter. The post-Newtonian (PN) theory provides an analytic expression for the inspiral phase incorporating various physical effects such as the spin-orbit effects, self-spin effects, cubic and higher order spin-effects, spin-precession effects, orbital eccentricity effects, etc. (see, for instance, \cite{Blanchet:2002av, Mishra:2016whh, Henry:2022dzx}). On the other hand, one needs to invoke numerical relativity techniques to model the highly non-linear relativistic merger phase (see \cite{Lehner:2001wq} for a review on numerical relativity modeling techniques). Further, the ringdown part can be modeled perturbatively using the BH perturbation theory techniques~\cite{Sasaki:2003xr, Pretorius:2007nq}.

In the inspiral phase, the effect of spin-induced quadrupole moment starts to appear at 2 PN, together with the other spin-spin terms. More precisely, the leading order PN coefficient is of the schematic form~\cite{Poisson:1995ef} $Q=-\kappa \chi^2 m^3$, where the negative sign indicates the \emph{oblate} deformation due to the spinning motion. The proportionality constant, $\kappa$, can take different values for different compact objects. For black holes, $\kappa_\text{BH}$ is 1. Slowly spinning neutron stars can have $\kappa$ values in the range $\rm{\kappa_{NS}}\sim 2-14$~\cite{Laarakkers:1997hb, Pappas:2012ns, Pappas:2012qg}, whereas for more exotic stars like boson stars, this range can be $\rm{\kappa_{BS}\sim}10-100$~\cite{PhysRevD.55.6081} depending on internal composition. There also exist gravastar proposals where the value $\rm{\kappa_{GS}}$ %\ckm{range?} \nvk{I think it is safe to not qoute any specific ranges for gravastars, see Uchikata+ 2015 and discussions there in} 
can match the BH value but also allows for negative values and prolate deformations~\cite{Uchikata:2015yma}. Measuring the SIQM parameter from GW observations can thus provide unique information about the nature of the compact object. 

For a binary system composed of two BHs with $\kappa_i$, following \cite{Krishnendu:2019tjp}, we define $\kappa_i=1+\delta\kappa_i$, where $i=1,2$, and $\delta\kappa_i=0$ gives the BH limit. Since the simultaneous measurements of both $\delta\kappa_i$ lead to uninformative results,\footnote{This is with reference to the current detector sensitivities and dominant mode waveform models. A detailed investigation can be carried out using future detector sensitivities and/or higher mode waveform models, but it is beyond the scope of this paper.} we stick to the proposal of~\cite{Krishnendu:2019tjp}, where a symmetric combination of $\delta\kappa_i$ is measured keeping the anti-symmetric combination ($\delta\kappa_a$) to zero. We call this symmetric combination SIQM deviation parameter henceforth and use the definition $\delta\kappa_s =(\delta\kappa_1 + \delta\kappa_2)/2$.

\subsection{Waveform models used for the test}
\label{sec:WF models}

%\ckm{May be you can move the general discussion about the test above as part of the main section, and discuss the waveform details under this subsection. (you may also want to then change the subsection title to something more suitable that represents waveform and the parametrizations involved.)}

GW data analysis has routinely employed phenomenological waveform models with varying properties. For instance, the SIQM analysis in the past was carried out using the precessing dominant mode phenomenological waveform model {\tt IMRPhenomPv2}~\cite{Khan:2019kot}. The two spin parameters, one for the out-of-plane spin effects -- the effective spin parameter ($\chi_\mathrm{eff}$) \cite{Ajith:2009bn, Santamaria:2010yb}, and another for the in-plane spin effects -- the spin-precession parameter ($\chi_\mathrm{p}$) \cite{Schmidt:2012rh, Hannam:2013oca, Schmidt:2014iyl}, are among the best-measured spin parameters in the parameter estimation of a GW signal. It has been demonstrated that $\chi_\mathrm{eff}$ captures the spin effects along the direction of the angular momentum axis~\cite{Ajith:2009bn}, and $\chi_\mathrm{p}$ measures the spin effects in the orbital plane of the binary~\cite{Schmidt:2014iyl}. The effective spin parameter for a binary with dimensionless spin components, $\chi_i = (\vec{S_{i}} \dotproduct \hat{L})/ m_i^2$, can be defined as
\begin{equation}
    \chi_{\rm{eff}}=\frac{\chi_1 m_1 + \chi_2 m_2}{m_1+m_2}.
    \label{eq:chieff}
\end{equation}
Here, $\vec{S_i}$ is the individual spin angular momentum vector of the compact object in the binary with mass $m_{i}$, and $\hat{L}$ represents the unit vector along the angular momentum axis of the binary. In terms of the perpendicular spin vectors, $S_{i\perp}=|\hat{L}\times (\vec{S_i}\times \hat{L})|$, the effective spin-precession parameter can be written as
\begin{equation}
    {\chi_\mathrm{p}}=\frac{1}{A_{1}m_{1}^2} \max(A_{1}S_{1\perp}, A_{2} S_{2\perp}),
\end{equation}
where $A_{1}=2+(3/2q)$ and $A_{2}=2+(3q/2)$ are mass parameters defined in terms of the mass ratio $q=m_1/m_2>1$.

With recent developments in phenomenological modeling, it is now possible to describe quasi-circular binaries involving generic spin components. Specifically, the waveform {\tt IMRPhenomXP} is the current state-of-the-art phenomenological model where the two-spin effects are introduced~\cite{Pratten:2020fqn}. 
Further, the {\tt IMRPhenomXPHM} model (which is also fully precessing) includes higher order modes $(\ell, |m|)= (3, 3), (4, 4), (2, 1), (3, 2)$ in the co-precessing frame, in addition to the dominant mode $(\ell, |m|)= (2, 2)$~\cite{Garcia-Quiros:2020qpx, Pratten:2020ceb}. 
We modify the inspiral phase coefficients of these waveform models at 2~PN and 3~PN orders by introducing explicit dependence on SIQM parameters, and study the measurement probabilities for simulated GW signals from BBHs and the detected GW events.

%\subsection{Brief overview of Bayesian parameter estimation and hypothesis testing}
\subsection{Brief overview of Bayesian parameter estimation}

Bayesian stochastic sampling algorithms are routinely employed to perform parameter estimation of detected compact binary signals~\cite{Veitch:2009hd, Veitch:2014wba, thrane2019introduction}. 
To review the details of parameter estimation, we start with the definition of likelihood function assuming Gaussian noise,
\begin{equation}
    \mathcal{L}(d|\theta) \propto  \exp \qty(-\frac{1}{2} \ip{d-h\qty(\theta)}{d-h\qty(\theta)} ),
    \label{eq:likelihood}
\end{equation}
where the inner product $\ip{a}{b}$ is defined as

\begin{equation}
    \ip{a}{b} = 4\int_0^\infty \frac{a^*(f) b(f)}{S_n(f)}.
\end{equation}
Here, $S_n$ is the noise power spectral density of the detector.
The posterior distribution on model parameters ($\theta$), given the data ($d$) and likelihood (${\mathcal{L}}$), is
\begin{equation}
\label{posterior}
    p(\theta|d) = \frac{\mathcal{L}(d|\theta) \mathcal{\pi}(\theta)}{p(d)} \,.
\end{equation}
Here $p(d)$ is the Bayesian evidence or marginalized likelihood and is a normalization constant obtained by integrating the likelihood over the entire prior region for the set of parameters, $\theta$. A generic binary is characterised by a set of intrinsic and extrinsic parameters. Adding to this parameter set, in the current analysis, we include $\delta\kappa_s$ as a free parameter to be constrained from the data. We can extract the $\delta\kappa_s$ posterior by marginalizing over all other parameters $\theta_{\rm BBH}$ from the multi-dimensional posterior samples as 
\begin{equation}
     p(\delta\kappa_s|d) = \int p(\theta | d) \text{d}\theta_\text{BBH}.
\end{equation}

We use {\tt LALSimulation} \cite{lalsuite} for generating all waveforms and the \emph{nested} sampling algorithm~\cite{2004AIPC..735..395S} implemented through {\tt dynesty}~\cite{speagle2020dynesty} sampler in {\tt bilby}~\cite{Ashton:2018jfp} and {\tt bilby\_pipe}~\cite{Romero-Shaw:2020owr} for parameter estimation throughout the analysis. 

\section{%Posteriors on $\delta\kappa_s$ from simulated binaries}
Results from the simulations}
\label{res:simulation}
%\ckm{Imagine a better title for the section} {\magenta KGA: Now?}

\begin{figure*}
    \centering
    \includegraphics[trim=0 0 0 0, clip, width=0.9\linewidth]{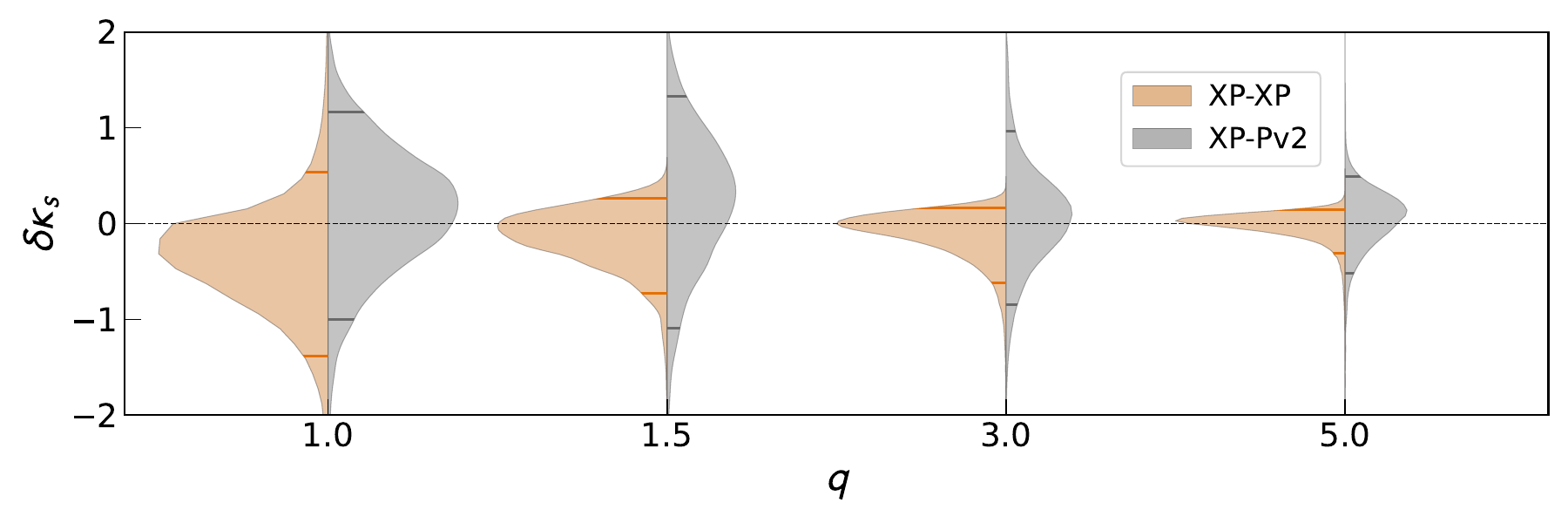}\hfill
    \includegraphics[trim=0 0 0 0, clip, width=0.49\linewidth]{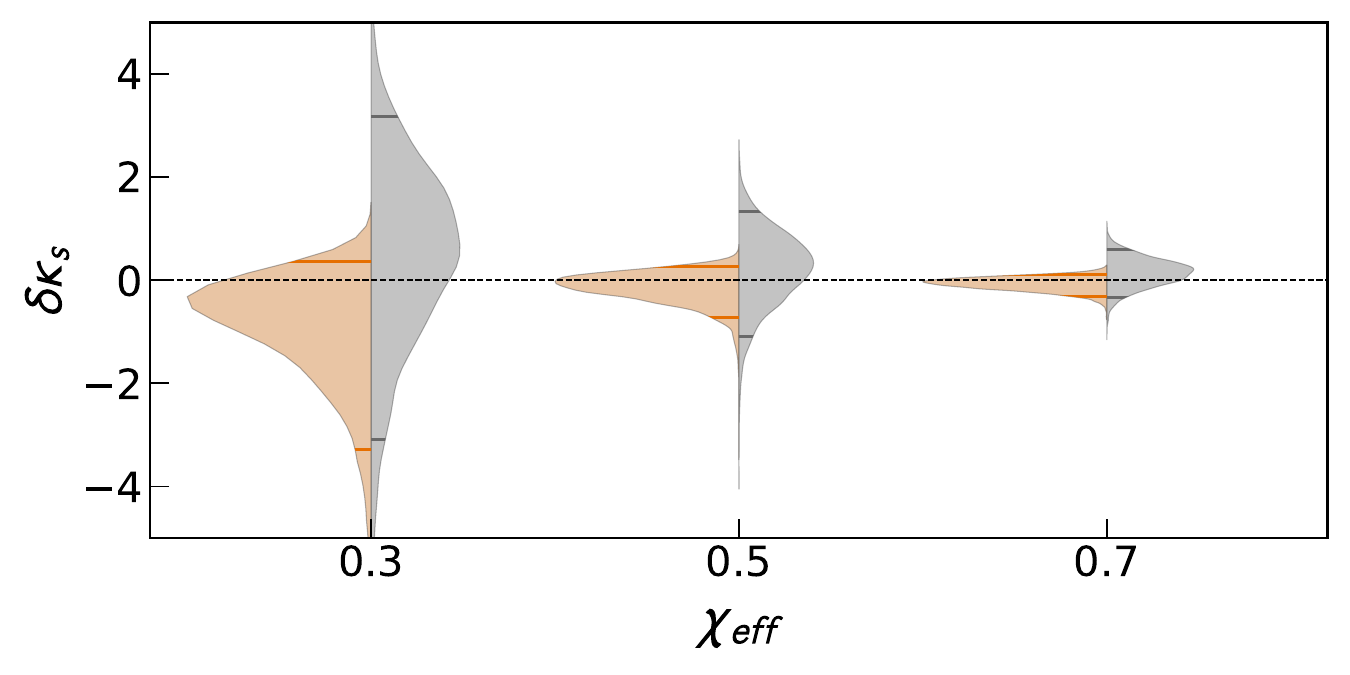}
    \includegraphics[trim=0 0 0 0, clip, width=0.5\linewidth]{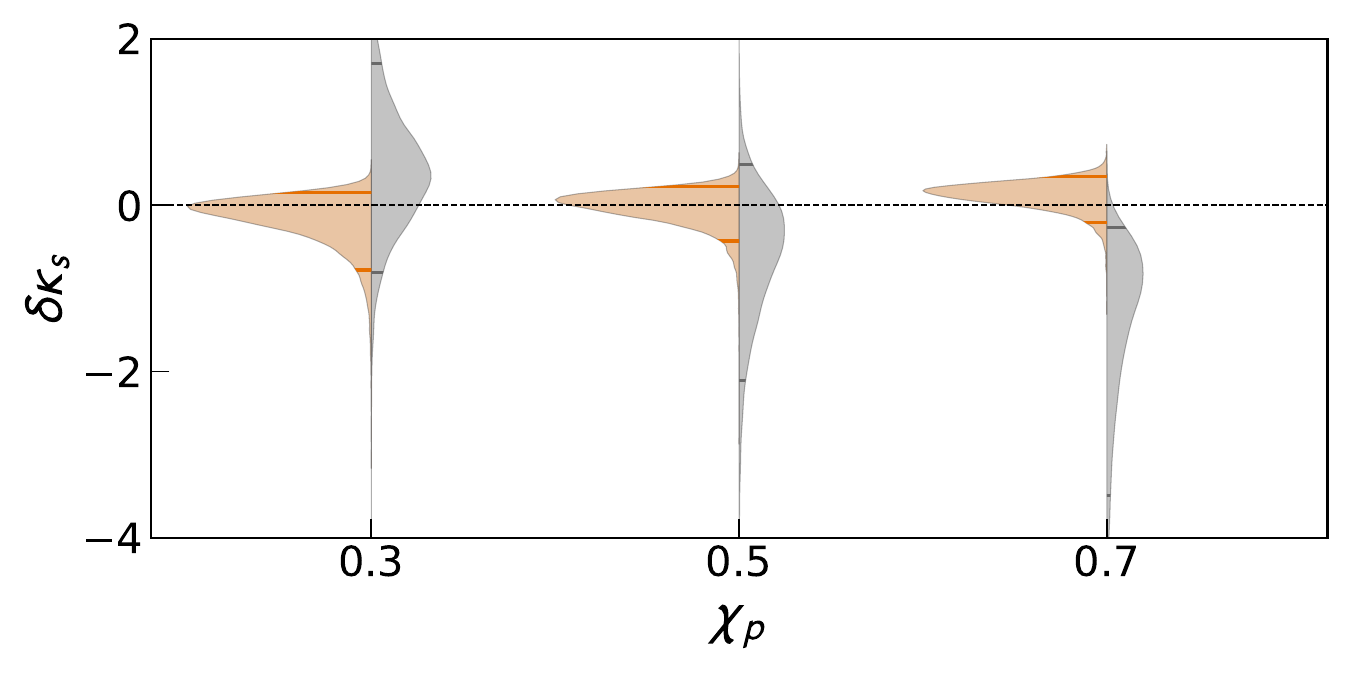}
    \caption{The violin plots show the posterior distributions for the SIQM-deviation parameter for various injection studies. Top: Four different mass ratio cases are chosen, $q=(1, 1.5, 3, 5)$; the spin magnitudes and related angles are fixed to the values included in Table~\ref{table:fixed_spin_angles}. Bottom Left: Three different values for the effective spin parameter are chosen, $\chi_\text{eff}$ = $0.3, 0.5, 0.7$ (see Table \ref{table:fixed_chip_q} for complete information); mass ratio is taken to be $q=1.5$ and $\rm{\chi_\mathrm{p}=0.3}$. Bottom right: Three different values for spin-precession parameter are chosen, $\chi_\mathrm{p}$ = 0.3, 0.5, 0.7 (see Table \ref{table:fixed_chieff_q} for information on other spin parameters); mass ratio is taken to be $q=3$ and $\chi_\text{eff}=0.5$. The total mass is fixed to $M=30~M_\odot$, and the network SNR is 40 for all cases. The injections are performed using the fully spin-precessing dominant mode waveform (\texttt{IMRPhenomXP}) and recovered with the same (orange) as well as with single spin-precessing dominant mode waveform \texttt{IMRPhenomPv2} (grey). The horizontal black-dashed lines denote the injected value, and the coloured lines inside the violins indicate the 90\% credible intervals for the respective posterior distributions. The legend follows the pattern ``injected waveform -- recovery waveform".}
    \label{fig:violins}
\end{figure*}

We demonstrate the importance of using a waveform model with double spin-precession and higher modes for analysing binary black hole signals with varying properties. Specifically, we look into different binaries of varying mass asymmetries and in-plane \& out-of-plane spin parameters, while fixing all other parameters, to look into the effect of mass and spin variations in the $\delta\kappa_s$ measurements. We consider four waveform models, {\tt IMRPhenomXPHM} (waveform with two-spin effects and higher modes), {\tt IMRPhenomXP} (waveform with two-spin effects and dominant mode),\footnote{ The solutions employed in \texttt{IMRPhenomX} family are precession-averaged.} {\tt IMRPhenomXHM} (waveform with no spin-precession effects but higher modes), and {\tt IMRPhenomPv2} (waveform model with single-spin precession approximation and dominant mode).

For injections, we fix the total mass of the binary to be $30~M_{\odot}$ and vary mass ratios and spins. All the binaries are placed in such a way that the network signal-to-noise ratio (SNR) is 40 in a three-detector network consisting of two advanced LIGO~\cite{KAGRA:2013rdx, AdvancedLIGO2010, TheLIGOScientific:2014jea} and one advanced Virgo~\cite{TheVirgo:2014hva, TheVirgostatus} detectors with advanced sensitivity~\cite{H1L1V1Dcc}. All the injections considered in this section are zero-noise injections and represent BBH mergers (i.e., $\delta\kappa_s = 0$). 
The GW signal from a binary system with spin-precession is characterized by component masses ($m_1, m_2$), dimensionless spin magnitudes ($a_1, a_2$), spin angles ($\Phi_1, \Phi_2, \Phi_{12}, \Phi_{\mathrm{j}l}$), luminosity distance ($d_L$), angles measuring the location of the source in the sky and the orientation of the source with respect to the line of sight ($\iota, \theta, \phi, \psi$), and the SIQM deviation parameter ($\delta\kappa_s$). We assume a uniform prior ranging [-500, 500] on $\delta\kappa_s$. More details about the definition of these parameters and corresponding prior ranges are given in Appendix~\ref{appendix1:priors} and Table~\ref{table:priors}. 

\subsection{Comparison of ${\delta\kappa_s}$ estimates: {\tt IMRPhenomXP} and {\tt IMRPhenomPv2}}
\label{subsec:comp-pv2-xp}

The effect of $\rm{\chi_{eff}}$ on the posteriors of $\delta\kappa_s$ is well established and was explored in Ref.~\cite{Krishnendu:2019tjp}, albeit using \texttt{IMRPhenomPv2}. In this section, we wish to compare the bounds on $\delta\kappa_s$ obtained from \texttt{IMRPhenomXP} (doubly spin-precessing waveform) and \texttt{IMRPhenomPv2} (single spin-precession waveform). Furthermore, while the study in~\cite{Krishnendu:2019tjp} was performed for aligned-spin systems, here, we choose systems with precessing spins and hence also explore the effect of varying $\chi_\mathrm{p}$ on $\delta\kappa_s$. We consider three cases: 
\begin{itemize}
    \item Fixed spins with varying mass ratio to investigate the effect of mass ratio on $\delta\kappa_s$. 
    \item Fixed masses and spin-precession parameter ($\chi_\mathrm{p}$), varying effective spin parameter ($\rm{\chi_\text{eff}}$). 
    \item Different spin-precession values, keeping the masses and aligned-spin components fixed.
\end{itemize} 

For all cases, we inject binary black hole ($\delta\kappa_s=0$) signals with precessing spins using \texttt{IMRPhenomXP} model and recover these with \texttt{IMRPhenomXP} and \texttt{IMRPhenomPv2}. We present the results in the form of violin plots in Fig.
\ref{fig:violins}.

\subsubsection{Effect of mass ratio variation on $\delta\kappa_s$}
\label{subsec:q_variation}
To see the effect of mass ratio on the $\delta\kappa_s$ measurements, we inject binaries with fixed spin magnitudes and angles, and vary the mass ratio as $q=1,1.5,3,5$. The spin parameter values are listed in Table~\ref{table:fixed_spin_angles}. The injection signals are created using \texttt{IMRPhenomXP} and are analysed with both {\tt IMRPhenomPv2} and {\tt IMRPhenomXP}.

For all the cases, we observe that {\tt IMRPhenomXP} outperforms {\tt IMRPhenomPv2}, especially for binaries with higher mass asymmetry. Moreover, we observe from the top panel of Fig.~\ref{fig:violins} that an increase in mass ratio results in better constraints on $\delta\kappa_s$. This is consistent with the findings of  Ref.~\cite{Krishnendu:2018nqa} where the dependence of mass ratio on the errors of SIQM parameter ($\Delta\kappa_s$) are discussed in detail [see Fig. 2 and the discussion around Eqs.~(4.2)--(4.4) there]. 
%which describes the dependence of leading order spin-spin term in phasing on mass ratio and $\Delta\kappa_s$ (see Fig. 2 and Eq. 4.2-4.4 of \cite{Krishnendu:2018nqa} along with the discussion around it).} %\divya{Need to say that they are reproduced here for clarity or don't use them at all, just write it in words.} \ckm{Explain the underlying cause} \divya{Do you mean that I should explain why spin parameters are better constrained for high $q$? I don't know why that happens.}
%\ckm{Why?} \adityac{I guess it is because the phasing term has $\chi \times \kappa_s$ or something like that. Since the SNR is constant across your injections, the phasing is measured to a precision of $1/40$ radian, ie. $\chi \times \kappa_s \sim 1/40$. Hence, if $\chi$ is high, $\kappa_s$ would be constrained to a lower value.} \divya{I think Chandra was asking about the reason for mass ratio trend not $\chi$ since the statement prior to the comment talks about mass ratio improving spins not spins improving $\delta\kappa_s$}\ckm{Yes. I hope you know the answer now.} \divya{Nope. Can someone explain this? Aditya's comment doesn't say anything about mass ratio.}\ckm{I agreed with your response to Aditya that I was referring to the effect of mass ratio and wanted you to explain that in the above text. I also discussed this effect in person with you. You can find proper justification in Krishnendu's second paper on SIQM test.} 
Moreover, the values of $\chi_\text{eff}$ also increase gradually as we go from equal mass case to unequal mass while keeping spin magnitudes and angles fixed. Both of these effects result in improvement of $\delta\kappa_s$ bounds with increasing mass ratio. We also show the corresponding corner plots on various parameters in Fig. \ref{fig:corner_fixed_spin_anlges_vary_q} in the Appendix. %{\blue [NVK: if Fig. 5 is only described here, it is better to keep the ordering right and make it Fig.2 by moving up inside the tex file. Suddenly going from Fig. 1 to 5 is not looking good.]} \divya{Moving it up will mean adding it to the main text instead of appendix. I would like to keep it in appendix. I've added the word `appendix' now in the sentence.}

\begin{figure*}[ht!]
    \centering
    \includegraphics[width=0.48\linewidth]{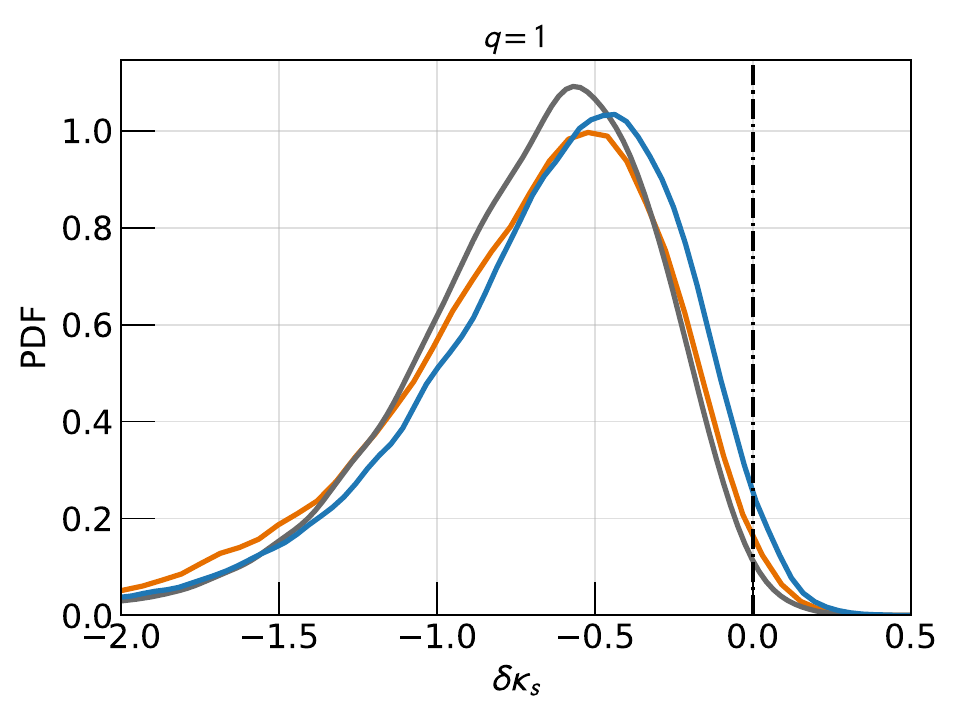}
    \includegraphics[width=0.48\linewidth]{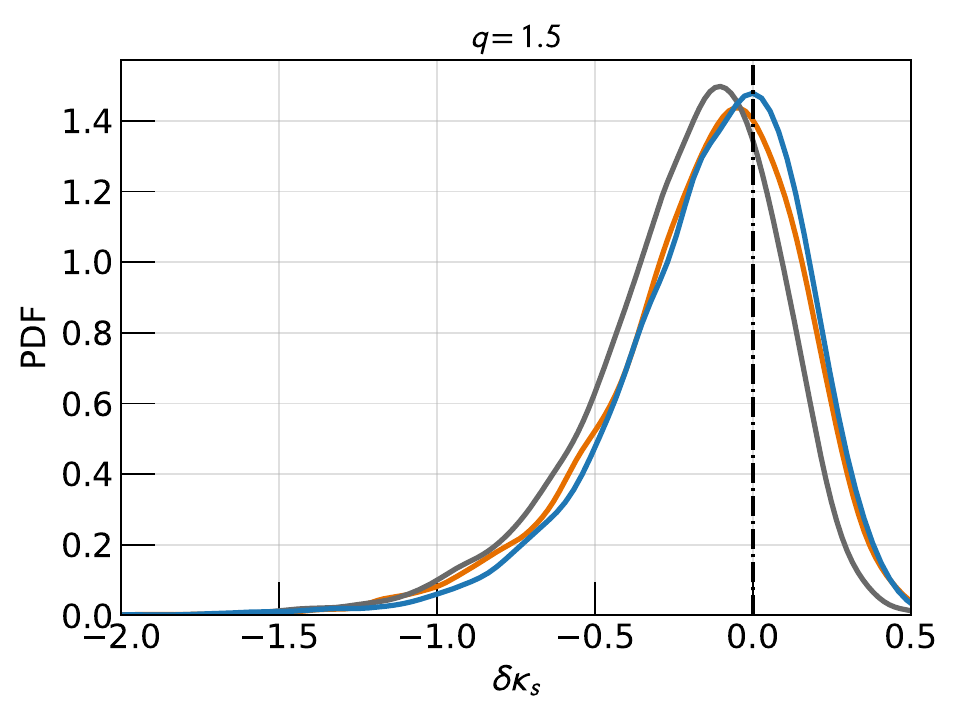}
    \includegraphics[width=0.48\linewidth]{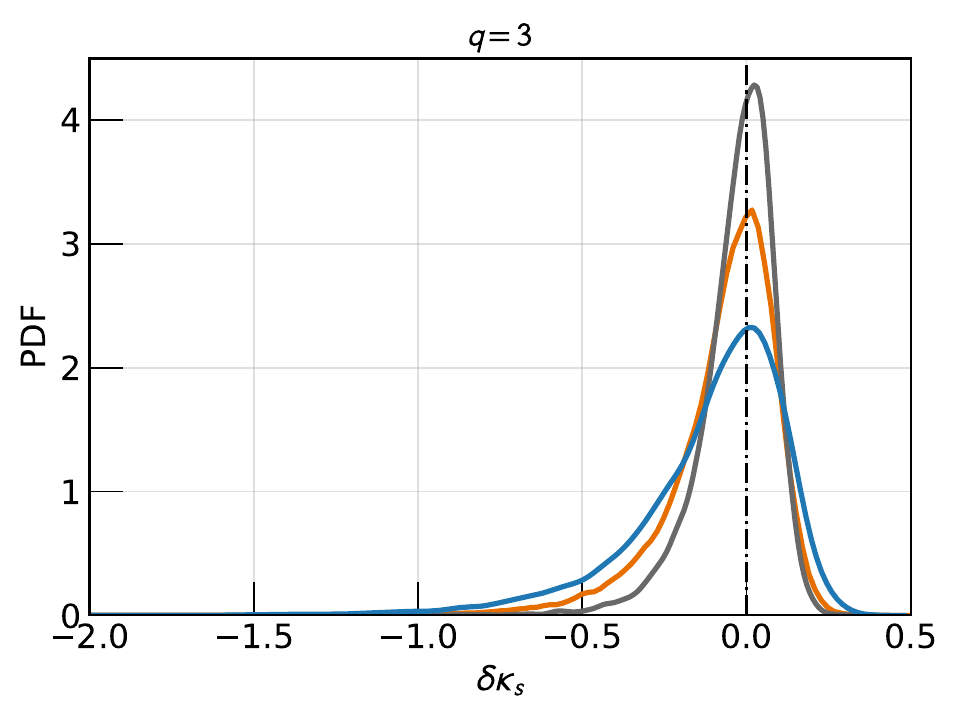}
    \includegraphics[width=0.48\linewidth]{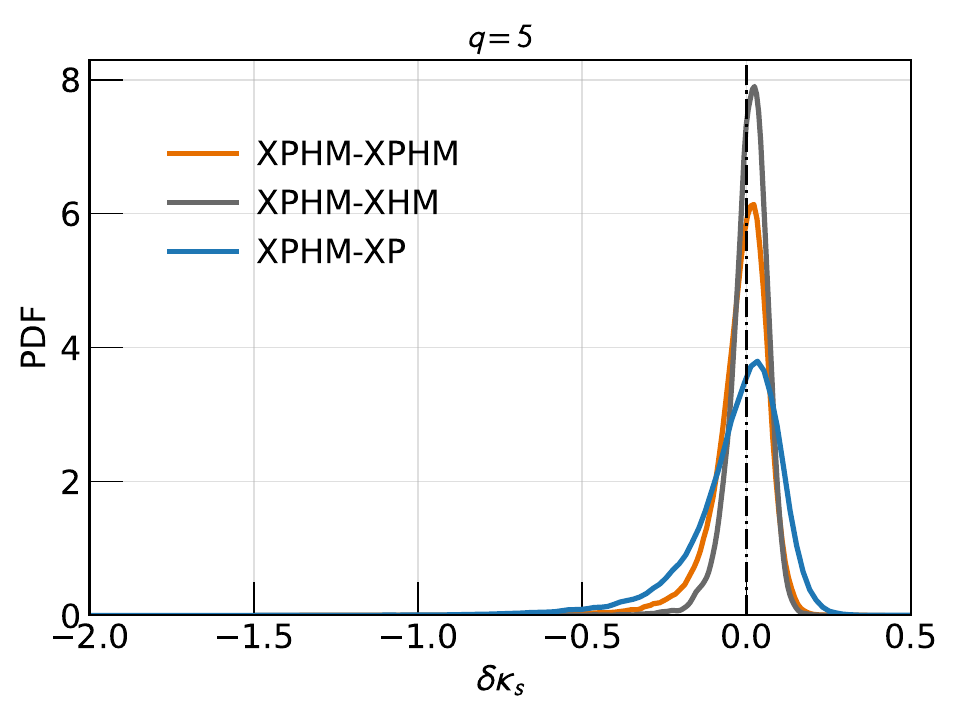}
    \caption{Posterior distributions of the SIQM deviation parameter for various mass ratios $q=1, 1.5, 3, 5$. The total mass is fixed to $M=30~M_{\odot}$, and network SNR is 40 for all cases. The spin magnitudes and angles are fixed (see Table \ref{table:fixed_spin_angles}). We use fully spin-precessing, higher mode waveform model {\tt{IMRPhenomXPHM}} for injections and recover using the same (orange), aligned spin higher mode waveform {\tt{IMRPhenomXHM}} (grey), and fully spin-precessing dominant mode waveform model {\tt{IMRPhenomXP}} (blue) to study the effect of higher modes on $\delta\kappa_s$ measurements. The vertical black-dashed lines denote the injected value. The legend follows the pattern ``injected waveform -- recovery waveform".}
    \label{fig:hm-inj-post}
\end{figure*}

\subsubsection{Effect of $\chi_\mathrm{eff}$ variation on $\delta\kappa_s$}
\label{subsec:chi_eff_variation}

Here we choose a nearly equal-mass binary, with mass ratio $q=1.5$ and $\rm{\chi_\mathrm{p}=0.3}$. Since we are not exploring the effect of HMs in this injection set, a nearly equal-mass system serves the purpose well. %\ckm{Explain why you did not choose $q=1$ case.} \divya{I'm leaning more and more towards dropping $q=1$ altogether.}\ckm{Well you may drop it (if everyone agree) but you still need to explain why you are not showing equal mass case.} \divya{I honestly don't know why $\delta\kappa_s$ is not recovered for $q=1$. The prior railing doesn't seem to affect the XP and Pv2 runs and I don't think it's just because of the HMs because the corresponding BBH run is showing a chirp mass and $\chi_\text{eff}$ shift to the right of the injected value whereas for the $\delta\kappa_s$ run, that shift is towards the left. I don't know why that is. See \href{https://ldas-jobs.ligo.caltech.edu/~divyajyoti.nln/SIQM_HM/runs_kappa_bilby_2Mar23_xp/Prod_runs_set_1/M_30_q_1/k1_0_k2_0/run01/comparison_inj_xphm_rec_xphm_bbh_ks/html/Comparison_chi_eff.html}{this plot}}
We vary the $\chi_\mathrm{eff}$ parameter as $\rm{\chi_\mathrm{eff}=0.3, 0.5, 0.7}$. This is done by fixing the $\rm{x-}$ and $\rm{y-}$ components of the two spin vectors and varying the $\rm{z-}$ components $\chi_\text{1z}$ and $\chi_\text{2z}$ to obtain three values of $\chi_\text{eff}$ as $0.3$, $0.5$, and $0.7$ (see Table~\ref{table:fixed_chip_q}).

As observed in Ref.~\cite{Krishnendu:2019tjp}, the estimates on $\delta\kappa_s$ improve as we choose large positive $\chi_\mathrm{eff}$ values. Also, for all values of $\chi_\mathrm{eff}$, the bounds obtained using \texttt{IMRPhenomXP} are better than \texttt{IMRPhenomPv2}. These improvements can be explained by looking at correlations between $\chi_\text{eff}$ and $\delta\kappa_s$ shown in Fig. \ref{fig:corner_vary_chieff} in the Appendix. %{\blue [NVK: make this as Fig. 3?]} \divya{Again, I would like to keep this in appendix.}

\subsubsection{Effect of $\chi_\mathrm{p}$ variation on $\delta\kappa_s$}
\label{subsec:chi_p_variation}

Here we choose a mass ratio of $q=3$ and a moderate value of $\chi_\text{eff}=0.5$. A slightly larger mass ratio is chosen here compared to Sec~\ref{subsec:chi_eff_variation} to avoid the uninformative inference on the analyses due to unconstrained spin-precession effects for near-equal mass binaries. Keeping the $\rm{z-}$ component of spin vectors the same, we vary the $\rm{x-}$ and $\rm{y-}$ components to obtain three distinct values of $\chi_\mathrm{p}$ as 0.3, 0.5, and 0.7 (see Table \ref{table:fixed_chieff_q}). 

As the values of $\chi_\mathrm{p}$ increase, the bounds on $\delta\kappa_s$ with \texttt{IMRPhenomXP} become tighter, enhancing the differences between \texttt{IMRPhenomXP} and \texttt{IMRPhenomPv2} waveforms. The \texttt{IMRPhenomPv2} bounds shift away from the injected value as we move from low to high $\chi_\mathrm{p}$ values, excluding 0 from the 90\% credible interval for $\chi_\mathrm{p}=0.7$. Additionally, they become increasingly worse (the posteriors become broader) compared to \texttt{IMRPhenomXP} as we go to higher values of $\chi_\mathrm{p}$. We suspect that the doubly spin-precessing model \texttt{IMRPhenomXP} is helping to break certain degeneracies between the SIQM parameter and the spins leading to a more symmetric estimate of $\delta\kappa_s$ for all the $\chi_\mathrm{p}$ values compared to the \texttt{IMRPhenomPv2} waveform model. %\nvk{same conclusion is obtained by Lyu et al. for their case with one object is in the mass gap and carries non-BBH SIQM value.} \divya{Since they are dealing with different systems and more importantly non-BBH injections, I would like to keep this out of discussion here. We have already referenced their study in introduction.}

\begin{table}[ht]
\def\arraystretch{1.7}
\begin{tabular}{|c|c|c|c|c|c|c|c|c|}
\hline
\textbf{$q=\frac{m_1}{m_2}$} & \textbf{$\rm{\chi_{1x}}$} & \textbf{$\rm{\chi_{1y}}$} & \textbf{$\rm{\chi_{1z}}$} & \textbf{$\rm{\chi_{2x}}$} & \textbf{$\rm{\chi_{2y}}$} & \textbf{$\rm{\chi_{2z}}$} & $\chi_\text{eff}$ & \textbf{$\chi_\mathrm{p}$} \\ \hline
\textbf{1} & 0.0992 & 0.1008 & 0.6 & 0.3343 & 0.3397 & 0.35 & 0.48 & 0.48 \\ \hline
\textbf{1.5} & 0.1013 & 0.0987 & 0.6 & 0.3414 & 0.3326 & 0.35 & 0.5 & 0.3 \\ \hline
\textbf{3} & 0.1015 & 0.0984 & 0.6 & 0.3422 & 0.3318 & 0.35 & 0.54 & 0.14 \\ \hline
\textbf{5} & 0.0997 & 0.1003 & 0.6 & 0.3359 & 0.3381 & 0.35 & 0.56 & 0.14 \\ \hline
\end{tabular}
\caption{Values of dimensionless spin components ($\rm{\chi_{1x}...\chi_{2z}}$), effective spin parameter ($\chi_\text{eff}$), and precessing-spin parameter ($\chi_\mathrm{p}$) for different mass ratios when the dimensionless spin magnitude and spin angles have been fixed as follows: $a_1=0.6164, ~a_2=0.5913, ~\theta_\text{jn}=0.4606, ~\Phi_{\text{j}l}=3.7926, ~\Phi_1=0.2315, ~\Phi_2=0.9374, ~\Phi_{12}=0.0$. This has been used for studying the effect of higher modes (Sec.~\ref{subsec:effect_of_hm}) and the effect of mass ratio variation (Sec.~\ref{subsec:q_variation}).}
\label{table:fixed_spin_angles}
\end{table}

\begin{table}[ht]
\def\arraystretch{1.5}
\begin{tabular}{|c|c|c|c|c|c|c|}
\hline
\textbf{$\chi_\mathrm{eff}$} & \textbf{$\rm{\chi_{1x}}$} & \textbf{$\rm{\chi_{2x}}$} & \textbf{$\rm{\chi_{1y}}$} & \textbf{$\rm{\chi_{2y}}$} & \textbf{$\rm{\chi_{1z}}$} & \textbf{$\rm{\chi_{2z}}$} \\ \hline
\textbf{0.3} & \multirow{3}{*}{0.1013} & \multirow{3}{*}{0.3414} & \multirow{3}{*}{0.0987} & \multirow{3}{*}{0.3326} & 0.4 & 0.15 \\ \cline{1-1} \cline{6-7} 
\textbf{0.5} &  &  &  &  & 0.6 & 0.35 \\ \cline{1-1} \cline{6-7} 
\textbf{0.7} &  &  &  &  & 0.8 & 0.55 \\ \hline
\end{tabular}
\caption{Here we fix the mass ratio $q=1.5$, inclination angle $\iota=0.5162$ rad, and $\rm{x-}$ and $\rm{y-}$ components of dimensionless spin vectors while varying the $\rm{z-}$ components in order to obtain different values of $\chi_\text{eff}$. The value of $\chi_\mathrm{p}$ remains constant at 0.3. This has been used to study the effect of $\chi_\text{eff}$ on $\delta\kappa_s$ (Sec.~\ref{subsec:chi_eff_variation}).}
\label{table:fixed_chip_q}
\end{table}

\begin{table}[ht!]
\def\arraystretch{1.5}
\begin{tabular}{|c|c|c|c|c|c|c|}
\hline
\textbf{$\chi_\mathrm{p}$} & \textbf{$\rm{\chi_{1x}}$} & \textbf{$\rm{\chi_{2x}}$} & \textbf{$\rm{\chi_{1y}}$} & \textbf{$\rm{\chi_{2y}}$} & \textbf{$\rm{\chi_{1z}}$} & \textbf{$\rm{\chi_{2z}}$} \\ \hline
\textbf{0.3} & 0.2792 & 0.1 & 0.11 & 0.1 & \multirow{3}{*}{0.56} & \multirow{3}{*}{0.32} \\ \cline{1-5}
\textbf{0.5} & 0.4 & 0.2 & 0.3 & 0.2 &  &  \\ \cline{1-5}
\textbf{0.7} & 0.5524 & 0.3 & 0.43 & 0.3 &  &  \\ \hline
\end{tabular}
\caption{Here we fix the mass ratio $q=3$, inclination angle $\iota=0.5149$ rad, and $\rm{z-}$ components of the dimensionless spin vectors while varying the $\rm{x-}$ and $\rm{y-}$ components in order to obtain different values of $\chi_\mathrm{p}$. The value of $\chi_\text{eff}$ remains constant at 0.5. This has been used to study the effect of precession on $\delta\kappa_s$ (Sec.~\ref{subsec:chi_p_variation}).}
\label{table:fixed_chieff_q}
\end{table}

\subsection{Effect of higher modes  and possible systematic biases}
\label{subsec:effect_of_hm}

In Fig.~\ref{fig:hm-inj-post}, the $\delta\kappa_s$ posteriors are shown for simulated binary signals with a total mass of $30~M_{\odot}$ and mass ratios $q=1, 1.5, 3, 5$. We fix the spin magnitudes ($a_1=0.6164, ~a_2=0.5913$), spin angles ($\Phi_{\text{j}l}=3.7926, ~\Phi_1=0.2315, ~\Phi_2=0.9374, ~\Phi_{12}=0.0$), and inclination angle ($\theta_\text{jn}=0.4606$), taking a different mass ratio in each case, leading to different values for the dimensionless spin components ($\rm{\chi_{1x}, \chi_{1y}, \chi_{1z}, \chi_{2x}, \chi_{2y}, \chi_{2z}}$) and hence different values of $\chi_\text{eff}$ and $\chi_\mathrm{p}$ as listed in Table~\ref{table:fixed_spin_angles}.
The histograms in each plot correspond to the difference in the waveform model used in analysing these signals. For instance, we generate simulations assuming {\tt IMRPhenomXPHM} model and analyse them using the same (orange), {\tt{IMRPhenomXHM}} (grey), and {\tt{IMRPhenomXP}} (blue) as shown in Fig.~\ref{fig:hm-inj-post}. The aim is to demonstrate the importance of using a waveform model with higher modes in measuring $\delta\kappa_s$ and to examine the possible biases that could arise by not including them. 

%effect of mass ratio XPHM-XPHM and other cases
We see the significance of higher modes when we go to higher mass ratio binaries. More precisely, the posteriors are tightly constrained to the true value ($\delta\kappa_s=0$) as the binary becomes asymmetric, and the $90\%$ bound on $\delta\kappa_s$ improves from 1.3 to 0.18 (nearly seven times) when moving from near-equal mass binary ($q=1.5$) to the most asymmetric binary ($q=5$) when using the higher mode waveform \texttt{IMRPhenomXHM}. For comparison, the improvement is only three times when we use the dominant mode waveform \texttt{IMRPhenomXP}. For both $q=3$ and $q=5$ cases, {\tt IMRPhenomXHM} constraints are better than the other models. This is because the relative contribution from the higher modes increases for higher mass ratios, and hence \texttt{IMRPhenomXHM} performs better than \texttt{IMRPhenomXP}. While \texttt{IMRPhenomXPHM} is also a higher mode waveform, the six spin parameters of the binary system increase the dimensionality of the parameter space. This is not the case with \texttt{IMRPhenomXHM} as it is an aligned-spin waveform model and has only two spin parameters. Since the system is only mildly precessing ($\rm{\chi_\mathrm{p} \sim 0.14}$),\footnote{See Table \ref{table:fixed_spin_angles} for values of other spin parameters.} the additional spin parameters do not contribute much toward improving the bounds on $\delta\kappa_s$, and the four extra parameters in \texttt{IMRPhenomXPHM} add a disadvantage 
%\ckm{Are you saying the recovery is poor with complete WF?} \divya{Yes. The parameter space is too bloated and the additional parameters are not adding much value for this case.} 
compared to recovery with \texttt{IMRPhenomXHM}. We have already discussed the effects of spin-precession in detail in Sec. \ref{subsec:chi_p_variation}.\\
\\
While all three waveforms are unable to recover the injected $\delta\kappa_s=0$ value for $q=1$ case, we believe that this is because of the prior railing effect, which arises when the injection is exactly at $q=1$. Since nearly the entire posterior volume lies outside the injected value of mass ratio (see Fig.~\ref{fig:corner_q1_xphm-xphm}), this causes a bias in the recovery of the chirp mass parameter. Given that both $\chi_\text{eff}$ and $\delta\kappa_s$ are correlated with chirp mass, it causes a bias in both of these parameters, and the injected value of $\delta\kappa_s$ is not recovered. These correlations can be seen in the corner plot (Fig.~\ref{fig:corner_q1_xphm-xphm}) where we see biases in the values of $\mathcal{M}$ and $\chi_\text{eff}$, resulting in a biased $\delta\kappa_s$ posterior.\footnote{In fact, by fixing q=1 for recovery waveforms, we see that the \texttt{IMRPhenomXPHM} injections are recovered with both \texttt{IMRPhenomXP} and \texttt{IMRPhenomXPHM} templates.}
Note also that the $\delta\kappa_s$ posteriors are well constrained on the positive side compared to the poorly constrained (with a long tail) negative side. This is likely due to our choice of aligned spin systems (i.e, with $\chi_\text{eff}>0$) in our investigations, and reverse trends are expected for anti-aligned systems ($\chi_\text{eff}<0$) as was observed in Ref.~\cite{Krishnendu:2019tjp} (see Sec.~III~B~1 there for a detailed discussion).

\begin{figure*}[ht!]
    \centering
    \includegraphics[width=0.48\linewidth]{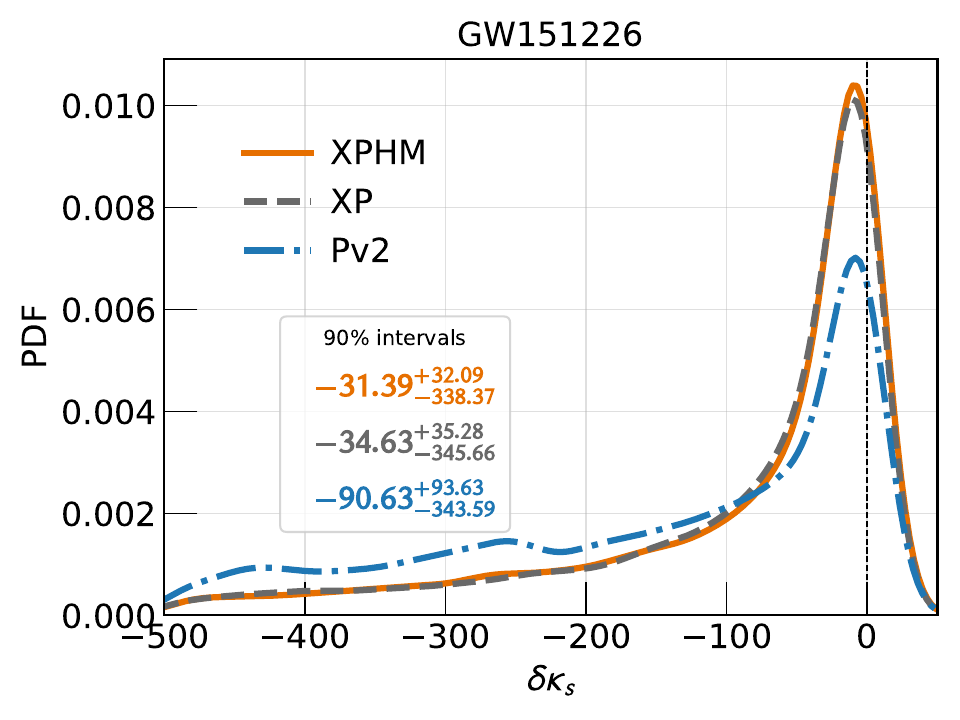}    \includegraphics[width=0.48\linewidth]{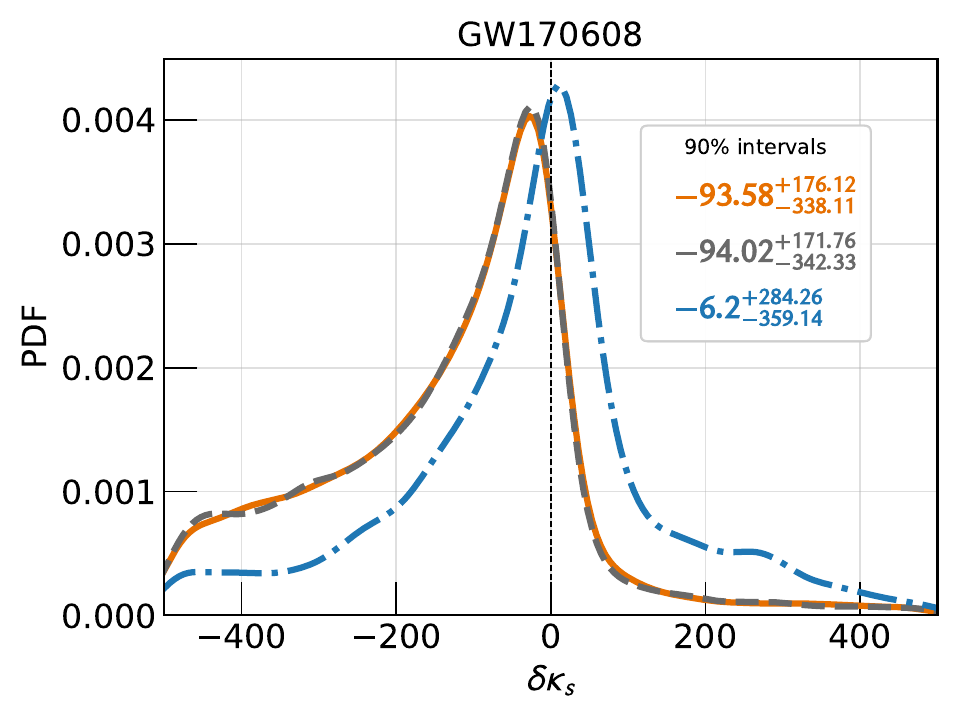}    \includegraphics[width=0.48\linewidth]{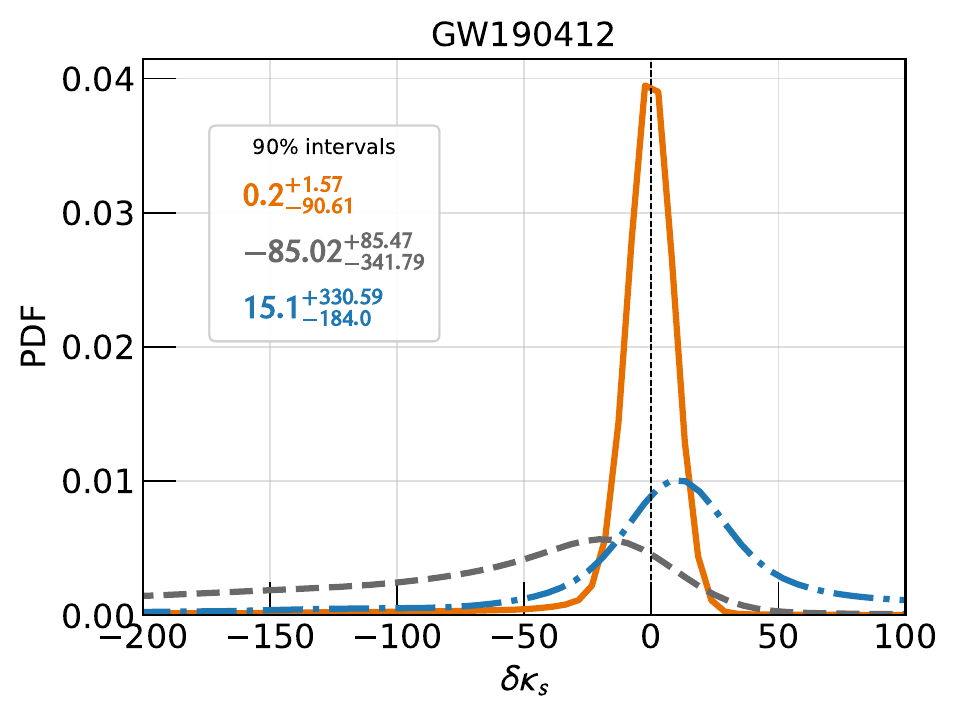}    \includegraphics[width=0.48\linewidth]{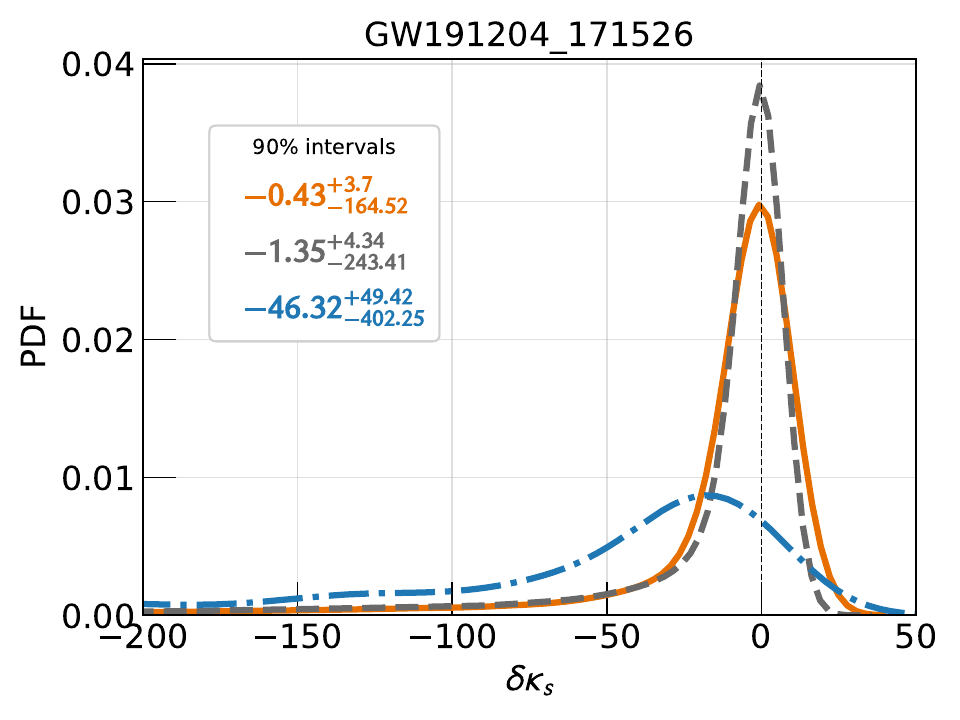}
    \caption{Posterior distributions on SIQM-deviation parameter for observed GW events. The curves labeled with {\tt{IMRPhenomPv2}} (blue) correspond to the previous results \cite{LIGOScientific:2020tif, LIGOScientific:2021sio} using {\tt{IMRPhenomPv2}} waveform model. These are being compared with the new models \texttt{IMRPhenomXP} (grey) and \texttt{IMRPhenomXPHM} (orange). The vertical dotted line indicates the BBH limit $\delta\kappa_s=0$, and the numbers written inside the plots denote the 50\% quantiles with error bounds at 5\% and 95\% quantiles for different waveform model recoveries in respective colours.}
\label{fig:posteriors_realevents}
\end{figure*}

\begin{figure*}
    \centering
    \includegraphics[width=0.8\linewidth]{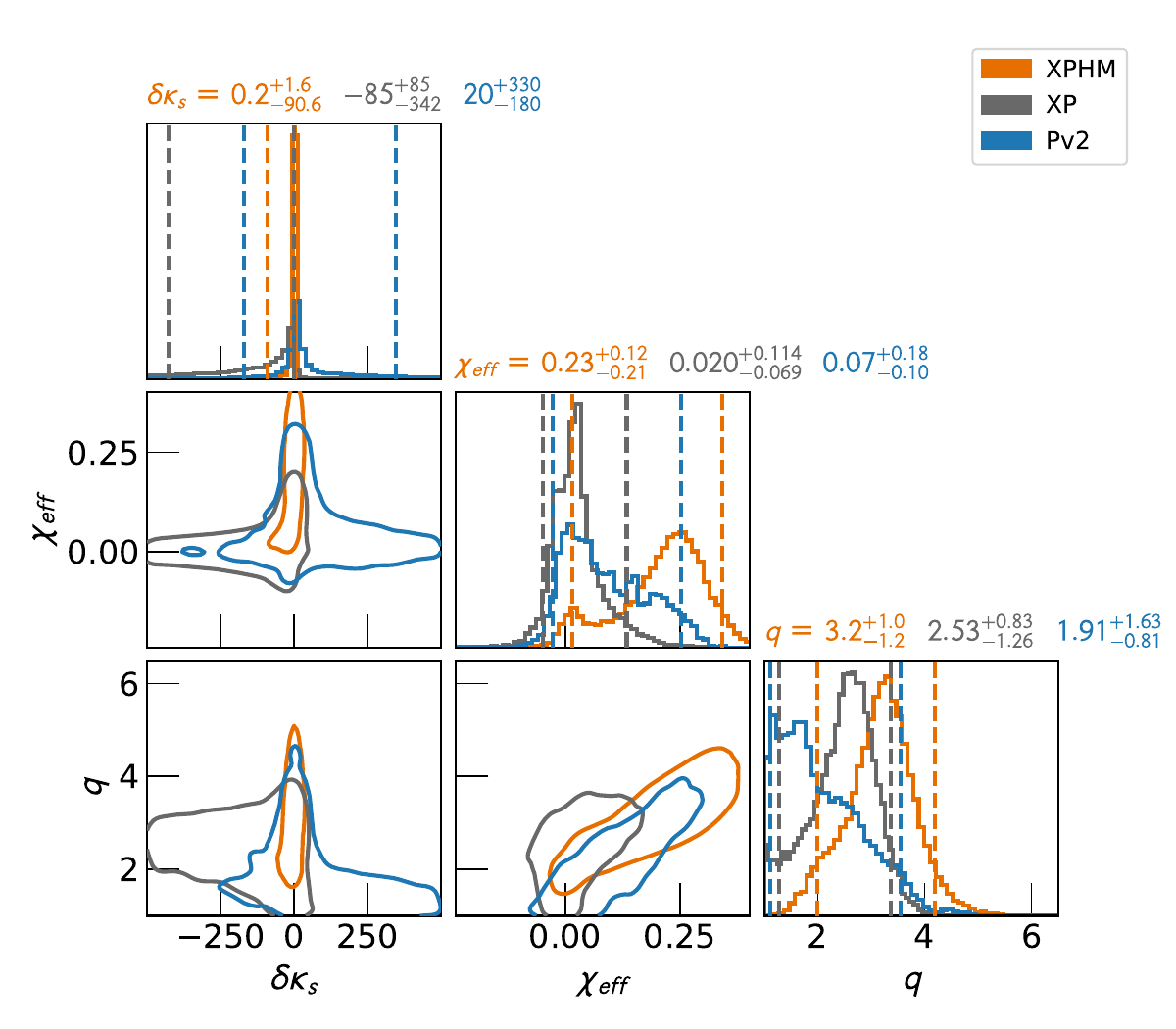}
    \caption{Corner plot for GW190412 showing posteriors on the SIQM deviation parameter ($\delta\kappa_s$), effective spin ($\chi_\text{eff}$), and mass ratio ($q=m_1/m_2$). We show on the same plot results from single spin-precessing dominant mode waveform model \texttt{IMRPhenomPv2} (blue), fully spin-precessing dominant mode waveform model \texttt{IMRPhenomXP} (grey), and fully spin-precessing HM waveform model \texttt{IMRPhenomXPHM} (orange). The histograms shown on the diagonal of the plot are 1D marginalized posteriors for the respective parameters with vertical dashed lines denoting $90\%$ credible intervals. The contours in the 2D plots are also drawn for 90\% credible interval. The titles on the 1D marginalized posteriors for respective parameters and recoveries indicate 50\% quantiles with error bounds at 5\% and 95\% quantiles.}
    \label{fig:corner_190412}
\end{figure*}

\section{Effect of spin-induced orbital precession and higher modes on real events}
\label{res:realevent}

We compare the performance of different waveform models in constraining  $\delta\kappa_s$ from the GW transient catalogs in Fig.~\ref{fig:posteriors_realevents}. We choose the events with best bounds on $\delta\kappa_s$ from GWTC-1, 2, and 3 to demonstrate the effect of waveform models on constraining the $\delta\kappa_s$ parameter. The {\tt IMRPhenomPv2} results~\cite{LIGOScientific:2020tif, LIGOScientific:2021sio} are shown in blue dot-dashed lines along with the {\tt IMRPhemomXPHM} (orange) and {\tt IMRPhemomXP} (grey) results.  

For events with nearly equal mass ($q \approx 1.4$ - 1.7) and slow spins ($\chi_\text{eff} \approx 0.05$ - 0.2), GW151226 and GW170608, even if we use the more informed models {\tt IMRPhemomXPHM} and {\tt IMRPhemomXP}, the bounds do not alter considerably compared to {\tt IMRPhenomPv2}. %\nvk{with the assumption of symmetric prior on the SIQM parameter. If we restrict to positive prior on $\delta\kappa_s$, the constraints show improvement when employing {\tt IMRPhemomXPHM} \footnote{Ref.~\cite{Lyu:2023zxv}, reports SIQM estimated from two GWTC-3 events but find that {\tt IMRPhenomPv2} better compared to {\tt IMRPhemomXPHM} with aligned-spin-induced and precession-induced SIQM effects, considering positive-only prior on the SIQM parameter.}}. 
On the other hand, the posteriors show a considerable difference from the {\tt IMRPhenomPv2} counterpart for GW190412. Note that GW190412 is the first asymmetric BBH event ($q \approx 3.75$) with an indication for moderate spins~\cite{LIGOScientific:2020stg} and higher modes. Hence, we expect the most noticeable effect on the bounds of $\delta\kappa_s$ from this event. The requirement of using waveform models with higher modes is also evident from the {\tt IMRPhemomXPHM} and  {\tt IMRPhemomXP} comparison for GW190412 as shown in Fig.~\ref{fig:posteriors_realevents}. This can also be seen in the corner plot shown in Fig.~\ref{fig:corner_190412}, where the bounds on mass ratio and $\chi_\text{eff}$ are considerably different for \texttt{IMRPhenomXPHM} compared to the dominant mode waveform models. As discussed in the previous sections, bounds on $\delta\kappa_s$ strongly depend on these parameters, and hence, for an event with non-negligible higher mode content, we observe that \texttt{IMRPhenomXPHM} performs much better. In fact, the bounds for GW190412 obtained using \texttt{IMRPhenomXPHM} exclude boson star binaries as the source, subject to the assumptions made in our study (such as neglect of the tidal corrections, assumption that $\delta\kappa_a=0$, and that spin-induced effects are accounted for only in the inspiral part of the waveform).

\section{Summary}
\label{sec:summary}

Spin-induced multipole moment-based tests were routinely employed to determine the nature of compact binary signal during the first three observing runs of the advanced LIGO and advanced Virgo detectors~\cite{LIGOScientific:2020tif, LIGOScientific:2021sio}. In this study, we extend the applicability of the test to binaries with spin-precession effects not considered in previous versions of the test and discuss the possible improvements in the measurement of the SIQM deviation parameter using a more informed waveform model containing two spin-precession effects and higher modes. 
Starting with a simulation study, we demonstrate the applicability of the SIQM test on binaries with large spin-precession and moderate mass asymmetries. We find that there are considerable differences in the bounds of $\delta\kappa_s$ obtained using \texttt{IMRPhenomPv2} compared to \texttt{IMRPhenomXP} for highly spinning systems. We also report on the improvements and biases observed in the $\delta\kappa_s$ bounds with the choice of different mass ratios and compare them between \texttt{IMRPhenomXP} and \texttt{IMRPhenomPv2} waveform models. Further, by injecting higher mode spin-precessing signals, we find that higher mode waveforms are essential when analysing GW signals with high mass asymmetries. Finally, we re-analyse selected events from GWTC-1, 2, and 3 with the most up-to-date waveform models including two-spin precession effects and higher modes. 
Our findings show that {\tt IMRPhemomXPHM} may be preferred for analysing GW events such as GW190412~\cite{LIGOScientific:2020stg}, where there is evidence for mass asymmetry and non-negligible spin effects. While the current paper studies waveform systematics on the SIQM tests for various spin and mass ratio configurations by injecting BBH waveforms consistent with GR, a detailed follow-up study with non-GR injections may be carried out in a future work. 
%\saleemc{Did we want to say non-BH waveforms instead of non-GR waveforms, to be specific? Or we meant non-GR in general?} \divya{Yes, we mean non-BH to be specific but since this is a test of GR, I think we can use the two inter-changeably here.}\ckm{Is it clear now?}
%While the aim of this paper was to study waveform systematics {\red in a limited sense}, a detailed follow-up analysis using higher mode waveforms maybe carried out for various spin configurations. Further, an analysis including non-GR injections may also be interesting.

\acknowledgments

We thank Michalis Agathos for his useful comments on the manuscript. DJ thanks Sayantani Datta, Pankaj Saini, and Sajad A. Bhat for useful discussions. We thank Anuradha Gupta, Archisman Ghosh, and Ish Gupta for the review of the waveform code, and Pankaj Saini and Sijil Jose for help with the review readiness. 
N.V.K. acknowledges the support from the Science and Engineering Research Board (SERB), Government of India, through the National Post Doctoral Fellowship Grant (Reg.~No.~PDF/2022/000379). 
M.S. acknowledges the support from the National Science Foundation with Grant Nos.~PHY-1806630, PHY-2010970, and PHY-2110238. 
A.V. is supported by the Department of Atomic Energy, Government of India, under Project No.~RTI4001. A.V. also acknowledges the support of the Natural Sciences and Engineering Research Council of Canada (NSERC) (funding reference number 568580). 
K.G.A. acknowledges support from the Department of Science and Technology and the Science and Engineering Research Board (SERB) of India via Swarnajayanti Fellowship Grant No.~DST/SJF/PSA-01/2017-18 and Core Research Grant No.~CRG/2021/004565. K.G.A. also acknowledges support from the Infosys Foundation. 
C.K.M. acknowledges the support of SERB's Core Research Grant No.~CRG/2022/007959. 
M.C. acknowledges funding from the Spanish Agencia Estatal de Investigación, Grant No.~IJC2019-041385. 
Computations were performed on the CIT cluster provided by the LIGO Laboratory. N.V.K. acknowledges Max Planck Computing and Data Facility computing cluster Cobra for early stages of the work. We acknowledge National Science Foundation Grants PHY-0757058 and PHY-0823459. This material is based upon work supported by NSF's LIGO Laboratory, which is a major facility fully funded by the National Science Foundation. We used the following software packages:
{\tt LALSuite}~\cite{lalsuite}, {\tt bilby}~\cite{Ashton:2018jfp}, {\tt bilby\textunderscore pipe}~\cite{Romero-Shaw:2020owr}, \texttt{PyCBC}~\cite{alex_nitz_2020_4134752}, {\tt NumPy}~\cite{2020Natur.585..357H}, {\tt PESummary}~\cite{Hoy:2020vys}, {\tt Matplotlib}~\cite{2007CSE.....9...90H}, {\tt Seaborn}~\cite{Waskom2021}, {\tt jupyter}~\cite{soton403913}, {\tt dynesty}~\cite{speagle2020dynesty}, and {\tt corner}~\cite{corner}.
This document has LIGO preprint No. {\tt LIGO-P2300337}. 

\appendix

\section{Additional information for injection runs}
\label{appendix1:priors}

The parameter space for precessing spin recoveries includes the following parameters: inverse mass ratio ($q_\text{inv} = m_2/m_1$),\footnote{We use the word \textit{inverse} here to indicate that this is inverse of the \textit{mass ratio} we have used throughout the paper. Please note that \texttt{bilby} uses the term \texttt{mass\_ratio} for this.} chirp mass ($\mathcal{M}$), luminosity distance ($d_L$), cos of inclination angle ($\cos{\theta_\text{jn}}$),\footnote{Please note that this is the inclination angle as observed in the detector frame, i.e. the angle between the line-of-sight $\vec{\text{n}}$ and total orbital momentum vector $\vec{\text{j}}$ of the binary. For precessing systems, value of $\theta_\text{jn}$ is different from $\iota$ which is the angle between the orbital angular momentum vector $\Vec{l}$ and normal vector $\Vec{\text{n}}.$} geocentric time ($t_c$), phase angle ($\phi_c$), dimensionless spin magnitudes ($a_1$, $a_2$), spin angles ($\Phi_1$, $\Phi_2$, $\Phi_{12}$, $\Phi_{\text{j}l}$), right ascension ($\alpha$), declination ($\delta$), polarization angle ($\psi$), and symmetric combination of SIQM deviation parameter ($\delta\kappa_s$). The priors for chirp mass have been modified according to the injection value such that the lower limit is $\sim 0.5$ times the injected value, whereas the upper limit is $\sim 3$ times the injected value. Accordingly, we have put constraints on component masses. The priors for the rest of the parameters are given in Table \ref{table:priors}. For aligned spin recovery, instead of the six spin parameters mentioned above, we use the parameters $\chi_1$ and $\chi_2$ using the \texttt{AlignedSpin} prior mentioned in \texttt{bilby}, which puts a uniform range of [0,0.99] on the spin magnitudes. For real event analyses, we have used the same priors as were used for the respective analyses of the events in GWTC-2~\cite{LIGOScientific:2020tif} and GWTC-3~\cite{LIGOScientific:2021sio} TGR papers. All the injection runs mentioned in this paper are for a binary black hole system ($\delta\kappa_s=0=\rm{\delta\kappa_a}$) with total mass of 30 $M_\odot$. The distance has been scaled in each case such that we get a fixed network SNR of 40. We have arbitrarily chosen the angles $\alpha$, $\delta$, $\psi$, and $\phi_c$ as 0, and $t_\text{gps}=1126259462$ s.

\begin{table}[ht]
\def\arraystretch{1.3}
\begin{tabular}{|c|c|c|}
\hline
\textbf{Parameter} & \textbf{Prior} & \textbf{Range} \\ \hline
$q_\text{inv}$ & Uniform & $0.125 \text{ - } 1$ \\ \hline
$d_L$ & \begin{tabular}[c]{@{}c@{}}Uniform \\ Source Frame\end{tabular} & $100 \text{ - } 1000$ Mpc \\ \hline
$\cos{\theta_\text{jn}}$ & Uniform & $-1 \text{ - } 1$ \\ \hline
$t_c$ & Uniform & $t_\text{gps}+(-2 \text{ - } 2)$~s \\ \hline
$\phi_c$ & Uniform & $0 \text{ - } 2\pi$ \\ \hline
$a_1$, $a_2$ & Uniform & $0 \text{ - } 0.99$ \\ \hline
$\Phi_1$, $\Phi_2$\textsuperscript{\ref{tilt footnote}} & Uniform sine & $0 \text{ - } \pi$ \\ \hline
$\Phi_{12}$ & Uniform & $0 \text{ - } 2\pi$ \\ \hline
$\Phi_{\text{j}l}$ & Uniform & $0 \text{ - } 2\pi$ \\ \hline
$\alpha$ & Uniform & $0 \text{ - } 2\pi$ \\ \hline
$\delta$ & Uniform cos & $-\pi/2 \text{ - } \pi/2$ \\ \hline
$\psi$ & Uniform & $0 \text{ - } \pi$ \\ \hline
$\delta\kappa_s$ & Uniform & $-500 \text{ - } 500$ \\ \hline
\end{tabular}
\caption{Priors for parameters used in precessing spin recoveries.}
\label{table:priors}
\end{table}

\footnotetext{\label{tilt footnote}Denoted by the spin angles \texttt{tilt1} and \texttt{tilt2} in \texttt{bilby} \cite{Ashton:2018jfp}.}

\begin{figure*}[ht]
    \centering
    \includegraphics[trim=40 40 70 60, clip, width=0.49\linewidth]{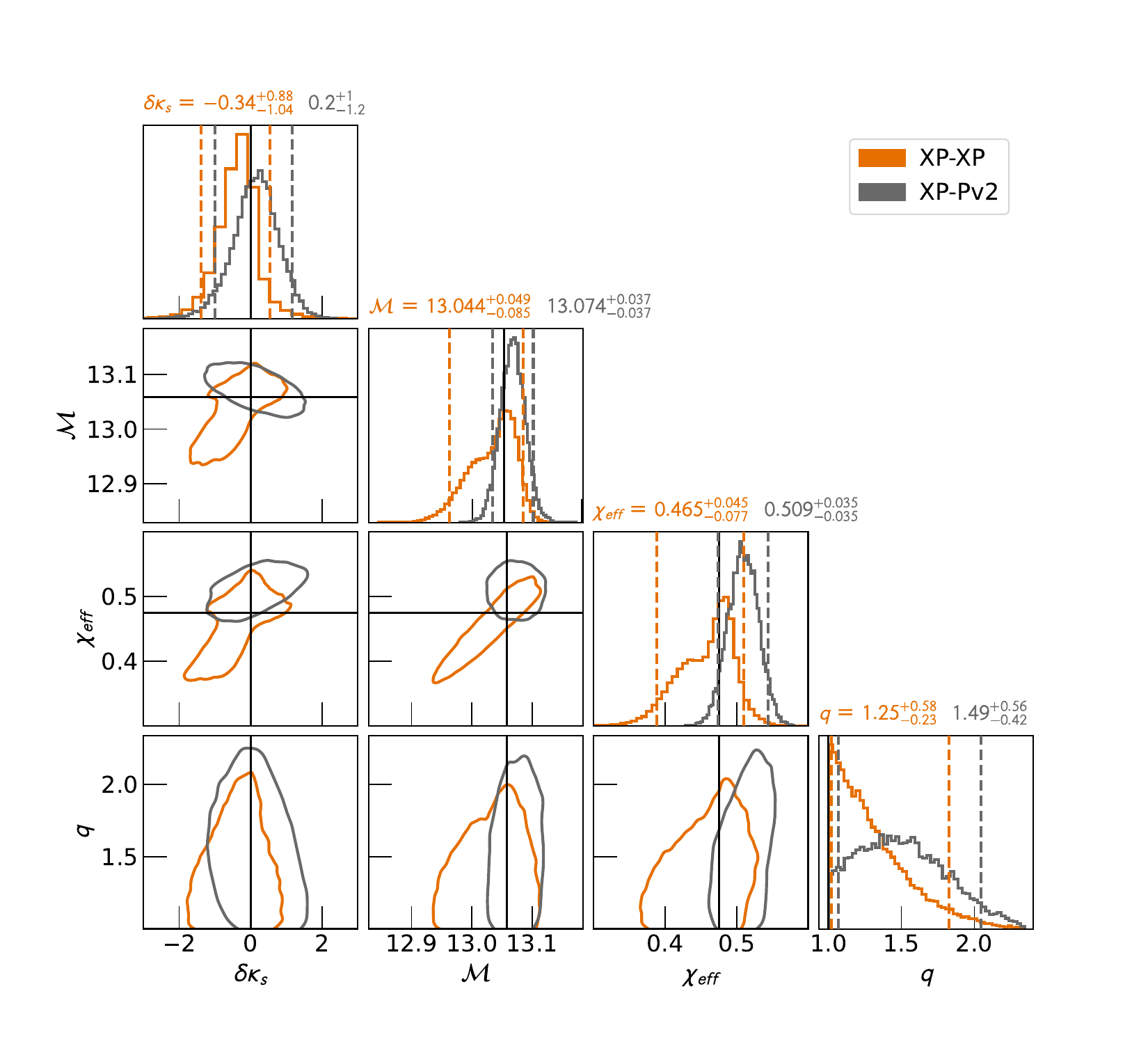}
    \includegraphics[trim=40 40 70 60, clip, width=0.49\linewidth]{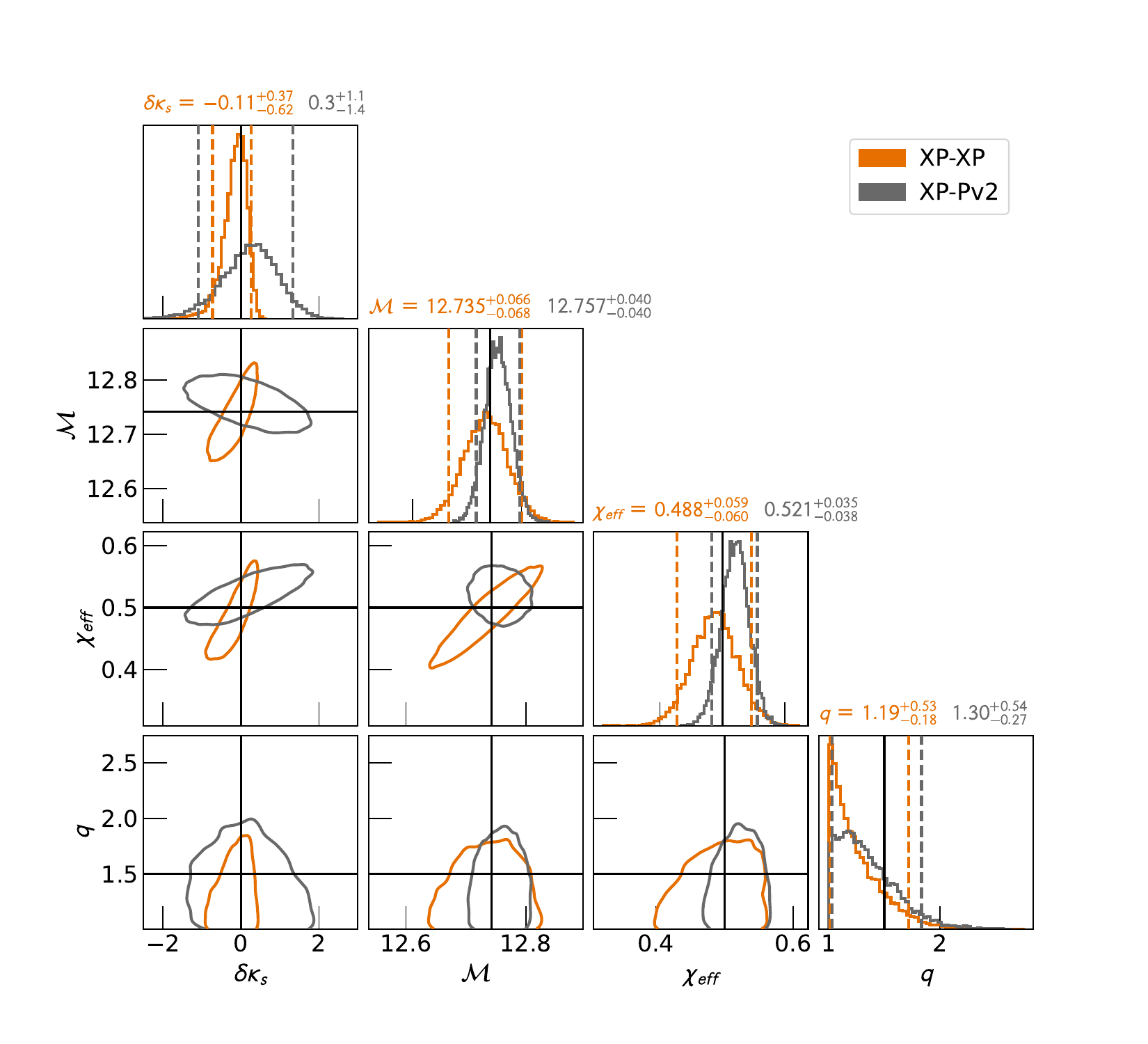}
    \includegraphics[trim=40 40 70 60, clip, width=0.49\linewidth]{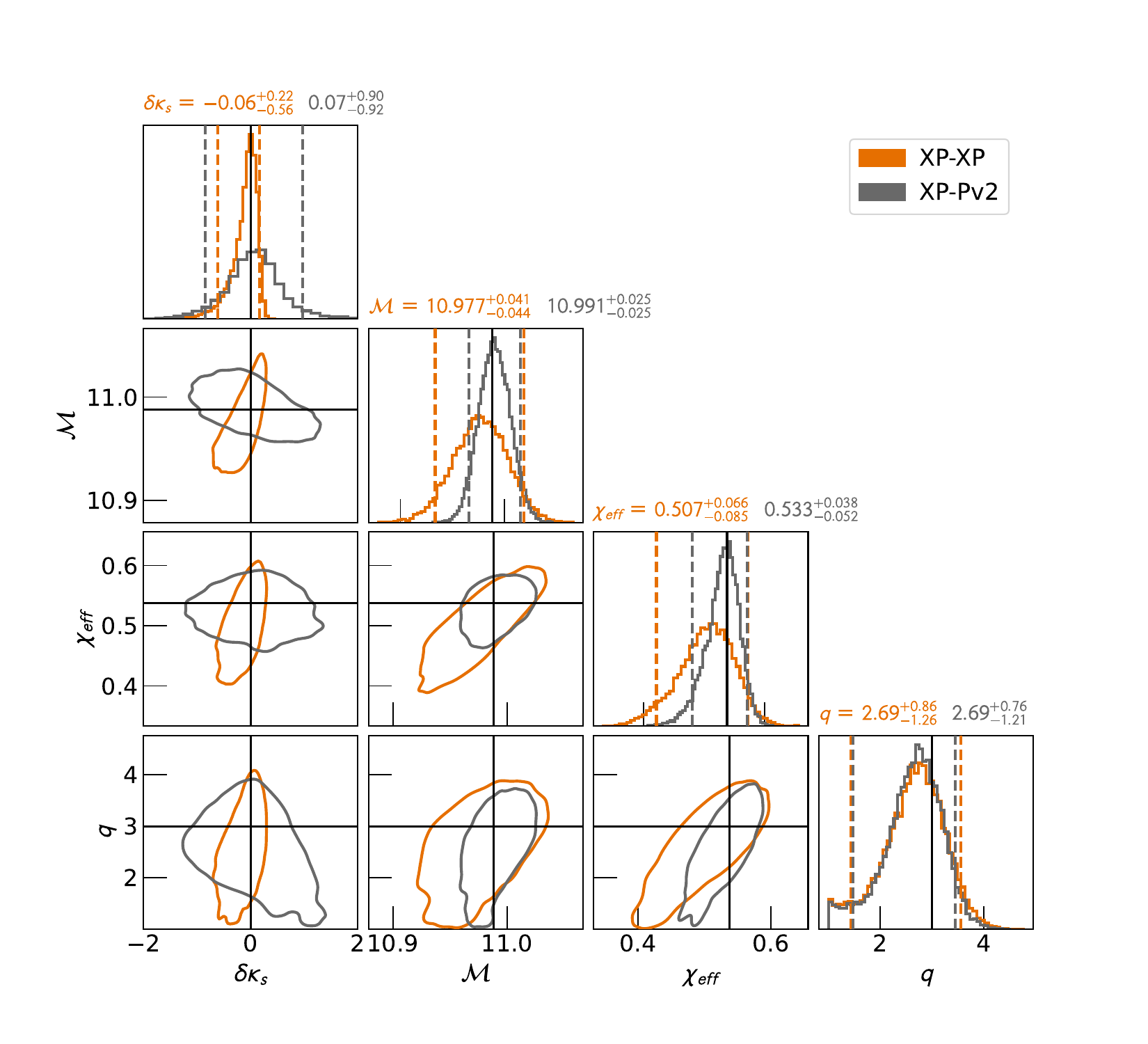}
    \includegraphics[trim=40 40 70 60, clip, width=0.49\linewidth]{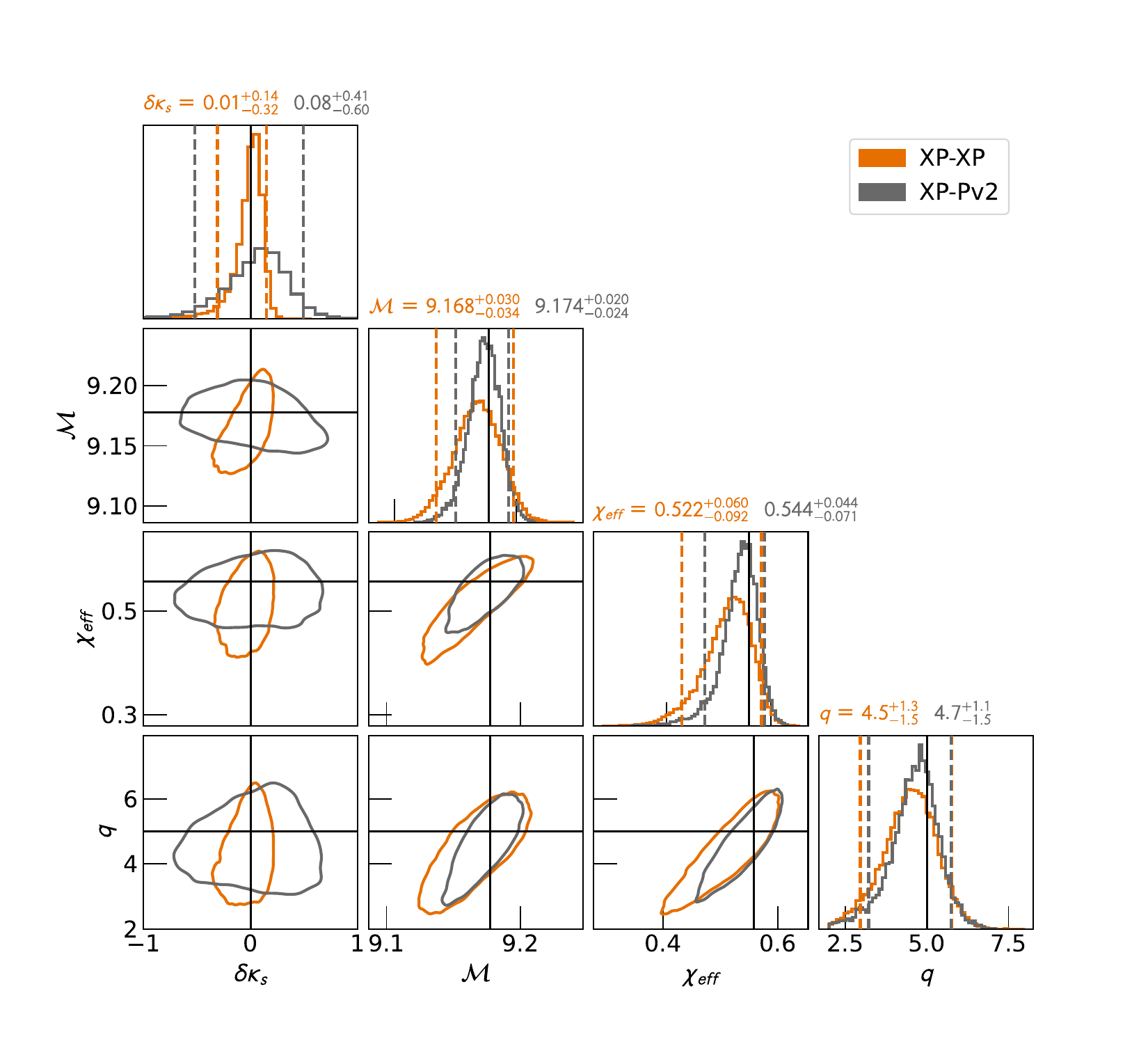}
    \caption{Corner plot for fixed spin magnitudes and angles given in Table \ref{table:fixed_spin_angles} when the mass ratio is varied as 1 (top-left), 1.5 (top-right), 3 (bottom-left), and 5 (bottom-right). The plots show 1D and 2D posteriors for the SIQM deviation parameter ($\delta\kappa_s$), chirp mass ($\mathcal{M}$), effective spin ($\chi_\text{eff}$), and mass ratio ($q$). The injections are performed using the fully spin-precessing dominant mode waveform (\texttt{IMRPhenomXP}) and recovered with the same (orange) as well as with single spin-precessing dominant mode waveform \texttt{IMRPhenomPv2} (grey). The histograms shown on the diagonal of the plots are 1D marginalized posteriors for the respective parameters with vertical dashed lines denoting $90\%$ credible intervals and black lines indicating the injected value of the parameters. The contours in the 2D plots are also drawn for 90\% credible interval. The titles on the 1D marginalized posteriors for respective parameters and recoveries indicate 50\% quantiles with error bounds at 5\% and 95\% quantiles.}
    \label{fig:corner_fixed_spin_anlges_vary_q}
\end{figure*}

\begin{figure*}[ht]
    \centering
    \includegraphics[trim=10 10 0 30, clip, width=0.49\linewidth]{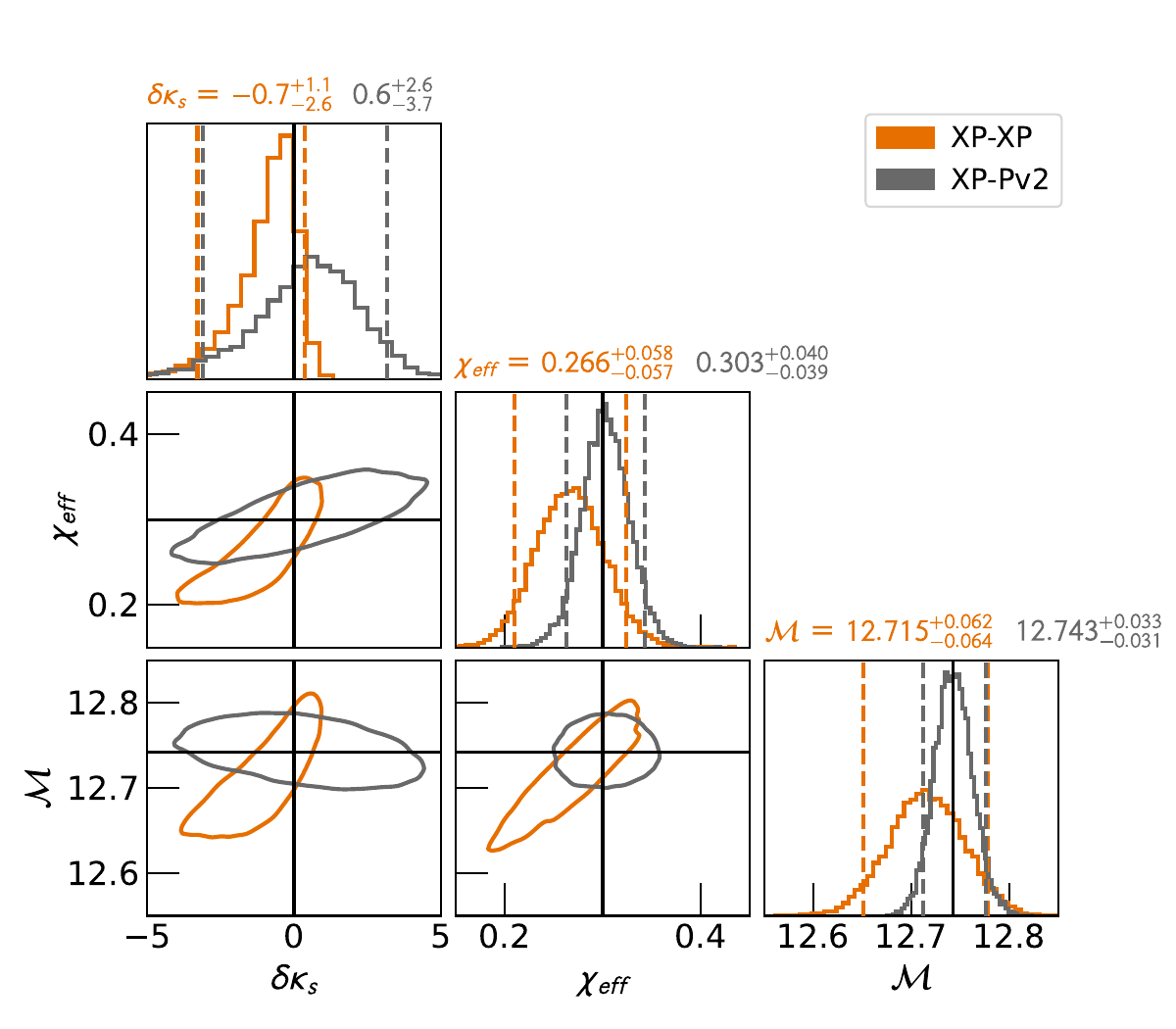}
    \includegraphics[trim=10 10 0 30, clip, width=0.49\linewidth]{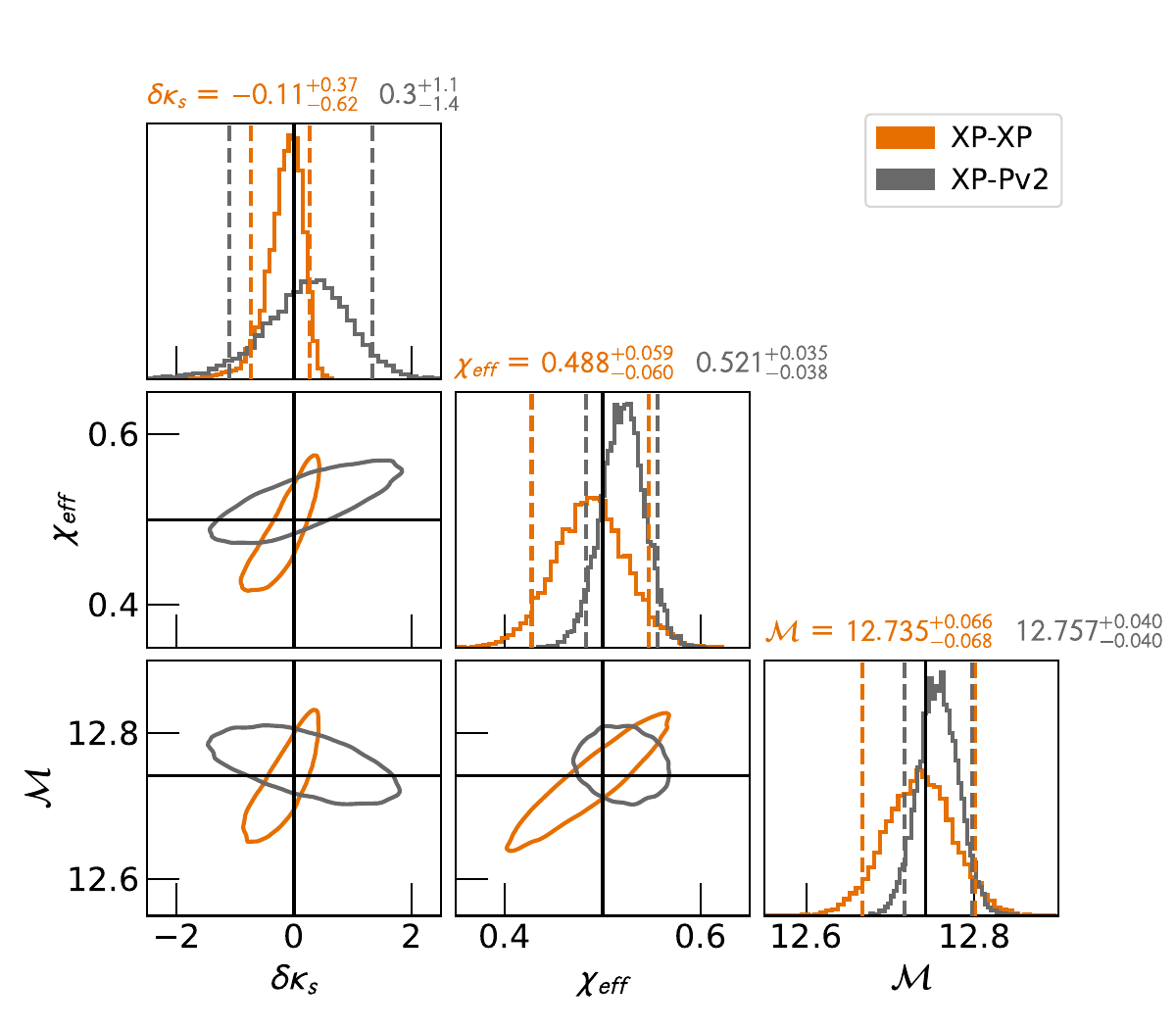}
    \includegraphics[trim=10 10 0 30, clip, width=0.49\linewidth]{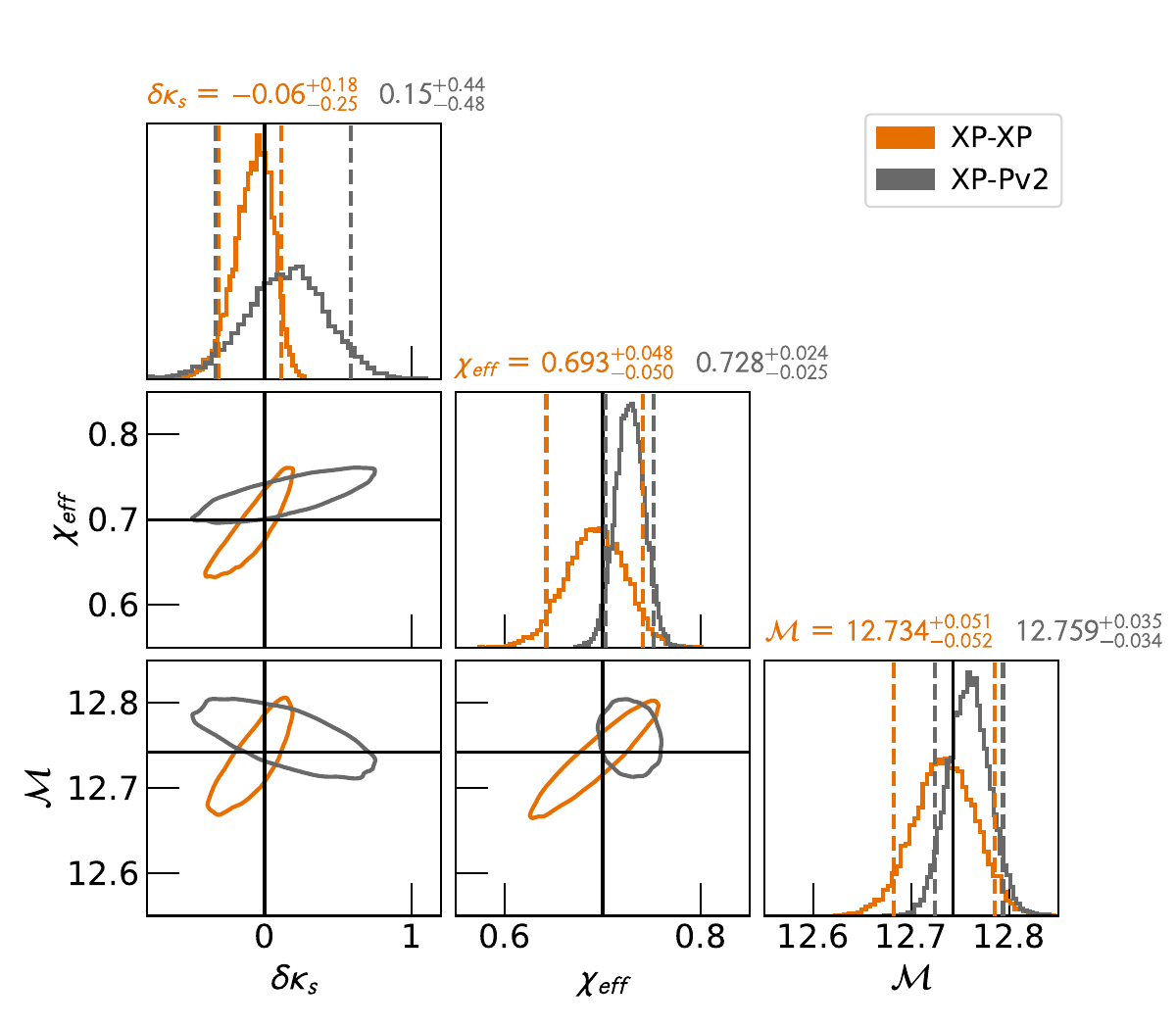}
    \caption{Same as Fig. \ref{fig:corner_fixed_spin_anlges_vary_q} but for fixed $q=1.5$, $\rm{\chi_\mathrm{p}=0.3}$ when $\chi_\text{eff}$ is varied as 0.3 (top-left), 0.5 (top-right), and 0.7 (bottom).}
    \label{fig:corner_vary_chieff}
\end{figure*}

\begin{figure*}
    \centering
    \includegraphics[width=\linewidth]{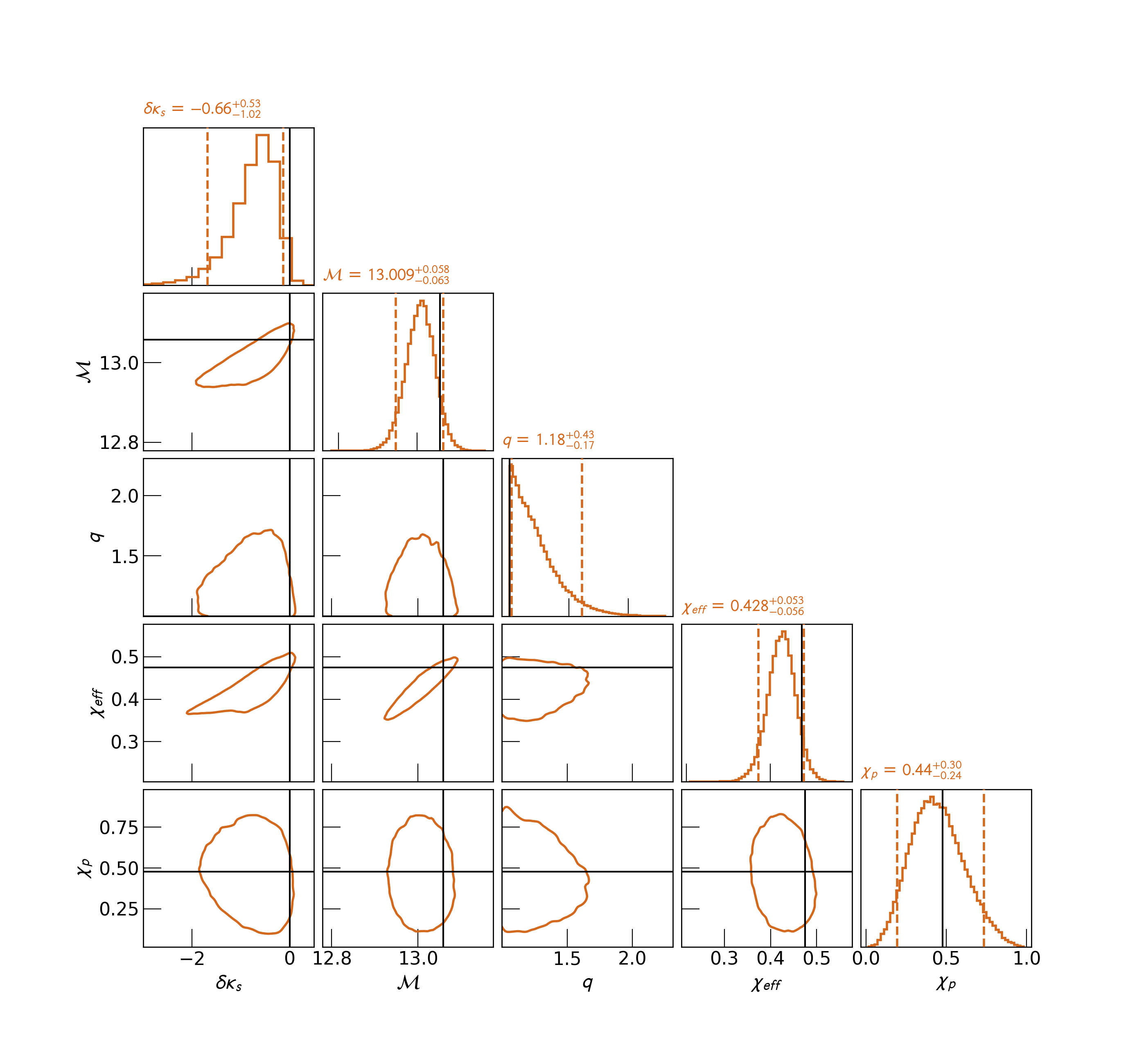}
    \caption{Corner plot for the equal-mass case ($q=1$) showing posterior on the SIQM deviation parameter ($\delta\kappa_s$), chirp mass ($\mathcal{M}$), mass ratio ($q$), the effective spin parameter ($\chi_\text{eff}$), and the spin-precession parameter ($\chi_\mathrm{p}$). We have used a fully spin-precessing waveform model including HMs (\texttt{IMRPhenomXPHM}) for both injection and recovery. The histograms shown on the diagonal of the plot are 1D marginalized posteriors for the respective parameters with vertical dashed lines denoting $90\%$ credible intervals. The contours in the 2D plots are also drawn for 90\% credible interval. The black lines denote the injected value of the parameters, and the titles on the 1D marginalized posteriors for respective parameters indicate 50\% quantiles with error bounds at 5\% and 95\% quantiles.}
    \label{fig:corner_q1_xphm-xphm}
\end{figure*}

\clearpage

\bibliography{master_refs}

\begin{thebibliography}{95}%
\makeatletter
\providecommand \@ifxundefined [1]{%
 \@ifx{#1\undefined}
}%
\providecommand \@ifnum [1]{%
 \ifnum #1\expandafter \@firstoftwo
 \else \expandafter \@secondoftwo
 \fi
}%
\providecommand \@ifx [1]{%
 \ifx #1\expandafter \@firstoftwo
 \else \expandafter \@secondoftwo
 \fi
}%
\providecommand \natexlab [1]{#1}%
\providecommand \enquote  [1]{``#1''}%
\providecommand \bibnamefont  [1]{#1}%
\providecommand \bibfnamefont [1]{#1}%
\providecommand \citenamefont [1]{#1}%
\providecommand \href@noop [0]{\@secondoftwo}%
\providecommand \href [0]{\begingroup \@sanitize@url \@href}%
\providecommand \@href[1]{\@@startlink{#1}\@@href}%
\providecommand \@@href[1]{\endgroup#1\@@endlink}%
\providecommand \@sanitize@url [0]{\catcode `\\12\catcode `\$12\catcode
  `\&12\catcode `\#12\catcode `\^12\catcode `\_12\catcode `\%12\relax}%
\providecommand \@@startlink[1]{}%
\providecommand \@@endlink[0]{}%
\providecommand \url  [0]{\begingroup\@sanitize@url \@url }%
\providecommand \@url [1]{\endgroup\@href {#1}{\urlprefix }}%
\providecommand \urlprefix  [0]{URL }%
\providecommand \Eprint [0]{\href }%
\providecommand \doibase [0]{http://dx.doi.org/}%
\providecommand \selectlanguage [0]{\@gobble}%
\providecommand \bibinfo  [0]{\@secondoftwo}%
\providecommand \bibfield  [0]{\@secondoftwo}%
\providecommand \translation [1]{[#1]}%
\providecommand \BibitemOpen [0]{}%
\providecommand \bibitemStop [0]{}%
\providecommand \bibitemNoStop [0]{.\EOS\space}%
\providecommand \EOS [0]{\spacefactor3000\relax}%
\providecommand \BibitemShut  [1]{\csname bibitem#1\endcsname}%
\let\auto@bib@innerbib\@empty
%</preamble>
\bibitem [{\citenamefont {Abbott}\ \emph
  {et~al.}(2016{\natexlab{a}})\citenamefont {Abbott} \emph
  {et~al.}}]{Abbott:2016blz}%
  \BibitemOpen
  \bibfield  {author} {\bibinfo {author} {\bibfnamefont {B.}~\bibnamefont
  {Abbott}} \emph {et~al.} (\bibinfo {collaboration} {LIGO Scientific,
  Virgo}),\ }\href {\doibase 10.1103/PhysRevLett.116.061102} {\bibfield
  {journal} {\bibinfo  {journal} {Phys. Rev. Lett.}\ }\textbf {\bibinfo
  {volume} {116}},\ \bibinfo {eid} {061102} (\bibinfo {year}
  {2016}{\natexlab{a}})},\ \Eprint {http://arxiv.org/abs/1602.03837}
  {arXiv:1602.03837 [gr-qc]} \BibitemShut {NoStop}%
\bibitem [{\citenamefont {Abbott}\ \emph {et~al.}(2017)\citenamefont {Abbott}
  \emph {et~al.}}]{TheLIGOScientific:2017qsa}%
  \BibitemOpen
  \bibfield  {author} {\bibinfo {author} {\bibfnamefont {B.~P.}\ \bibnamefont
  {Abbott}} \emph {et~al.} (\bibinfo {collaboration} {LIGO Scientific,
  Virgo}),\ }\href {\doibase 10.1103/PhysRevLett.119.161101} {\bibfield
  {journal} {\bibinfo  {journal} {Phys. Rev. Lett.}\ }\textbf {\bibinfo
  {volume} {119}},\ \bibinfo {eid} {161101} (\bibinfo {year} {2017})},\ \Eprint
  {http://arxiv.org/abs/1710.05832} {arXiv:1710.05832 [gr-qc]} \BibitemShut
  {NoStop}%
\bibitem [{\citenamefont {Abbott}\ \emph {et~al.}(2019)\citenamefont {Abbott}
  \emph {et~al.}}]{LIGOScientific:2018mvr}%
  \BibitemOpen
  \bibfield  {author} {\bibinfo {author} {\bibfnamefont {B.~P.}\ \bibnamefont
  {Abbott}} \emph {et~al.} (\bibinfo {collaboration} {LIGO Scientific and Virgo
  Collaborations}),\ }\href {\doibase 10.1103/PhysRevX.9.031040} {\bibfield
  {journal} {\bibinfo  {journal} {Phys. Rev. X}\ }\textbf {\bibinfo {volume}
  {9}},\ \bibinfo {pages} {031040} (\bibinfo {year} {2019})},\ \Eprint
  {http://arxiv.org/abs/1811.12907} {arXiv:1811.12907 [astro-ph.HE]}
  \BibitemShut {NoStop}%
\bibitem [{\citenamefont {Abbott}\ \emph
  {et~al.}(2020{\natexlab{a}})\citenamefont {Abbott} \emph
  {et~al.}}]{LIGOScientific:2020aai}%
  \BibitemOpen
  \bibfield  {author} {\bibinfo {author} {\bibfnamefont {B.~P.}\ \bibnamefont
  {Abbott}} \emph {et~al.} (\bibinfo {collaboration} {LIGO Scientific,
  Virgo}),\ }\href {\doibase 10.3847/2041-8213/ab75f5} {\bibfield  {journal}
  {\bibinfo  {journal} {Astrophys. J. Lett.}\ }\textbf {\bibinfo {volume}
  {892}},\ \bibinfo {pages} {L3} (\bibinfo {year} {2020}{\natexlab{a}})},\
  \Eprint {http://arxiv.org/abs/2001.01761} {arXiv:2001.01761 [astro-ph.HE]}
  \BibitemShut {NoStop}%
\bibitem [{\citenamefont {Abbott}\ \emph
  {et~al.}(2021{\natexlab{a}})\citenamefont {Abbott} \emph
  {et~al.}}]{LIGOScientific:2020ibl}%
  \BibitemOpen
  \bibfield  {author} {\bibinfo {author} {\bibfnamefont {R.}~\bibnamefont
  {Abbott}} \emph {et~al.} (\bibinfo {collaboration} {LIGO Scientific and Virgo
  Collaborations}),\ }\href {\doibase 10.1103/PhysRevX.11.021053} {\bibfield
  {journal} {\bibinfo  {journal} {Phys. Rev. X}\ }\textbf {\bibinfo {volume}
  {11}},\ \bibinfo {pages} {021053} (\bibinfo {year} {2021}{\natexlab{a}})},\
  \Eprint {http://arxiv.org/abs/2010.14527} {arXiv:2010.14527 [gr-qc]}
  \BibitemShut {NoStop}%
\bibitem [{\citenamefont {Aasi}\ \emph
  {et~al.}(2015{\natexlab{a}})\citenamefont {Aasi} \emph
  {et~al.}}]{LIGOScientific:2014pky}%
  \BibitemOpen
  \bibfield  {author} {\bibinfo {author} {\bibfnamefont {J.}~\bibnamefont
  {Aasi}} \emph {et~al.} (\bibinfo {collaboration} {LIGO Scientific}),\ }\href
  {\doibase 10.1088/0264-9381/32/7/074001} {\bibfield  {journal} {\bibinfo
  {journal} {Class. Quant. Grav.}\ }\textbf {\bibinfo {volume} {32}},\ \bibinfo
  {pages} {074001} (\bibinfo {year} {2015}{\natexlab{a}})},\ \Eprint
  {http://arxiv.org/abs/1411.4547} {arXiv:1411.4547 [gr-qc]} \BibitemShut
  {NoStop}%
\bibitem [{\citenamefont {{Buikema}}\ \emph {et~al.}(2020)\citenamefont
  {{Buikema}} \emph {et~al.}}]{2020PhRvD.102f2003B}%
  \BibitemOpen
  \bibfield  {author} {\bibinfo {author} {\bibfnamefont {A.}~\bibnamefont
  {{Buikema}}} \emph {et~al.},\ }\href {\doibase 10.1103/PhysRevD.102.062003}
  {\bibfield  {journal} {\bibinfo  {journal} {\prd}\ }\textbf {\bibinfo
  {volume} {102}},\ \bibinfo {eid} {062003} (\bibinfo {year} {2020})},\ \Eprint
  {http://arxiv.org/abs/2008.01301} {arXiv:2008.01301 [astro-ph.IM]}
  \BibitemShut {NoStop}%
\bibitem [{\citenamefont {Acernese}\ \emph
  {et~al.}(2015{\natexlab{a}})\citenamefont {Acernese} \emph
  {et~al.}}]{VIRGO:2014yos}%
  \BibitemOpen
  \bibfield  {author} {\bibinfo {author} {\bibfnamefont {F.}~\bibnamefont
  {Acernese}} \emph {et~al.} (\bibinfo {collaboration} {VIRGO}),\ }\href
  {\doibase 10.1088/0264-9381/32/2/024001} {\bibfield  {journal} {\bibinfo
  {journal} {Class. Quant. Grav.}\ }\textbf {\bibinfo {volume} {32}},\ \bibinfo
  {pages} {024001} (\bibinfo {year} {2015}{\natexlab{a}})},\ \Eprint
  {http://arxiv.org/abs/1408.3978} {arXiv:1408.3978 [gr-qc]} \BibitemShut
  {NoStop}%
\bibitem [{\citenamefont {Acernese}\ \emph {et~al.}(2019)\citenamefont
  {Acernese} \emph {et~al.}}]{PhysRevLett.123.231108}%
  \BibitemOpen
  \bibfield  {author} {\bibinfo {author} {\bibfnamefont {F.}~\bibnamefont
  {Acernese}} \emph {et~al.} (\bibinfo {collaboration} {Virgo Collaboration}),\
  }\href {\doibase 10.1103/PhysRevLett.123.231108} {\bibfield  {journal}
  {\bibinfo  {journal} {Phys. Rev. Lett.}\ }\textbf {\bibinfo {volume} {123}},\
  \bibinfo {pages} {231108} (\bibinfo {year} {2019})}\BibitemShut {NoStop}%
\bibitem [{\citenamefont {Akutsu}\ \emph {et~al.}(2021)\citenamefont {Akutsu}
  \emph {et~al.}}]{KAGRA:2020tym}%
  \BibitemOpen
  \bibfield  {author} {\bibinfo {author} {\bibfnamefont {T.}~\bibnamefont
  {Akutsu}} \emph {et~al.} (\bibinfo {collaboration} {KAGRA}),\ }\href
  {\doibase 10.1093/ptep/ptaa125} {\bibfield  {journal} {\bibinfo  {journal}
  {PTEP}\ }\textbf {\bibinfo {volume} {2021}},\ \bibinfo {pages} {05A101}
  (\bibinfo {year} {2021})},\ \Eprint {http://arxiv.org/abs/2005.05574}
  {arXiv:2005.05574 [physics.ins-det]} \BibitemShut {NoStop}%
\bibitem [{\citenamefont {Aso}\ \emph {et~al.}(2013)\citenamefont {Aso},
  \citenamefont {Michimura}, \citenamefont {Somiya}, \citenamefont {Ando},
  \citenamefont {Miyakawa}, \citenamefont {Sekiguchi}, \citenamefont
  {Tatsumi},\ and\ \citenamefont {Yamamoto}}]{Aso:2013eba}%
  \BibitemOpen
  \bibfield  {author} {\bibinfo {author} {\bibfnamefont {Y.}~\bibnamefont
  {Aso}}, \bibinfo {author} {\bibfnamefont {Y.}~\bibnamefont {Michimura}},
  \bibinfo {author} {\bibfnamefont {K.}~\bibnamefont {Somiya}}, \bibinfo
  {author} {\bibfnamefont {M.}~\bibnamefont {Ando}}, \bibinfo {author}
  {\bibfnamefont {O.}~\bibnamefont {Miyakawa}}, \bibinfo {author}
  {\bibfnamefont {T.}~\bibnamefont {Sekiguchi}}, \bibinfo {author}
  {\bibfnamefont {D.}~\bibnamefont {Tatsumi}}, \ and\ \bibinfo {author}
  {\bibfnamefont {H.}~\bibnamefont {Yamamoto}} (\bibinfo {collaboration}
  {KAGRA}),\ }\href {\doibase 10.1103/PhysRevD.88.043007} {\bibfield  {journal}
  {\bibinfo  {journal} {Phys. Rev. D}\ }\textbf {\bibinfo {volume} {88}},\
  \bibinfo {pages} {043007} (\bibinfo {year} {2013})},\ \Eprint
  {http://arxiv.org/abs/1306.6747} {arXiv:1306.6747 [gr-qc]} \BibitemShut
  {NoStop}%
\bibitem [{\citenamefont {Abbott}\ \emph
  {et~al.}(2016{\natexlab{b}})\citenamefont {Abbott} \emph
  {et~al.}}]{LIGOScientific:2016vlm}%
  \BibitemOpen
  \bibfield  {author} {\bibinfo {author} {\bibfnamefont {B.~P.}\ \bibnamefont
  {Abbott}} \emph {et~al.} (\bibinfo {collaboration} {LIGO Scientific,
  Virgo}),\ }\href {\doibase 10.1103/PhysRevLett.116.241102} {\bibfield
  {journal} {\bibinfo  {journal} {Phys. Rev. Lett.}\ }\textbf {\bibinfo
  {volume} {116}},\ \bibinfo {pages} {241102} (\bibinfo {year}
  {2016}{\natexlab{b}})},\ \Eprint {http://arxiv.org/abs/1602.03840}
  {arXiv:1602.03840 [gr-qc]} \BibitemShut {NoStop}%
\bibitem [{\citenamefont {Abbott}\ \emph
  {et~al.}(2020{\natexlab{b}})\citenamefont {Abbott} \emph
  {et~al.}}]{LIGOScientific:2020zkf}%
  \BibitemOpen
  \bibfield  {author} {\bibinfo {author} {\bibfnamefont {R.}~\bibnamefont
  {Abbott}} \emph {et~al.} (\bibinfo {collaboration} {LIGO Scientific,
  Virgo}),\ }\href {\doibase 10.3847/2041-8213/ab960f} {\bibfield  {journal}
  {\bibinfo  {journal} {Astrophys. J. Lett.}\ }\textbf {\bibinfo {volume}
  {896}},\ \bibinfo {pages} {L44} (\bibinfo {year} {2020}{\natexlab{b}})},\
  \Eprint {http://arxiv.org/abs/2006.12611} {arXiv:2006.12611 [astro-ph.HE]}
  \BibitemShut {NoStop}%
\bibitem [{\citenamefont {Clesse}\ and\ \citenamefont
  {Garcia-Bellido}(2022)}]{Clesse:2020ghq}%
  \BibitemOpen
  \bibfield  {author} {\bibinfo {author} {\bibfnamefont {S.}~\bibnamefont
  {Clesse}}\ and\ \bibinfo {author} {\bibfnamefont {J.}~\bibnamefont
  {Garcia-Bellido}},\ }\href {\doibase 10.1016/j.dark.2022.101111} {\bibfield
  {journal} {\bibinfo  {journal} {Phys. Dark Univ.}\ }\textbf {\bibinfo
  {volume} {38}},\ \bibinfo {pages} {101111} (\bibinfo {year} {2022})},\
  \Eprint {http://arxiv.org/abs/2007.06481} {arXiv:2007.06481 [astro-ph.CO]}
  \BibitemShut {NoStop}%
\bibitem [{\citenamefont {Vattis}\ \emph {et~al.}(2020)\citenamefont {Vattis},
  \citenamefont {Goldstein},\ and\ \citenamefont
  {Koushiappas}}]{Vattis:2020iuz}%
  \BibitemOpen
  \bibfield  {author} {\bibinfo {author} {\bibfnamefont {K.}~\bibnamefont
  {Vattis}}, \bibinfo {author} {\bibfnamefont {I.~S.}\ \bibnamefont
  {Goldstein}}, \ and\ \bibinfo {author} {\bibfnamefont {S.~M.}\ \bibnamefont
  {Koushiappas}},\ }\href {\doibase 10.1103/PhysRevD.102.061301} {\bibfield
  {journal} {\bibinfo  {journal} {Phys. Rev. D}\ }\textbf {\bibinfo {volume}
  {102}},\ \bibinfo {pages} {061301} (\bibinfo {year} {2020})},\ \Eprint
  {http://arxiv.org/abs/2006.15675} {arXiv:2006.15675 [astro-ph.HE]}
  \BibitemShut {NoStop}%
\bibitem [{\citenamefont {Huang}\ \emph {et~al.}(2020)\citenamefont {Huang},
  \citenamefont {Hu}, \citenamefont {Zhang},\ and\ \citenamefont
  {Shen}}]{Huang:2020cab}%
  \BibitemOpen
  \bibfield  {author} {\bibinfo {author} {\bibfnamefont {K.}~\bibnamefont
  {Huang}}, \bibinfo {author} {\bibfnamefont {J.}~\bibnamefont {Hu}}, \bibinfo
  {author} {\bibfnamefont {Y.}~\bibnamefont {Zhang}}, \ and\ \bibinfo {author}
  {\bibfnamefont {H.}~\bibnamefont {Shen}},\ }\href {\doibase
  10.3847/1538-4357/abbb37} {\bibfield  {journal} {\bibinfo  {journal}
  {Astrophys. J.}\ }\textbf {\bibinfo {volume} {904}},\ \bibinfo {pages} {39}
  (\bibinfo {year} {2020})},\ \Eprint {http://arxiv.org/abs/2008.04491}
  {arXiv:2008.04491 [nucl-th]} \BibitemShut {NoStop}%
\bibitem [{\citenamefont {Roupas}(2021)}]{Roupas:2020jyv}%
  \BibitemOpen
  \bibfield  {author} {\bibinfo {author} {\bibfnamefont {Z.}~\bibnamefont
  {Roupas}},\ }\href {\doibase 10.1007/s10509-021-03919-5} {\bibfield
  {journal} {\bibinfo  {journal} {Astrophys. Space Sci.}\ }\textbf {\bibinfo
  {volume} {366}},\ \bibinfo {pages} {9} (\bibinfo {year} {2021})},\ \Eprint
  {http://arxiv.org/abs/2007.10679} {arXiv:2007.10679 [gr-qc]} \BibitemShut
  {NoStop}%
\bibitem [{\citenamefont {Biswas}\ \emph {et~al.}(2021)\citenamefont {Biswas},
  \citenamefont {Nandi}, \citenamefont {Char}, \citenamefont {Bose},\ and\
  \citenamefont {Stergioulas}}]{Biswas:2020xna}%
  \BibitemOpen
  \bibfield  {author} {\bibinfo {author} {\bibfnamefont {B.}~\bibnamefont
  {Biswas}}, \bibinfo {author} {\bibfnamefont {R.}~\bibnamefont {Nandi}},
  \bibinfo {author} {\bibfnamefont {P.}~\bibnamefont {Char}}, \bibinfo {author}
  {\bibfnamefont {S.}~\bibnamefont {Bose}}, \ and\ \bibinfo {author}
  {\bibfnamefont {N.}~\bibnamefont {Stergioulas}},\ }\href {\doibase
  10.1093/mnras/stab1383} {\bibfield  {journal} {\bibinfo  {journal} {Mon. Not.
  Roy. Astron. Soc.}\ }\textbf {\bibinfo {volume} {505}},\ \bibinfo {pages}
  {1600} (\bibinfo {year} {2021})},\ \Eprint {http://arxiv.org/abs/2010.02090}
  {arXiv:2010.02090 [astro-ph.HE]} \BibitemShut {NoStop}%
\bibitem [{\citenamefont {Tews}\ \emph {et~al.}(2021)\citenamefont {Tews},
  \citenamefont {Pang}, \citenamefont {Dietrich}, \citenamefont {Coughlin},
  \citenamefont {Antier}, \citenamefont {Bulla}, \citenamefont {Heinzel},\ and\
  \citenamefont {Issa}}]{Tews:2020ylw}%
  \BibitemOpen
  \bibfield  {author} {\bibinfo {author} {\bibfnamefont {I.}~\bibnamefont
  {Tews}}, \bibinfo {author} {\bibfnamefont {P.~T.~H.}\ \bibnamefont {Pang}},
  \bibinfo {author} {\bibfnamefont {T.}~\bibnamefont {Dietrich}}, \bibinfo
  {author} {\bibfnamefont {M.~W.}\ \bibnamefont {Coughlin}}, \bibinfo {author}
  {\bibfnamefont {S.}~\bibnamefont {Antier}}, \bibinfo {author} {\bibfnamefont
  {M.}~\bibnamefont {Bulla}}, \bibinfo {author} {\bibfnamefont
  {J.}~\bibnamefont {Heinzel}}, \ and\ \bibinfo {author} {\bibfnamefont
  {L.}~\bibnamefont {Issa}},\ }\href {\doibase 10.3847/2041-8213/abdaae}
  {\bibfield  {journal} {\bibinfo  {journal} {Astrophys. J. Lett.}\ }\textbf
  {\bibinfo {volume} {908}},\ \bibinfo {pages} {L1} (\bibinfo {year} {2021})},\
  \Eprint {http://arxiv.org/abs/2007.06057} {arXiv:2007.06057 [astro-ph.HE]}
  \BibitemShut {NoStop}%
\bibitem [{\citenamefont {Laarakkers}\ and\ \citenamefont
  {Poisson}(1999)}]{Laarakkers:1997hb}%
  \BibitemOpen
  \bibfield  {author} {\bibinfo {author} {\bibfnamefont {W.~G.}\ \bibnamefont
  {Laarakkers}}\ and\ \bibinfo {author} {\bibfnamefont {E.}~\bibnamefont
  {Poisson}},\ }\href {\doibase 10.1086/306732} {\bibfield  {journal} {\bibinfo
   {journal} {Astrophys. J.}\ }\textbf {\bibinfo {volume} {512}},\ \bibinfo
  {pages} {282} (\bibinfo {year} {1999})},\ \Eprint
  {http://arxiv.org/abs/gr-qc/9709033} {arXiv:gr-qc/9709033} \BibitemShut
  {NoStop}%
\bibitem [{\citenamefont {Ryan}(1997)}]{PhysRevD.55.6081}%
  \BibitemOpen
  \bibfield  {author} {\bibinfo {author} {\bibfnamefont {F.~D.}\ \bibnamefont
  {Ryan}},\ }\href {\doibase 10.1103/PhysRevD.55.6081} {\bibfield  {journal}
  {\bibinfo  {journal} {Phys. Rev. D}\ }\textbf {\bibinfo {volume} {55}},\
  \bibinfo {pages} {6081} (\bibinfo {year} {1997})}\BibitemShut {NoStop}%
\bibitem [{\citenamefont {Poisson}(1998)}]{Poisson:1997ha}%
  \BibitemOpen
  \bibfield  {author} {\bibinfo {author} {\bibfnamefont {E.}~\bibnamefont
  {Poisson}},\ }\href {\doibase 10.1103/PhysRevD.57.5287} {\bibfield  {journal}
  {\bibinfo  {journal} {Phys. Rev. D}\ }\textbf {\bibinfo {volume} {57}},\
  \bibinfo {pages} {5287} (\bibinfo {year} {1998})}\BibitemShut {NoStop}%
\bibitem [{\citenamefont {Pacilio}\ \emph {et~al.}(2020)\citenamefont
  {Pacilio}, \citenamefont {Vaglio}, \citenamefont {Maselli},\ and\
  \citenamefont {Pani}}]{Pacilio:2020jza}%
  \BibitemOpen
  \bibfield  {author} {\bibinfo {author} {\bibfnamefont {C.}~\bibnamefont
  {Pacilio}}, \bibinfo {author} {\bibfnamefont {M.}~\bibnamefont {Vaglio}},
  \bibinfo {author} {\bibfnamefont {A.}~\bibnamefont {Maselli}}, \ and\
  \bibinfo {author} {\bibfnamefont {P.}~\bibnamefont {Pani}},\ }\href {\doibase
  10.1103/PhysRevD.102.083002} {\bibfield  {journal} {\bibinfo  {journal}
  {Phys. Rev. D}\ }\textbf {\bibinfo {volume} {102}},\ \bibinfo {pages}
  {083002} (\bibinfo {year} {2020})},\ \Eprint
  {http://arxiv.org/abs/2007.05264} {arXiv:2007.05264 [gr-qc]} \BibitemShut
  {NoStop}%
\bibitem [{\citenamefont {G\"urlebeck}(2015)}]{Gurlebeck:2015xpa}%
  \BibitemOpen
  \bibfield  {author} {\bibinfo {author} {\bibfnamefont {N.}~\bibnamefont
  {G\"urlebeck}},\ }\href {\doibase 10.1103/PhysRevLett.114.151102} {\bibfield
  {journal} {\bibinfo  {journal} {Phys. Rev. Lett.}\ }\textbf {\bibinfo
  {volume} {114}},\ \bibinfo {pages} {151102} (\bibinfo {year} {2015})},\
  \Eprint {http://arxiv.org/abs/1503.03240} {arXiv:1503.03240 [gr-qc]}
  \BibitemShut {NoStop}%
\bibitem [{\citenamefont {Le~Tiec}\ and\ \citenamefont
  {Casals}(2021)}]{LeTiec:2020spy}%
  \BibitemOpen
  \bibfield  {author} {\bibinfo {author} {\bibfnamefont {A.}~\bibnamefont
  {Le~Tiec}}\ and\ \bibinfo {author} {\bibfnamefont {M.}~\bibnamefont
  {Casals}},\ }\href {\doibase 10.1103/PhysRevLett.126.131102} {\bibfield
  {journal} {\bibinfo  {journal} {Phys. Rev. Lett.}\ }\textbf {\bibinfo
  {volume} {126}},\ \bibinfo {pages} {131102} (\bibinfo {year} {2021})},\
  \Eprint {http://arxiv.org/abs/2007.00214} {arXiv:2007.00214 [gr-qc]}
  \BibitemShut {NoStop}%
\bibitem [{\citenamefont {Mendes}\ and\ \citenamefont
  {Yang}(2017)}]{Mendes:2016vdr}%
  \BibitemOpen
  \bibfield  {author} {\bibinfo {author} {\bibfnamefont {R.~F.~P.}\
  \bibnamefont {Mendes}}\ and\ \bibinfo {author} {\bibfnamefont
  {H.}~\bibnamefont {Yang}},\ }\href {\doibase 10.1088/1361-6382/aa842d}
  {\bibfield  {journal} {\bibinfo  {journal} {Classical Quantum Gravity}\
  }\textbf {\bibinfo {volume} {34}},\ \bibinfo {pages} {185001} (\bibinfo
  {year} {2017})},\ \Eprint {http://arxiv.org/abs/1606.03035} {arXiv:1606.03035
  [astro-ph.CO]} \BibitemShut {NoStop}%
\bibitem [{\citenamefont {Uchikata}\ \emph {et~al.}(2016)\citenamefont
  {Uchikata}, \citenamefont {Yoshida},\ and\ \citenamefont
  {Pani}}]{Uchikata:2016qku}%
  \BibitemOpen
  \bibfield  {author} {\bibinfo {author} {\bibfnamefont {N.}~\bibnamefont
  {Uchikata}}, \bibinfo {author} {\bibfnamefont {S.}~\bibnamefont {Yoshida}}, \
  and\ \bibinfo {author} {\bibfnamefont {P.}~\bibnamefont {Pani}},\ }\href
  {\doibase 10.1103/PhysRevD.94.064015} {\bibfield  {journal} {\bibinfo
  {journal} {Phys. Rev. D}\ }\textbf {\bibinfo {volume} {94}},\ \bibinfo
  {pages} {064015} (\bibinfo {year} {2016})},\ \Eprint
  {http://arxiv.org/abs/1607.03593} {arXiv:1607.03593 [gr-qc]} \BibitemShut
  {NoStop}%
\bibitem [{\citenamefont {Johnson-Mcdaniel}\ \emph {et~al.}(2020)\citenamefont
  {Johnson-Mcdaniel}, \citenamefont {Mukherjee}, \citenamefont {Kashyap},
  \citenamefont {Ajith}, \citenamefont {Del~Pozzo},\ and\ \citenamefont
  {Vitale}}]{Johnson-Mcdaniel:2018cdu}%
  \BibitemOpen
  \bibfield  {author} {\bibinfo {author} {\bibfnamefont {N.~K.}\ \bibnamefont
  {Johnson-Mcdaniel}}, \bibinfo {author} {\bibfnamefont {A.}~\bibnamefont
  {Mukherjee}}, \bibinfo {author} {\bibfnamefont {R.}~\bibnamefont {Kashyap}},
  \bibinfo {author} {\bibfnamefont {P.}~\bibnamefont {Ajith}}, \bibinfo
  {author} {\bibfnamefont {W.}~\bibnamefont {Del~Pozzo}}, \ and\ \bibinfo
  {author} {\bibfnamefont {S.}~\bibnamefont {Vitale}},\ }\href {\doibase
  10.1103/PhysRevD.102.123010} {\bibfield  {journal} {\bibinfo  {journal}
  {Phys. Rev. D}\ }\textbf {\bibinfo {volume} {102}},\ \bibinfo {pages}
  {123010} (\bibinfo {year} {2020})},\ \Eprint
  {http://arxiv.org/abs/1804.08026} {arXiv:1804.08026 [gr-qc]} \BibitemShut
  {NoStop}%
%%CITATION = ARXIV:1804.08026;%%
\bibitem [{\citenamefont {Jim\'enez~Forteza}\ \emph {et~al.}(2018)\citenamefont
  {Jim\'enez~Forteza}, \citenamefont {Abdelsalhin}, \citenamefont {Pani},\ and\
  \citenamefont {Gualtieri}}]{JimenezForteza:2018rwr}%
  \BibitemOpen
  \bibfield  {author} {\bibinfo {author} {\bibfnamefont {X.}~\bibnamefont
  {Jim\'enez~Forteza}}, \bibinfo {author} {\bibfnamefont {T.}~\bibnamefont
  {Abdelsalhin}}, \bibinfo {author} {\bibfnamefont {P.}~\bibnamefont {Pani}}, \
  and\ \bibinfo {author} {\bibfnamefont {L.}~\bibnamefont {Gualtieri}},\ }\href
  {\doibase 10.1103/PhysRevD.98.124014} {\bibfield  {journal} {\bibinfo
  {journal} {Phys. Rev. D}\ }\textbf {\bibinfo {volume} {98}},\ \bibinfo
  {pages} {124014} (\bibinfo {year} {2018})},\ \Eprint
  {http://arxiv.org/abs/1807.08016} {arXiv:1807.08016 [gr-qc]} \BibitemShut
  {NoStop}%
%%CITATION = ARXIV:1807.08016;%%
\bibitem [{\citenamefont {Abdelsalhin}\ \emph {et~al.}(2018)\citenamefont
  {Abdelsalhin}, \citenamefont {Gualtieri},\ and\ \citenamefont
  {Pani}}]{Abdelsalhin:2018reg}%
  \BibitemOpen
  \bibfield  {author} {\bibinfo {author} {\bibfnamefont {T.}~\bibnamefont
  {Abdelsalhin}}, \bibinfo {author} {\bibfnamefont {L.}~\bibnamefont
  {Gualtieri}}, \ and\ \bibinfo {author} {\bibfnamefont {P.}~\bibnamefont
  {Pani}},\ }\href {\doibase 10.1103/PhysRevD.98.104046} {\bibfield  {journal}
  {\bibinfo  {journal} {Phys. Rev.}\ }\textbf {\bibinfo {volume} {D98}},\
  \bibinfo {pages} {104046} (\bibinfo {year} {2018})},\ \Eprint
  {http://arxiv.org/abs/1805.01487} {arXiv:1805.01487 [gr-qc]} \BibitemShut
  {NoStop}%
%%CITATION = ARXIV:1805.01487;%%
\bibitem [{\citenamefont {Datta}\ and\ \citenamefont
  {Bose}(2019)}]{Datta:2019euh}%
  \BibitemOpen
  \bibfield  {author} {\bibinfo {author} {\bibfnamefont {S.}~\bibnamefont
  {Datta}}\ and\ \bibinfo {author} {\bibfnamefont {S.}~\bibnamefont {Bose}},\
  }\href {\doibase 10.1103/PhysRevD.99.084001} {\bibfield  {journal} {\bibinfo
  {journal} {Phys. Rev. D}\ }\textbf {\bibinfo {volume} {99}},\ \bibinfo
  {pages} {084001} (\bibinfo {year} {2019})},\ \Eprint
  {http://arxiv.org/abs/1902.01723} {arXiv:1902.01723 [gr-qc]} \BibitemShut
  {NoStop}%
\bibitem [{\citenamefont {Hartle}(1973)}]{Hartle:1973zz}%
  \BibitemOpen
  \bibfield  {author} {\bibinfo {author} {\bibfnamefont {J.~B.}\ \bibnamefont
  {Hartle}},\ }\href {\doibase 10.1103/PhysRevD.8.1010} {\bibfield  {journal}
  {\bibinfo  {journal} {Phys. Rev. D}\ }\textbf {\bibinfo {volume} {8}},\
  \bibinfo {pages} {1010} (\bibinfo {year} {1973})}\BibitemShut {NoStop}%
%%CITATION = PHRVA,D8,1010;%%
\bibitem [{\citenamefont {Chatziioannou}\ \emph {et~al.}(2016)\citenamefont
  {Chatziioannou}, \citenamefont {Poisson},\ and\ \citenamefont
  {Yunes}}]{Chatziioannou:2016kem}%
  \BibitemOpen
  \bibfield  {author} {\bibinfo {author} {\bibfnamefont {K.}~\bibnamefont
  {Chatziioannou}}, \bibinfo {author} {\bibfnamefont {E.}~\bibnamefont
  {Poisson}}, \ and\ \bibinfo {author} {\bibfnamefont {N.}~\bibnamefont
  {Yunes}},\ }\href {\doibase 10.1103/PhysRevD.94.084043} {\bibfield  {journal}
  {\bibinfo  {journal} {Phys. Rev. D}\ }\textbf {\bibinfo {volume} {94}},\
  \bibinfo {pages} {084043} (\bibinfo {year} {2016})},\ \Eprint
  {http://arxiv.org/abs/1608.02899} {arXiv:1608.02899 [gr-qc]} \BibitemShut
  {NoStop}%
\bibitem [{\citenamefont {Datta}\ \emph {et~al.}(2021)\citenamefont {Datta},
  \citenamefont {Phukon},\ and\ \citenamefont {Bose}}]{Datta:2020gem}%
  \BibitemOpen
  \bibfield  {author} {\bibinfo {author} {\bibfnamefont {S.}~\bibnamefont
  {Datta}}, \bibinfo {author} {\bibfnamefont {K.~S.}\ \bibnamefont {Phukon}}, \
  and\ \bibinfo {author} {\bibfnamefont {S.}~\bibnamefont {Bose}},\ }\href
  {\doibase 10.1103/PhysRevD.104.084006} {\bibfield  {journal} {\bibinfo
  {journal} {Phys. Rev. D}\ }\textbf {\bibinfo {volume} {104}},\ \bibinfo
  {pages} {084006} (\bibinfo {year} {2021})},\ \Eprint
  {http://arxiv.org/abs/2004.05974} {arXiv:2004.05974 [gr-qc]} \BibitemShut
  {NoStop}%
\bibitem [{\citenamefont {Krishnendu}\ \emph {et~al.}(2017)\citenamefont
  {Krishnendu}, \citenamefont {Arun},\ and\ \citenamefont
  {Mishra}}]{Krishnendu:2017shb}%
  \BibitemOpen
  \bibfield  {author} {\bibinfo {author} {\bibfnamefont {N.~V.}\ \bibnamefont
  {Krishnendu}}, \bibinfo {author} {\bibfnamefont {K.~G.}\ \bibnamefont
  {Arun}}, \ and\ \bibinfo {author} {\bibfnamefont {C.~K.}\ \bibnamefont
  {Mishra}},\ }\href {\doibase 10.1103/PhysRevLett.119.091101} {\bibfield
  {journal} {\bibinfo  {journal} {Phys. Rev. Lett.}\ }\textbf {\bibinfo
  {volume} {119}},\ \bibinfo {pages} {091101} (\bibinfo {year} {2017})},\
  \Eprint {http://arxiv.org/abs/1701.06318} {arXiv:1701.06318 [gr-qc]}
  \BibitemShut {NoStop}%
\bibitem [{\citenamefont {Krishnendu}\ \emph
  {et~al.}(2019{\natexlab{a}})\citenamefont {Krishnendu}, \citenamefont
  {Mishra},\ and\ \citenamefont {Arun}}]{Krishnendu:2018nqa}%
  \BibitemOpen
  \bibfield  {author} {\bibinfo {author} {\bibfnamefont {N.~V.}\ \bibnamefont
  {Krishnendu}}, \bibinfo {author} {\bibfnamefont {C.~K.}\ \bibnamefont
  {Mishra}}, \ and\ \bibinfo {author} {\bibfnamefont {K.~G.}\ \bibnamefont
  {Arun}},\ }\href {\doibase 10.1103/PhysRevD.99.064008} {\bibfield  {journal}
  {\bibinfo  {journal} {Phys. Rev. D}\ }\textbf {\bibinfo {volume} {99}},\
  \bibinfo {pages} {064008} (\bibinfo {year} {2019}{\natexlab{a}})},\ \Eprint
  {http://arxiv.org/abs/1811.00317} {arXiv:1811.00317 [gr-qc]} \BibitemShut
  {NoStop}%
\bibitem [{\citenamefont {Krishnendu}\ \emph
  {et~al.}(2019{\natexlab{b}})\citenamefont {Krishnendu}, \citenamefont
  {Saleem}, \citenamefont {Samajdar}, \citenamefont {Arun}, \citenamefont
  {Del~Pozzo},\ and\ \citenamefont {Mishra}}]{Krishnendu:2019tjp}%
  \BibitemOpen
  \bibfield  {author} {\bibinfo {author} {\bibfnamefont {N.~V.}\ \bibnamefont
  {Krishnendu}}, \bibinfo {author} {\bibfnamefont {M.}~\bibnamefont {Saleem}},
  \bibinfo {author} {\bibfnamefont {A.}~\bibnamefont {Samajdar}}, \bibinfo
  {author} {\bibfnamefont {K.~G.}\ \bibnamefont {Arun}}, \bibinfo {author}
  {\bibfnamefont {W.}~\bibnamefont {Del~Pozzo}}, \ and\ \bibinfo {author}
  {\bibfnamefont {C.~K.}\ \bibnamefont {Mishra}},\ }\href {\doibase
  10.1103/PhysRevD.100.104019} {\bibfield  {journal} {\bibinfo  {journal}
  {Phys. Rev. D}\ }\textbf {\bibinfo {volume} {100}},\ \bibinfo {pages}
  {104019} (\bibinfo {year} {2019}{\natexlab{b}})},\ \Eprint
  {http://arxiv.org/abs/1908.02247} {arXiv:1908.02247 [gr-qc]} \BibitemShut
  {NoStop}%
\bibitem [{\citenamefont {Krishnendu}\ and\ \citenamefont
  {Yelikar}(2020)}]{Krishnendu:2019ebd}%
  \BibitemOpen
  \bibfield  {author} {\bibinfo {author} {\bibfnamefont {N.~V.}\ \bibnamefont
  {Krishnendu}}\ and\ \bibinfo {author} {\bibfnamefont {A.~B.}\ \bibnamefont
  {Yelikar}},\ }\href {\doibase 10.1088/1361-6382/ababb1} {\bibfield  {journal}
  {\bibinfo  {journal} {Classical Quantum Gravity}\ }\textbf {\bibinfo {volume}
  {37}},\ \bibinfo {pages} {205019} (\bibinfo {year} {2020})},\ \Eprint
  {http://arxiv.org/abs/1904.12712} {arXiv:1904.12712 [gr-qc]} \BibitemShut
  {NoStop}%
\bibitem [{\citenamefont {Saleem}\ \emph {et~al.}(2022)\citenamefont {Saleem},
  \citenamefont {Krishnendu}, \citenamefont {Ghosh}, \citenamefont {Gupta},
  \citenamefont {Del~Pozzo}, \citenamefont {Ghosh},\ and\ \citenamefont
  {Arun}}]{Saleem:2021vph}%
  \BibitemOpen
  \bibfield  {author} {\bibinfo {author} {\bibfnamefont {M.}~\bibnamefont
  {Saleem}}, \bibinfo {author} {\bibfnamefont {N.~V.}\ \bibnamefont
  {Krishnendu}}, \bibinfo {author} {\bibfnamefont {A.}~\bibnamefont {Ghosh}},
  \bibinfo {author} {\bibfnamefont {A.}~\bibnamefont {Gupta}}, \bibinfo
  {author} {\bibfnamefont {W.}~\bibnamefont {Del~Pozzo}}, \bibinfo {author}
  {\bibfnamefont {A.}~\bibnamefont {Ghosh}}, \ and\ \bibinfo {author}
  {\bibfnamefont {K.~G.}\ \bibnamefont {Arun}},\ }\href {\doibase
  10.1103/PhysRevD.105.104066} {\bibfield  {journal} {\bibinfo  {journal}
  {Phys. Rev. D}\ }\textbf {\bibinfo {volume} {105}},\ \bibinfo {pages}
  {104066} (\bibinfo {year} {2022})},\ \Eprint
  {http://arxiv.org/abs/2111.04135} {arXiv:2111.04135 [gr-qc]} \BibitemShut
  {NoStop}%
\bibitem [{\citenamefont {Abbott}\ \emph
  {et~al.}(2021{\natexlab{b}})\citenamefont {Abbott} \emph
  {et~al.}}]{LIGOScientific:2020tif}%
  \BibitemOpen
  \bibfield  {author} {\bibinfo {author} {\bibfnamefont {R.}~\bibnamefont
  {Abbott}} \emph {et~al.} (\bibinfo {collaboration} {LIGO Scientific and Virgo
  Collaborations}),\ }\href {\doibase 10.1103/PhysRevD.103.122002} {\bibfield
  {journal} {\bibinfo  {journal} {Phys. Rev. D}\ }\textbf {\bibinfo {volume}
  {103}},\ \bibinfo {pages} {122002} (\bibinfo {year} {2021}{\natexlab{b}})},\
  \Eprint {http://arxiv.org/abs/2010.14529} {arXiv:2010.14529 [gr-qc]}
  \BibitemShut {NoStop}%
\bibitem [{\citenamefont {Abbott}\ \emph
  {et~al.}(2021{\natexlab{c}})\citenamefont {Abbott} \emph
  {et~al.}}]{LIGOScientific:2021sio}%
  \BibitemOpen
  \bibfield  {author} {\bibinfo {author} {\bibfnamefont {R.}~\bibnamefont
  {Abbott}} \emph {et~al.} (\bibinfo {collaboration} {LIGO Scientific, VIRGO,
  and KAGRA Collaborations}),\ }\href@noop {} {\  (\bibinfo {year}
  {2021}{\natexlab{c}})},\ \Eprint {http://arxiv.org/abs/2112.06861}
  {arXiv:2112.06861 [gr-qc]} \BibitemShut {NoStop}%
\bibitem [{\citenamefont {Saini}\ and\ \citenamefont
  {Krishnendu}(2023)}]{Saini:2023gaw}%
  \BibitemOpen
  \bibfield  {author} {\bibinfo {author} {\bibfnamefont {P.}~\bibnamefont
  {Saini}}\ and\ \bibinfo {author} {\bibfnamefont {N.~V.}\ \bibnamefont
  {Krishnendu}},\ }\href@noop {} {\  (\bibinfo {year} {2023})},\ \Eprint
  {http://arxiv.org/abs/2308.01309} {arXiv:2308.01309 [gr-qc]} \BibitemShut
  {NoStop}%
\bibitem [{\citenamefont {Chia}\ \emph {et~al.}(2022)\citenamefont {Chia},
  \citenamefont {Edwards}, \citenamefont {George}, \citenamefont {Zimmerman},
  \citenamefont {Coogan}, \citenamefont {Freese}, \citenamefont {Messick},\
  and\ \citenamefont {Setzer}}]{Chia:2022rwc}%
  \BibitemOpen
  \bibfield  {author} {\bibinfo {author} {\bibfnamefont {H.~S.}\ \bibnamefont
  {Chia}}, \bibinfo {author} {\bibfnamefont {T.~D.~P.}\ \bibnamefont
  {Edwards}}, \bibinfo {author} {\bibfnamefont {R.~N.}\ \bibnamefont {George}},
  \bibinfo {author} {\bibfnamefont {A.}~\bibnamefont {Zimmerman}}, \bibinfo
  {author} {\bibfnamefont {A.}~\bibnamefont {Coogan}}, \bibinfo {author}
  {\bibfnamefont {K.}~\bibnamefont {Freese}}, \bibinfo {author} {\bibfnamefont
  {C.}~\bibnamefont {Messick}}, \ and\ \bibinfo {author} {\bibfnamefont
  {C.~N.}\ \bibnamefont {Setzer}},\ }\href@noop {} {\  (\bibinfo {year}
  {2022})},\ \Eprint {http://arxiv.org/abs/2211.00039} {arXiv:2211.00039
  [gr-qc]} \BibitemShut {NoStop}%
\bibitem [{\citenamefont {Husa}\ \emph {et~al.}(2016)\citenamefont {Husa},
  \citenamefont {Khan}, \citenamefont {Hannam}, \citenamefont {P\"urrer},
  \citenamefont {Ohme}, \citenamefont {Jim\'enez~Forteza},\ and\ \citenamefont
  {Boh\'e}}]{Husa:2015iqa}%
  \BibitemOpen
  \bibfield  {author} {\bibinfo {author} {\bibfnamefont {S.}~\bibnamefont
  {Husa}}, \bibinfo {author} {\bibfnamefont {S.}~\bibnamefont {Khan}}, \bibinfo
  {author} {\bibfnamefont {M.}~\bibnamefont {Hannam}}, \bibinfo {author}
  {\bibfnamefont {M.}~\bibnamefont {P\"urrer}}, \bibinfo {author}
  {\bibfnamefont {F.}~\bibnamefont {Ohme}}, \bibinfo {author} {\bibfnamefont
  {X.}~\bibnamefont {Jim\'enez~Forteza}}, \ and\ \bibinfo {author}
  {\bibfnamefont {A.}~\bibnamefont {Boh\'e}},\ }\href {\doibase
  10.1103/PhysRevD.93.044006} {\bibfield  {journal} {\bibinfo  {journal} {Phys.
  Rev. D}\ }\textbf {\bibinfo {volume} {93}},\ \bibinfo {pages} {044006}
  (\bibinfo {year} {2016})},\ \Eprint {http://arxiv.org/abs/1508.07250}
  {arXiv:1508.07250 [gr-qc]} \BibitemShut {NoStop}%
\bibitem [{\citenamefont {Khan}\ \emph {et~al.}(2016)\citenamefont {Khan},
  \citenamefont {Husa}, \citenamefont {Hannam}, \citenamefont {Ohme},
  \citenamefont {P\"urrer}, \citenamefont {Jim\'enez~Forteza},\ and\
  \citenamefont {Boh\'e}}]{Khan:2015jqa}%
  \BibitemOpen
  \bibfield  {author} {\bibinfo {author} {\bibfnamefont {S.}~\bibnamefont
  {Khan}}, \bibinfo {author} {\bibfnamefont {S.}~\bibnamefont {Husa}}, \bibinfo
  {author} {\bibfnamefont {M.}~\bibnamefont {Hannam}}, \bibinfo {author}
  {\bibfnamefont {F.}~\bibnamefont {Ohme}}, \bibinfo {author} {\bibfnamefont
  {M.}~\bibnamefont {P\"urrer}}, \bibinfo {author} {\bibfnamefont
  {X.}~\bibnamefont {Jim\'enez~Forteza}}, \ and\ \bibinfo {author}
  {\bibfnamefont {A.}~\bibnamefont {Boh\'e}},\ }\href {\doibase
  10.1103/PhysRevD.93.044007} {\bibfield  {journal} {\bibinfo  {journal} {Phys.
  Rev. D}\ }\textbf {\bibinfo {volume} {93}},\ \bibinfo {pages} {044007}
  (\bibinfo {year} {2016})},\ \Eprint {http://arxiv.org/abs/1508.07253}
  {arXiv:1508.07253 [gr-qc]} \BibitemShut {NoStop}%
\bibitem [{\citenamefont {Hannam}\ \emph {et~al.}(2014)\citenamefont {Hannam},
  \citenamefont {Schmidt}, \citenamefont {Boh\'e}, \citenamefont {Haegel},
  \citenamefont {Husa}, \citenamefont {Ohme}, \citenamefont {Pratten},\ and\
  \citenamefont {P\"urrer}}]{Hannam:2013oca}%
  \BibitemOpen
  \bibfield  {author} {\bibinfo {author} {\bibfnamefont {M.}~\bibnamefont
  {Hannam}}, \bibinfo {author} {\bibfnamefont {P.}~\bibnamefont {Schmidt}},
  \bibinfo {author} {\bibfnamefont {A.}~\bibnamefont {Boh\'e}}, \bibinfo
  {author} {\bibfnamefont {L.}~\bibnamefont {Haegel}}, \bibinfo {author}
  {\bibfnamefont {S.}~\bibnamefont {Husa}}, \bibinfo {author} {\bibfnamefont
  {F.}~\bibnamefont {Ohme}}, \bibinfo {author} {\bibfnamefont {G.}~\bibnamefont
  {Pratten}}, \ and\ \bibinfo {author} {\bibfnamefont {M.}~\bibnamefont
  {P\"urrer}},\ }\href {\doibase 10.1103/PhysRevLett.113.151101} {\bibfield
  {journal} {\bibinfo  {journal} {Phys. Rev. Lett.}\ }\textbf {\bibinfo
  {volume} {113}},\ \bibinfo {pages} {151101} (\bibinfo {year} {2014})},\
  \Eprint {http://arxiv.org/abs/1308.3271} {arXiv:1308.3271 [gr-qc]}
  \BibitemShut {NoStop}%
\bibitem [{\citenamefont {LaHaye}\ \emph {et~al.}(2023)\citenamefont {LaHaye},
  \citenamefont {Yang}, \citenamefont {Bonga},\ and\ \citenamefont
  {Lyu}}]{LaHaye:2022yxa}%
  \BibitemOpen
  \bibfield  {author} {\bibinfo {author} {\bibfnamefont {M.}~\bibnamefont
  {LaHaye}}, \bibinfo {author} {\bibfnamefont {H.}~\bibnamefont {Yang}},
  \bibinfo {author} {\bibfnamefont {B.}~\bibnamefont {Bonga}}, \ and\ \bibinfo
  {author} {\bibfnamefont {Z.}~\bibnamefont {Lyu}},\ }\href {\doibase
  10.1103/PhysRevD.108.043018} {\bibfield  {journal} {\bibinfo  {journal}
  {Phys. Rev. D}\ }\textbf {\bibinfo {volume} {108}},\ \bibinfo {pages}
  {043018} (\bibinfo {year} {2023})},\ \Eprint
  {http://arxiv.org/abs/2212.04657} {arXiv:2212.04657 [gr-qc]} \BibitemShut
  {NoStop}%
\bibitem [{\citenamefont {Lyu}\ \emph {et~al.}(2023)\citenamefont {Lyu},
  \citenamefont {LaHaye}, \citenamefont {Yang},\ and\ \citenamefont
  {Bonga}}]{Lyu:2023zxv}%
  \BibitemOpen
  \bibfield  {author} {\bibinfo {author} {\bibfnamefont {Z.}~\bibnamefont
  {Lyu}}, \bibinfo {author} {\bibfnamefont {M.}~\bibnamefont {LaHaye}},
  \bibinfo {author} {\bibfnamefont {H.}~\bibnamefont {Yang}}, \ and\ \bibinfo
  {author} {\bibfnamefont {B.}~\bibnamefont {Bonga}},\ }\href@noop {} {\
  (\bibinfo {year} {2023})},\ \Eprint {http://arxiv.org/abs/2308.09032}
  {arXiv:2308.09032 [gr-qc]} \BibitemShut {NoStop}%
\bibitem [{\citenamefont {Calder\'on~Bustillo}\ \emph
  {et~al.}(2017)\citenamefont {Calder\'on~Bustillo}, \citenamefont {Laguna},\
  and\ \citenamefont {Shoemaker}}]{Bustillo:2016gid}%
  \BibitemOpen
  \bibfield  {author} {\bibinfo {author} {\bibfnamefont {J.}~\bibnamefont
  {Calder\'on~Bustillo}}, \bibinfo {author} {\bibfnamefont {P.}~\bibnamefont
  {Laguna}}, \ and\ \bibinfo {author} {\bibfnamefont {D.}~\bibnamefont
  {Shoemaker}},\ }\href {\doibase 10.1103/PhysRevD.95.104038} {\bibfield
  {journal} {\bibinfo  {journal} {Phys. Rev. D}\ }\textbf {\bibinfo {volume}
  {95}},\ \bibinfo {pages} {104038} (\bibinfo {year} {2017})},\ \Eprint
  {http://arxiv.org/abs/1612.02340} {arXiv:1612.02340 [gr-qc]} \BibitemShut
  {NoStop}%
\bibitem [{\citenamefont {Krishnendu}\ and\ \citenamefont
  {Ohme}(2022)}]{Krishnendu:2021cyi}%
  \BibitemOpen
  \bibfield  {author} {\bibinfo {author} {\bibfnamefont {N.~V.}\ \bibnamefont
  {Krishnendu}}\ and\ \bibinfo {author} {\bibfnamefont {F.}~\bibnamefont
  {Ohme}},\ }\href {\doibase 10.1103/PhysRevD.105.064012} {\bibfield  {journal}
  {\bibinfo  {journal} {Phys. Rev. D}\ }\textbf {\bibinfo {volume} {105}},\
  \bibinfo {pages} {064012} (\bibinfo {year} {2022})},\ \Eprint
  {http://arxiv.org/abs/2110.00766} {arXiv:2110.00766 [gr-qc]} \BibitemShut
  {NoStop}%
\bibitem [{\citenamefont {Krishnendu}\ and\ \citenamefont
  {Ohme}(2021)}]{Krishnendu:2021fga}%
  \BibitemOpen
  \bibfield  {author} {\bibinfo {author} {\bibfnamefont {N.~V.}\ \bibnamefont
  {Krishnendu}}\ and\ \bibinfo {author} {\bibfnamefont {F.}~\bibnamefont
  {Ohme}},\ }\href {\doibase 10.3390/universe7120497} {\bibfield  {journal}
  {\bibinfo  {journal} {Universe}\ }\textbf {\bibinfo {volume} {7}},\ \bibinfo
  {pages} {497} (\bibinfo {year} {2021})},\ \Eprint
  {http://arxiv.org/abs/2201.05418} {arXiv:2201.05418 [gr-qc]} \BibitemShut
  {NoStop}%
\bibitem [{\citenamefont {Puecher}\ \emph {et~al.}(2022)\citenamefont
  {Puecher}, \citenamefont {Kalaghatgi}, \citenamefont {Roy}, \citenamefont
  {Setyawati}, \citenamefont {Gupta}, \citenamefont {Sathyaprakash},\ and\
  \citenamefont {Van Den~Broeck}}]{Puecher:2022sfm}%
  \BibitemOpen
  \bibfield  {author} {\bibinfo {author} {\bibfnamefont {A.}~\bibnamefont
  {Puecher}}, \bibinfo {author} {\bibfnamefont {C.}~\bibnamefont {Kalaghatgi}},
  \bibinfo {author} {\bibfnamefont {S.}~\bibnamefont {Roy}}, \bibinfo {author}
  {\bibfnamefont {Y.}~\bibnamefont {Setyawati}}, \bibinfo {author}
  {\bibfnamefont {I.}~\bibnamefont {Gupta}}, \bibinfo {author} {\bibfnamefont
  {B.~S.}\ \bibnamefont {Sathyaprakash}}, \ and\ \bibinfo {author}
  {\bibfnamefont {C.}~\bibnamefont {Van Den~Broeck}},\ }\href {\doibase
  10.1103/PhysRevD.106.082003} {\bibfield  {journal} {\bibinfo  {journal}
  {Phys. Rev. D}\ }\textbf {\bibinfo {volume} {106}},\ \bibinfo {pages}
  {082003} (\bibinfo {year} {2022})},\ \Eprint
  {http://arxiv.org/abs/2205.09062} {arXiv:2205.09062 [gr-qc]} \BibitemShut
  {NoStop}%
\bibitem [{\citenamefont {Mehta}\ \emph {et~al.}(2023)\citenamefont {Mehta},
  \citenamefont {Buonanno}, \citenamefont {Cotesta}, \citenamefont {Ghosh},
  \citenamefont {Sennett},\ and\ \citenamefont {Steinhoff}}]{Mehta:2022pcn}%
  \BibitemOpen
  \bibfield  {author} {\bibinfo {author} {\bibfnamefont {A.~K.}\ \bibnamefont
  {Mehta}}, \bibinfo {author} {\bibfnamefont {A.}~\bibnamefont {Buonanno}},
  \bibinfo {author} {\bibfnamefont {R.}~\bibnamefont {Cotesta}}, \bibinfo
  {author} {\bibfnamefont {A.}~\bibnamefont {Ghosh}}, \bibinfo {author}
  {\bibfnamefont {N.}~\bibnamefont {Sennett}}, \ and\ \bibinfo {author}
  {\bibfnamefont {J.}~\bibnamefont {Steinhoff}},\ }\href {\doibase
  10.1103/PhysRevD.107.044020} {\bibfield  {journal} {\bibinfo  {journal}
  {Phys. Rev. D}\ }\textbf {\bibinfo {volume} {107}},\ \bibinfo {pages}
  {044020} (\bibinfo {year} {2023})},\ \Eprint
  {http://arxiv.org/abs/2203.13937} {arXiv:2203.13937 [gr-qc]} \BibitemShut
  {NoStop}%
\bibitem [{\citenamefont {Islam}(2021)}]{Islam:2021pbd}%
  \BibitemOpen
  \bibfield  {author} {\bibinfo {author} {\bibfnamefont {T.}~\bibnamefont
  {Islam}},\ }\href@noop {} {\  (\bibinfo {year} {2021})},\ \Eprint
  {http://arxiv.org/abs/2111.00111} {arXiv:2111.00111 [gr-qc]} \BibitemShut
  {NoStop}%
\bibitem [{\citenamefont {Breschi}\ \emph {et~al.}(2019)\citenamefont
  {Breschi}, \citenamefont {O'Shaughnessy}, \citenamefont {Lange},\ and\
  \citenamefont {Birnholtz}}]{Breschi:2019wki}%
  \BibitemOpen
  \bibfield  {author} {\bibinfo {author} {\bibfnamefont {M.}~\bibnamefont
  {Breschi}}, \bibinfo {author} {\bibfnamefont {R.}~\bibnamefont
  {O'Shaughnessy}}, \bibinfo {author} {\bibfnamefont {J.}~\bibnamefont
  {Lange}}, \ and\ \bibinfo {author} {\bibfnamefont {O.}~\bibnamefont
  {Birnholtz}},\ }\href {\doibase 10.1088/1361-6382/ab5629} {\bibfield
  {journal} {\bibinfo  {journal} {Class. Quant. Grav.}\ }\textbf {\bibinfo
  {volume} {36}},\ \bibinfo {pages} {245019} (\bibinfo {year} {2019})},\
  \Eprint {http://arxiv.org/abs/1903.05982} {arXiv:1903.05982 [gr-qc]}
  \BibitemShut {NoStop}%
\bibitem [{\citenamefont {Pratten}\ \emph {et~al.}(2021)\citenamefont {Pratten}
  \emph {et~al.}}]{Pratten:2020ceb}%
  \BibitemOpen
  \bibfield  {author} {\bibinfo {author} {\bibfnamefont {G.}~\bibnamefont
  {Pratten}} \emph {et~al.},\ }\href {\doibase 10.1103/PhysRevD.103.104056}
  {\bibfield  {journal} {\bibinfo  {journal} {Phys. Rev. D}\ }\textbf {\bibinfo
  {volume} {103}},\ \bibinfo {pages} {104056} (\bibinfo {year} {2021})},\
  \Eprint {http://arxiv.org/abs/2004.06503} {arXiv:2004.06503 [gr-qc]}
  \BibitemShut {NoStop}%
\bibitem [{\citenamefont {Pratten}\ \emph {et~al.}(2020)\citenamefont
  {Pratten}, \citenamefont {Husa}, \citenamefont {Garcia-Quiros}, \citenamefont
  {Colleoni}, \citenamefont {Ramos-Buades}, \citenamefont {Estelles},\ and\
  \citenamefont {Jaume}}]{Pratten:2020fqn}%
  \BibitemOpen
  \bibfield  {author} {\bibinfo {author} {\bibfnamefont {G.}~\bibnamefont
  {Pratten}}, \bibinfo {author} {\bibfnamefont {S.}~\bibnamefont {Husa}},
  \bibinfo {author} {\bibfnamefont {C.}~\bibnamefont {Garcia-Quiros}}, \bibinfo
  {author} {\bibfnamefont {M.}~\bibnamefont {Colleoni}}, \bibinfo {author}
  {\bibfnamefont {A.}~\bibnamefont {Ramos-Buades}}, \bibinfo {author}
  {\bibfnamefont {H.}~\bibnamefont {Estelles}}, \ and\ \bibinfo {author}
  {\bibfnamefont {R.}~\bibnamefont {Jaume}},\ }\href {\doibase
  10.1103/PhysRevD.102.064001} {\bibfield  {journal} {\bibinfo  {journal}
  {Phys. Rev. D}\ }\textbf {\bibinfo {volume} {102}},\ \bibinfo {pages}
  {064001} (\bibinfo {year} {2020})},\ \Eprint
  {http://arxiv.org/abs/2001.11412} {arXiv:2001.11412 [gr-qc]} \BibitemShut
  {NoStop}%
\bibitem [{\citenamefont {Garc\'\i{}a-Quir\'os}\ \emph
  {et~al.}(2020)\citenamefont {Garc\'\i{}a-Quir\'os}, \citenamefont {Colleoni},
  \citenamefont {Husa}, \citenamefont {Estell\'es}, \citenamefont {Pratten},
  \citenamefont {Ramos-Buades}, \citenamefont {Mateu-Lucena},\ and\
  \citenamefont {Jaume}}]{Garcia-Quiros:2020qpx}%
  \BibitemOpen
  \bibfield  {author} {\bibinfo {author} {\bibfnamefont {C.}~\bibnamefont
  {Garc\'\i{}a-Quir\'os}}, \bibinfo {author} {\bibfnamefont {M.}~\bibnamefont
  {Colleoni}}, \bibinfo {author} {\bibfnamefont {S.}~\bibnamefont {Husa}},
  \bibinfo {author} {\bibfnamefont {H.}~\bibnamefont {Estell\'es}}, \bibinfo
  {author} {\bibfnamefont {G.}~\bibnamefont {Pratten}}, \bibinfo {author}
  {\bibfnamefont {A.}~\bibnamefont {Ramos-Buades}}, \bibinfo {author}
  {\bibfnamefont {M.}~\bibnamefont {Mateu-Lucena}}, \ and\ \bibinfo {author}
  {\bibfnamefont {R.}~\bibnamefont {Jaume}},\ }\href {\doibase
  10.1103/PhysRevD.102.064002} {\bibfield  {journal} {\bibinfo  {journal}
  {Phys. Rev. D}\ }\textbf {\bibinfo {volume} {102}},\ \bibinfo {pages}
  {064002} (\bibinfo {year} {2020})},\ \Eprint
  {http://arxiv.org/abs/2001.10914} {arXiv:2001.10914 [gr-qc]} \BibitemShut
  {NoStop}%
\bibitem [{\citenamefont {Abbott}\ \emph
  {et~al.}(2020{\natexlab{c}})\citenamefont {Abbott} \emph
  {et~al.}}]{LIGOScientific:2020stg}%
  \BibitemOpen
  \bibfield  {author} {\bibinfo {author} {\bibfnamefont {R.}~\bibnamefont
  {Abbott}} \emph {et~al.} (\bibinfo {collaboration} {LIGO Scientific and Virgo
  Collaborations}),\ }\href {\doibase 10.1103/PhysRevD.102.043015} {\bibfield
  {journal} {\bibinfo  {journal} {Phys. Rev. D}\ }\textbf {\bibinfo {volume}
  {102}},\ \bibinfo {pages} {043015} (\bibinfo {year} {2020}{\natexlab{c}})},\
  \Eprint {http://arxiv.org/abs/2004.08342} {arXiv:2004.08342 [astro-ph.HE]}
  \BibitemShut {NoStop}%
\bibitem [{\citenamefont {Blanchet}(2002)}]{Blanchet:2002av}%
  \BibitemOpen
  \bibfield  {author} {\bibinfo {author} {\bibfnamefont {L.}~\bibnamefont
  {Blanchet}},\ }\href {\doibase 10.12942/lrr-2002-3} {\bibfield  {journal}
  {\bibinfo  {journal} {Living Rev. Rel.}\ }\textbf {\bibinfo {volume} {5}},\
  \bibinfo {pages} {3} (\bibinfo {year} {2002})},\ \Eprint
  {http://arxiv.org/abs/gr-qc/0202016} {arXiv:gr-qc/0202016} \BibitemShut
  {NoStop}%
\bibitem [{\citenamefont {Mishra}\ \emph {et~al.}(2016)\citenamefont {Mishra},
  \citenamefont {Kela}, \citenamefont {Arun},\ and\ \citenamefont
  {Faye}}]{Mishra:2016whh}%
  \BibitemOpen
  \bibfield  {author} {\bibinfo {author} {\bibfnamefont {C.~K.}\ \bibnamefont
  {Mishra}}, \bibinfo {author} {\bibfnamefont {A.}~\bibnamefont {Kela}},
  \bibinfo {author} {\bibfnamefont {K.~G.}\ \bibnamefont {Arun}}, \ and\
  \bibinfo {author} {\bibfnamefont {G.}~\bibnamefont {Faye}},\ }\href {\doibase
  10.1103/PhysRevD.93.084054} {\bibfield  {journal} {\bibinfo  {journal} {Phys.
  Rev. D}\ }\textbf {\bibinfo {volume} {93}},\ \bibinfo {pages} {084054}
  (\bibinfo {year} {2016})},\ \Eprint {http://arxiv.org/abs/1601.05588}
  {arXiv:1601.05588 [gr-qc]} \BibitemShut {NoStop}%
\bibitem [{\citenamefont {Henry}\ \emph {et~al.}(2022)\citenamefont {Henry},
  \citenamefont {Marsat},\ and\ \citenamefont {Khalil}}]{Henry:2022dzx}%
  \BibitemOpen
  \bibfield  {author} {\bibinfo {author} {\bibfnamefont {Q.}~\bibnamefont
  {Henry}}, \bibinfo {author} {\bibfnamefont {S.}~\bibnamefont {Marsat}}, \
  and\ \bibinfo {author} {\bibfnamefont {M.}~\bibnamefont {Khalil}},\
  }\href@noop {} {\  (\bibinfo {year} {2022})},\ \Eprint
  {http://arxiv.org/abs/2209.00374} {arXiv:2209.00374 [gr-qc]} \BibitemShut
  {NoStop}%
\bibitem [{\citenamefont {Lehner}(2001)}]{Lehner:2001wq}%
  \BibitemOpen
  \bibfield  {author} {\bibinfo {author} {\bibfnamefont {L.}~\bibnamefont
  {Lehner}},\ }\href {\doibase 10.1088/0264-9381/18/17/202} {\bibfield
  {journal} {\bibinfo  {journal} {Class. Quant. Grav.}\ }\textbf {\bibinfo
  {volume} {18}},\ \bibinfo {pages} {R25} (\bibinfo {year} {2001})},\ \Eprint
  {http://arxiv.org/abs/gr-qc/0106072} {arXiv:gr-qc/0106072} \BibitemShut
  {NoStop}%
\bibitem [{\citenamefont {Sasaki}\ and\ \citenamefont
  {Tagoshi}(2003)}]{Sasaki:2003xr}%
  \BibitemOpen
  \bibfield  {author} {\bibinfo {author} {\bibfnamefont {M.}~\bibnamefont
  {Sasaki}}\ and\ \bibinfo {author} {\bibfnamefont {H.}~\bibnamefont
  {Tagoshi}},\ }\href {\doibase 10.12942/lrr-2003-6} {\bibfield  {journal}
  {\bibinfo  {journal} {Living Rev. Rel.}\ }\textbf {\bibinfo {volume} {6}},\
  \bibinfo {pages} {6} (\bibinfo {year} {2003})},\ \Eprint
  {http://arxiv.org/abs/gr-qc/0306120} {arXiv:gr-qc/0306120} \BibitemShut
  {NoStop}%
\bibitem [{\citenamefont {Pretorius}(2007)}]{Pretorius:2007nq}%
  \BibitemOpen
  \bibfield  {author} {\bibinfo {author} {\bibfnamefont {F.}~\bibnamefont
  {Pretorius}},\ }\href@noop {} {\  (\bibinfo {year} {2007})},\ \Eprint
  {http://arxiv.org/abs/0710.1338} {arXiv:0710.1338 [gr-qc]} \BibitemShut
  {NoStop}%
\bibitem [{\citenamefont {Poisson}\ and\ \citenamefont
  {Will}(1995)}]{Poisson:1995ef}%
  \BibitemOpen
  \bibfield  {author} {\bibinfo {author} {\bibfnamefont {E.}~\bibnamefont
  {Poisson}}\ and\ \bibinfo {author} {\bibfnamefont {C.~M.}\ \bibnamefont
  {Will}},\ }\href {\doibase 10.1103/PhysRevD.52.848} {\bibfield  {journal}
  {\bibinfo  {journal} {Phys. Rev. D}\ }\textbf {\bibinfo {volume} {52}},\
  \bibinfo {pages} {848} (\bibinfo {year} {1995})},\ \Eprint
  {http://arxiv.org/abs/gr-qc/9502040} {arXiv:gr-qc/9502040} \BibitemShut
  {NoStop}%
\bibitem [{\citenamefont {Pappas}\ and\ \citenamefont
  {Apostolatos}(2012{\natexlab{a}})}]{Pappas:2012ns}%
  \BibitemOpen
  \bibfield  {author} {\bibinfo {author} {\bibfnamefont {G.}~\bibnamefont
  {Pappas}}\ and\ \bibinfo {author} {\bibfnamefont {T.~A.}\ \bibnamefont
  {Apostolatos}},\ }\href {\doibase 10.1103/PhysRevLett.108.231104} {\bibfield
  {journal} {\bibinfo  {journal} {Phys. Rev. Lett.}\ }\textbf {\bibinfo
  {volume} {108}},\ \bibinfo {pages} {231104} (\bibinfo {year}
  {2012}{\natexlab{a}})},\ \Eprint {http://arxiv.org/abs/1201.6067}
  {arXiv:1201.6067 [gr-qc]} \BibitemShut {NoStop}%
\bibitem [{\citenamefont {Pappas}\ and\ \citenamefont
  {Apostolatos}(2012{\natexlab{b}})}]{Pappas:2012qg}%
  \BibitemOpen
  \bibfield  {author} {\bibinfo {author} {\bibfnamefont {G.}~\bibnamefont
  {Pappas}}\ and\ \bibinfo {author} {\bibfnamefont {T.~A.}\ \bibnamefont
  {Apostolatos}},\ }\href@noop {} {\  (\bibinfo {year} {2012}{\natexlab{b}})},\
  \Eprint {http://arxiv.org/abs/1211.6299} {arXiv:1211.6299 [gr-qc]}
  \BibitemShut {NoStop}%
\bibitem [{\citenamefont {Uchikata}\ and\ \citenamefont
  {Yoshida}(2016)}]{Uchikata:2015yma}%
  \BibitemOpen
  \bibfield  {author} {\bibinfo {author} {\bibfnamefont {N.}~\bibnamefont
  {Uchikata}}\ and\ \bibinfo {author} {\bibfnamefont {S.}~\bibnamefont
  {Yoshida}},\ }\href {\doibase 10.1088/0264-9381/33/2/025005} {\bibfield
  {journal} {\bibinfo  {journal} {Class. Quant. Grav.}\ }\textbf {\bibinfo
  {volume} {33}},\ \bibinfo {pages} {025005} (\bibinfo {year} {2016})},\
  \Eprint {http://arxiv.org/abs/1506.06485} {arXiv:1506.06485 [gr-qc]}
  \BibitemShut {NoStop}%
\bibitem [{\citenamefont {Khan}\ \emph {et~al.}(2020)\citenamefont {Khan},
  \citenamefont {Ohme}, \citenamefont {Chatziioannou},\ and\ \citenamefont
  {Hannam}}]{Khan:2019kot}%
  \BibitemOpen
  \bibfield  {author} {\bibinfo {author} {\bibfnamefont {S.}~\bibnamefont
  {Khan}}, \bibinfo {author} {\bibfnamefont {F.}~\bibnamefont {Ohme}}, \bibinfo
  {author} {\bibfnamefont {K.}~\bibnamefont {Chatziioannou}}, \ and\ \bibinfo
  {author} {\bibfnamefont {M.}~\bibnamefont {Hannam}},\ }\href {\doibase
  10.1103/PhysRevD.101.024056} {\bibfield  {journal} {\bibinfo  {journal}
  {Phys. Rev. D}\ }\textbf {\bibinfo {volume} {101}},\ \bibinfo {pages}
  {024056} (\bibinfo {year} {2020})},\ \Eprint
  {http://arxiv.org/abs/1911.06050} {arXiv:1911.06050 [gr-qc]} \BibitemShut
  {NoStop}%
\bibitem [{\citenamefont {Ajith}\ \emph {et~al.}(2011)\citenamefont {Ajith}
  \emph {et~al.}}]{Ajith:2009bn}%
  \BibitemOpen
  \bibfield  {author} {\bibinfo {author} {\bibfnamefont {P.}~\bibnamefont
  {Ajith}} \emph {et~al.},\ }\href {\doibase 10.1103/PhysRevLett.106.241101}
  {\bibfield  {journal} {\bibinfo  {journal} {Phys. Rev. Lett.}\ }\textbf
  {\bibinfo {volume} {106}},\ \bibinfo {pages} {241101} (\bibinfo {year}
  {2011})},\ \Eprint {http://arxiv.org/abs/0909.2867} {arXiv:0909.2867 [gr-qc]}
  \BibitemShut {NoStop}%
\bibitem [{\citenamefont {Santamaria}\ \emph {et~al.}(2010)\citenamefont
  {Santamaria} \emph {et~al.}}]{Santamaria:2010yb}%
  \BibitemOpen
  \bibfield  {author} {\bibinfo {author} {\bibfnamefont {L.}~\bibnamefont
  {Santamaria}} \emph {et~al.},\ }\href {\doibase 10.1103/PhysRevD.82.064016}
  {\bibfield  {journal} {\bibinfo  {journal} {Phys. Rev. D}\ }\textbf {\bibinfo
  {volume} {82}},\ \bibinfo {pages} {064016} (\bibinfo {year} {2010})},\
  \Eprint {http://arxiv.org/abs/1005.3306} {arXiv:1005.3306 [gr-qc]}
  \BibitemShut {NoStop}%
\bibitem [{\citenamefont {Schmidt}\ \emph {et~al.}(2012)\citenamefont
  {Schmidt}, \citenamefont {Hannam},\ and\ \citenamefont
  {Husa}}]{Schmidt:2012rh}%
  \BibitemOpen
  \bibfield  {author} {\bibinfo {author} {\bibfnamefont {P.}~\bibnamefont
  {Schmidt}}, \bibinfo {author} {\bibfnamefont {M.}~\bibnamefont {Hannam}}, \
  and\ \bibinfo {author} {\bibfnamefont {S.}~\bibnamefont {Husa}},\ }\href
  {\doibase 10.1103/PhysRevD.86.104063} {\bibfield  {journal} {\bibinfo
  {journal} {Phys. Rev. D}\ }\textbf {\bibinfo {volume} {86}},\ \bibinfo
  {pages} {104063} (\bibinfo {year} {2012})},\ \Eprint
  {http://arxiv.org/abs/1207.3088} {arXiv:1207.3088 [gr-qc]} \BibitemShut
  {NoStop}%
\bibitem [{\citenamefont {Schmidt}\ \emph {et~al.}(2015)\citenamefont
  {Schmidt}, \citenamefont {Ohme},\ and\ \citenamefont
  {Hannam}}]{Schmidt:2014iyl}%
  \BibitemOpen
  \bibfield  {author} {\bibinfo {author} {\bibfnamefont {P.}~\bibnamefont
  {Schmidt}}, \bibinfo {author} {\bibfnamefont {F.}~\bibnamefont {Ohme}}, \
  and\ \bibinfo {author} {\bibfnamefont {M.}~\bibnamefont {Hannam}},\ }\href
  {\doibase 10.1103/PhysRevD.91.024043} {\bibfield  {journal} {\bibinfo
  {journal} {Phys. Rev. D}\ }\textbf {\bibinfo {volume} {91}},\ \bibinfo
  {pages} {024043} (\bibinfo {year} {2015})},\ \Eprint
  {http://arxiv.org/abs/1408.1810} {arXiv:1408.1810 [gr-qc]} \BibitemShut
  {NoStop}%
\bibitem [{\citenamefont {Veitch}\ and\ \citenamefont
  {Vecchio}(2010)}]{Veitch:2009hd}%
  \BibitemOpen
  \bibfield  {author} {\bibinfo {author} {\bibfnamefont {J.}~\bibnamefont
  {Veitch}}\ and\ \bibinfo {author} {\bibfnamefont {A.}~\bibnamefont
  {Vecchio}},\ }\href {\doibase 10.1103/PhysRevD.81.062003} {\bibfield
  {journal} {\bibinfo  {journal} {Phys. Rev. D}\ }\textbf {\bibinfo {volume}
  {81}},\ \bibinfo {pages} {062003} (\bibinfo {year} {2010})},\ \Eprint
  {http://arxiv.org/abs/0911.3820} {arXiv:0911.3820 [astro-ph.CO]} \BibitemShut
  {NoStop}%
\bibitem [{\citenamefont {Veitch}\ \emph {et~al.}(2015)\citenamefont {Veitch}
  \emph {et~al.}}]{Veitch:2014wba}%
  \BibitemOpen
  \bibfield  {author} {\bibinfo {author} {\bibfnamefont {J.}~\bibnamefont
  {Veitch}} \emph {et~al.},\ }\href {\doibase 10.1103/PhysRevD.91.042003}
  {\bibfield  {journal} {\bibinfo  {journal} {Phys. Rev. D}\ }\textbf {\bibinfo
  {volume} {91}},\ \bibinfo {pages} {042003} (\bibinfo {year} {2015})},\
  \Eprint {http://arxiv.org/abs/1409.7215} {arXiv:1409.7215 [gr-qc]}
  \BibitemShut {NoStop}%
\bibitem [{\citenamefont {Thrane}\ and\ \citenamefont
  {Talbot}(2019)}]{thrane2019introduction}%
  \BibitemOpen
  \bibfield  {author} {\bibinfo {author} {\bibfnamefont {E.}~\bibnamefont
  {Thrane}}\ and\ \bibinfo {author} {\bibfnamefont {C.}~\bibnamefont
  {Talbot}},\ }\href@noop {} {\bibfield  {journal} {\bibinfo  {journal}
  {Publications of the Astronomical Society of Australia}\ }\textbf {\bibinfo
  {volume} {36}},\ \bibinfo {pages} {e010} (\bibinfo {year}
  {2019})}\BibitemShut {NoStop}%
\bibitem [{\citenamefont {{LIGO Scientific Collaboration}}(2018)}]{lalsuite}%
  \BibitemOpen
  \bibfield  {author} {\bibinfo {author} {\bibnamefont {{LIGO Scientific
  Collaboration}}},\ }\href {\doibase 10.7935/GT1W-FZ16} {\enquote {\bibinfo
  {title} {{LIGO} {A}lgorithm {L}ibrary - {LALS}uite},}\ }\bibinfo
  {howpublished} {free software (GPL)} (\bibinfo {year} {2018})\BibitemShut
  {NoStop}%
\bibitem [{\citenamefont {{Skilling}}(2004)}]{2004AIPC..735..395S}%
  \BibitemOpen
  \bibfield  {author} {\bibinfo {author} {\bibfnamefont {J.}~\bibnamefont
  {{Skilling}}},\ }in\ \href {\doibase 10.1063/1.1835238} {\emph {\bibinfo
  {booktitle} {Bayesian Inference and Maximum Entropy Methods in Science and
  Engineering: 24th International Workshop on Bayesian Inference and Maximum
  Entropy Methods in Science and Engineering}}},\ \bibinfo {series} {American
  Institute of Physics Conference Series}, Vol.\ \bibinfo {volume} {735},\
  \bibinfo {editor} {edited by\ \bibinfo {editor} {\bibfnamefont
  {R.}~\bibnamefont {{Fischer}}}, \bibinfo {editor} {\bibfnamefont
  {R.}~\bibnamefont {{Preuss}}}, \ and\ \bibinfo {editor} {\bibfnamefont
  {U.~V.}\ \bibnamefont {{Toussaint}}}}\ (\bibinfo {year} {2004})\ pp.\
  \bibinfo {pages} {395--405}\BibitemShut {NoStop}%
\bibitem [{\citenamefont {Speagle}(2020)}]{speagle2020dynesty}%
  \BibitemOpen
  \bibfield  {author} {\bibinfo {author} {\bibfnamefont {J.~S.}\ \bibnamefont
  {Speagle}},\ }\href@noop {} {\bibfield  {journal} {\bibinfo  {journal}
  {Monthly Notices of the Royal Astronomical Society}\ }\textbf {\bibinfo
  {volume} {493}},\ \bibinfo {pages} {3132} (\bibinfo {year}
  {2020})}\BibitemShut {NoStop}%
\bibitem [{\citenamefont {Ashton}\ \emph {et~al.}(2019)\citenamefont {Ashton}
  \emph {et~al.}}]{Ashton:2018jfp}%
  \BibitemOpen
  \bibfield  {author} {\bibinfo {author} {\bibfnamefont {G.}~\bibnamefont
  {Ashton}} \emph {et~al.},\ }\href {\doibase 10.3847/1538-4365/ab06fc}
  {\bibfield  {journal} {\bibinfo  {journal} {Astrophys. J. Suppl.}\ }\textbf
  {\bibinfo {volume} {241}},\ \bibinfo {pages} {27} (\bibinfo {year} {2019})},\
  \Eprint {http://arxiv.org/abs/1811.02042} {arXiv:1811.02042 [astro-ph.IM]}
  \BibitemShut {NoStop}%
\bibitem [{\citenamefont {Romero-Shaw}\ \emph {et~al.}(2020)\citenamefont
  {Romero-Shaw} \emph {et~al.}}]{Romero-Shaw:2020owr}%
  \BibitemOpen
  \bibfield  {author} {\bibinfo {author} {\bibfnamefont {I.~M.}\ \bibnamefont
  {Romero-Shaw}} \emph {et~al.},\ }\href {\doibase 10.1093/mnras/staa2850}
  {\bibfield  {journal} {\bibinfo  {journal} {Mon. Not. Roy. Astron. Soc.}\
  }\textbf {\bibinfo {volume} {499}},\ \bibinfo {pages} {3295} (\bibinfo {year}
  {2020})},\ \Eprint {http://arxiv.org/abs/2006.00714} {arXiv:2006.00714
  [astro-ph.IM]} \BibitemShut {NoStop}%
\bibitem [{\citenamefont {Abbott}\ \emph {et~al.}(2018)\citenamefont {Abbott}
  \emph {et~al.}}]{KAGRA:2013rdx}%
  \BibitemOpen
  \bibfield  {author} {\bibinfo {author} {\bibfnamefont {B.~P.}\ \bibnamefont
  {Abbott}} \emph {et~al.} (\bibinfo {collaboration} {KAGRA, LIGO Scientific,
  Virgo, VIRGO}),\ }\href {\doibase 10.1007/s41114-020-00026-9} {\bibfield
  {journal} {\bibinfo  {journal} {Living Rev. Rel.}\ }\textbf {\bibinfo
  {volume} {21}},\ \bibinfo {pages} {3} (\bibinfo {year} {2018})},\ \Eprint
  {http://arxiv.org/abs/1304.0670} {arXiv:1304.0670 [gr-qc]} \BibitemShut
  {NoStop}%
\bibitem [{\citenamefont {{Harry}}\ and\ \citenamefont {{LIGO Scientific
  Collaboration}}(2010)}]{AdvancedLIGO2010}%
  \BibitemOpen
  \bibfield  {author} {\bibinfo {author} {\bibfnamefont {G.~M.}\ \bibnamefont
  {{Harry}}}\ and\ \bibinfo {author} {\bibnamefont {{LIGO Scientific
  Collaboration}}},\ }\href {\doibase 10.1088/0264-9381/27/8/084006} {\bibfield
   {journal} {\bibinfo  {journal} {Classical and Quantum Gravity}\ }\textbf
  {\bibinfo {volume} {27}},\ \bibinfo {eid} {084006} (\bibinfo {year}
  {2010})}\BibitemShut {NoStop}%
\bibitem [{\citenamefont {Aasi}\ \emph
  {et~al.}(2015{\natexlab{b}})\citenamefont {Aasi} \emph
  {et~al.}}]{TheLIGOScientific:2014jea}%
  \BibitemOpen
  \bibfield  {author} {\bibinfo {author} {\bibfnamefont {J.}~\bibnamefont
  {Aasi}} \emph {et~al.} (\bibinfo {collaboration} {LIGO Scientific}),\ }\href
  {\doibase 10.1088/0264-9381/32/7/074001} {\bibfield  {journal} {\bibinfo
  {journal} {Class. Quant. Grav.}\ }\textbf {\bibinfo {volume} {32}},\ \bibinfo
  {pages} {074001} (\bibinfo {year} {2015}{\natexlab{b}})},\ \Eprint
  {http://arxiv.org/abs/1411.4547} {arXiv:1411.4547 [gr-qc]} \BibitemShut
  {NoStop}%
\bibitem [{\citenamefont {Acernese}\ \emph
  {et~al.}(2015{\natexlab{b}})\citenamefont {Acernese} \emph
  {et~al.}}]{TheVirgo:2014hva}%
  \BibitemOpen
  \bibfield  {author} {\bibinfo {author} {\bibfnamefont {F.}~\bibnamefont
  {Acernese}} \emph {et~al.} (\bibinfo {collaboration} {VIRGO Collaboration}),\
  }\href {\doibase 10.1088/0264-9381/32/2/024001} {\bibfield  {journal}
  {\bibinfo  {journal} {Classical Quantum Gravity}\ }\textbf {\bibinfo {volume}
  {32}},\ \bibinfo {pages} {024001} (\bibinfo {year} {2015}{\natexlab{b}})},\
  \Eprint {http://arxiv.org/abs/1408.3978} {arXiv:1408.3978 [gr-qc]}
  \BibitemShut {NoStop}%
%%CITATION = ARXIV:1408.3978;%%
\bibitem [{\citenamefont {Acernese}\ and\ \citenamefont
  {et.al.}(2006)}]{TheVirgostatus}%
  \BibitemOpen
  \bibfield  {author} {\bibinfo {author} {\bibfnamefont {F.}~\bibnamefont
  {Acernese}}\ and\ \bibinfo {author} {\bibfnamefont {P.~A.}\ \bibnamefont
  {et.al.}},\ }\href@noop {} {\bibfield  {journal} {\bibinfo  {journal}
  {Classical and Quantum Gravity}\ }\textbf {\bibinfo {volume} {23}},\ \bibinfo
  {pages} {S635} (\bibinfo {year} {2006})}\BibitemShut {NoStop}%
\bibitem [{\citenamefont {{LIGO Scientific Collaboration}}(2022)}]{H1L1V1Dcc}%
  \BibitemOpen
  \bibfield  {author} {\bibinfo {author} {\bibnamefont {{LIGO Scientific
  Collaboration}}},\ }\href@noop {} {}\bibinfo {howpublished}
  {\url{https://dcc.ligo.org/LIGO-P1200087-v42/public}} (\bibinfo {year}
  {2022})\BibitemShut {NoStop}%
\bibitem [{\citenamefont {Nitz}\ \emph {et~al.}(2020)\citenamefont {Nitz},
  \citenamefont {Harry}, \citenamefont {Brown}, \citenamefont {Biwer},
  \citenamefont {Willis}, \citenamefont {Canton}, \citenamefont {Capano},
  \citenamefont {Pekowsky}, \citenamefont {Dent}, \citenamefont {Williamson},
  \citenamefont {Davies}, \citenamefont {De}, \citenamefont {Cabero},
  \citenamefont {Machenschalk}, \citenamefont {Kumar}, \citenamefont {Reyes},
  \citenamefont {Macleod}, \citenamefont {dfinstad}, \citenamefont {Pannarale},
  \citenamefont {Massinger}, \citenamefont {Kumar}, \citenamefont {Tápai},
  \citenamefont {Singer}, \citenamefont {Khan}, \citenamefont {Fairhurst},
  \citenamefont {Nielsen}, \citenamefont {Singh},\ and\ \citenamefont
  {shasvath}}]{alex_nitz_2020_4134752}%
  \BibitemOpen
  \bibfield  {author} {\bibinfo {author} {\bibfnamefont {A.}~\bibnamefont
  {Nitz}}, \bibinfo {author} {\bibfnamefont {I.}~\bibnamefont {Harry}},
  \bibinfo {author} {\bibfnamefont {D.}~\bibnamefont {Brown}}, \bibinfo
  {author} {\bibfnamefont {C.~M.}\ \bibnamefont {Biwer}}, \bibinfo {author}
  {\bibfnamefont {J.}~\bibnamefont {Willis}}, \bibinfo {author} {\bibfnamefont
  {T.~D.}\ \bibnamefont {Canton}}, \bibinfo {author} {\bibfnamefont
  {C.}~\bibnamefont {Capano}}, \bibinfo {author} {\bibfnamefont
  {L.}~\bibnamefont {Pekowsky}}, \bibinfo {author} {\bibfnamefont
  {T.}~\bibnamefont {Dent}}, \bibinfo {author} {\bibfnamefont {A.~R.}\
  \bibnamefont {Williamson}}, \bibinfo {author} {\bibfnamefont {G.~S.}\
  \bibnamefont {Davies}}, \bibinfo {author} {\bibfnamefont {S.}~\bibnamefont
  {De}}, \bibinfo {author} {\bibfnamefont {M.}~\bibnamefont {Cabero}}, \bibinfo
  {author} {\bibfnamefont {B.}~\bibnamefont {Machenschalk}}, \bibinfo {author}
  {\bibfnamefont {P.}~\bibnamefont {Kumar}}, \bibinfo {author} {\bibfnamefont
  {S.}~\bibnamefont {Reyes}}, \bibinfo {author} {\bibfnamefont
  {D.}~\bibnamefont {Macleod}}, \bibinfo {author} {\bibnamefont {dfinstad}},
  \bibinfo {author} {\bibfnamefont {F.}~\bibnamefont {Pannarale}}, \bibinfo
  {author} {\bibfnamefont {T.}~\bibnamefont {Massinger}}, \bibinfo {author}
  {\bibfnamefont {S.}~\bibnamefont {Kumar}}, \bibinfo {author} {\bibfnamefont
  {M.}~\bibnamefont {Tápai}}, \bibinfo {author} {\bibfnamefont
  {L.}~\bibnamefont {Singer}}, \bibinfo {author} {\bibfnamefont
  {S.}~\bibnamefont {Khan}}, \bibinfo {author} {\bibfnamefont {S.}~\bibnamefont
  {Fairhurst}}, \bibinfo {author} {\bibfnamefont {A.}~\bibnamefont {Nielsen}},
  \bibinfo {author} {\bibfnamefont {S.}~\bibnamefont {Singh}}, \ and\ \bibinfo
  {author} {\bibnamefont {shasvath}},\ }\href {\doibase 10.5281/zenodo.4134752}
  {\enquote {\bibinfo {title} {gwastro/pycbc: Pycbc release v1.16.11},}\ }
  (\bibinfo {year} {2020})\BibitemShut {NoStop}%
\bibitem [{\citenamefont {{Harris}}\ \emph {et~al.}(2020)\citenamefont
  {{Harris}}, \citenamefont {{Millman}}, \citenamefont {{van der Walt}},
  \citenamefont {{Gommers}}, \citenamefont {{Virtanen}}, \citenamefont
  {{Cournapeau}}, \citenamefont {{Wieser}}, \citenamefont {{Taylor}},
  \citenamefont {{Berg}}, \citenamefont {{Smith}}, \citenamefont {{Kern}},
  \citenamefont {{Picus}}, \citenamefont {{Hoyer}}, \citenamefont {{van
  Kerkwijk}}, \citenamefont {{Brett}}, \citenamefont {{Haldane}}, \citenamefont
  {{del R{\'\i}o}}, \citenamefont {{Wiebe}}, \citenamefont {{Peterson}},
  \citenamefont {{G{\'e}rard-Marchant}}, \citenamefont {{Sheppard}},
  \citenamefont {{Reddy}}, \citenamefont {{Weckesser}}, \citenamefont
  {{Abbasi}}, \citenamefont {{Gohlke}},\ and\ \citenamefont
  {{Oliphant}}}]{2020Natur.585..357H}%
  \BibitemOpen
  \bibfield  {author} {\bibinfo {author} {\bibfnamefont {C.~R.}\ \bibnamefont
  {{Harris}}}, \bibinfo {author} {\bibfnamefont {K.~J.}\ \bibnamefont
  {{Millman}}}, \bibinfo {author} {\bibfnamefont {S.~J.}\ \bibnamefont {{van
  der Walt}}}, \bibinfo {author} {\bibfnamefont {R.}~\bibnamefont {{Gommers}}},
  \bibinfo {author} {\bibfnamefont {P.}~\bibnamefont {{Virtanen}}}, \bibinfo
  {author} {\bibfnamefont {D.}~\bibnamefont {{Cournapeau}}}, \bibinfo {author}
  {\bibfnamefont {E.}~\bibnamefont {{Wieser}}}, \bibinfo {author}
  {\bibfnamefont {J.}~\bibnamefont {{Taylor}}}, \bibinfo {author}
  {\bibfnamefont {S.}~\bibnamefont {{Berg}}}, \bibinfo {author} {\bibfnamefont
  {N.~J.}\ \bibnamefont {{Smith}}}, \bibinfo {author} {\bibfnamefont
  {R.}~\bibnamefont {{Kern}}}, \bibinfo {author} {\bibfnamefont
  {M.}~\bibnamefont {{Picus}}}, \bibinfo {author} {\bibfnamefont
  {S.}~\bibnamefont {{Hoyer}}}, \bibinfo {author} {\bibfnamefont {M.~H.}\
  \bibnamefont {{van Kerkwijk}}}, \bibinfo {author} {\bibfnamefont
  {M.}~\bibnamefont {{Brett}}}, \bibinfo {author} {\bibfnamefont
  {A.}~\bibnamefont {{Haldane}}}, \bibinfo {author} {\bibfnamefont {J.~F.}\
  \bibnamefont {{del R{\'\i}o}}}, \bibinfo {author} {\bibfnamefont
  {M.}~\bibnamefont {{Wiebe}}}, \bibinfo {author} {\bibfnamefont
  {P.}~\bibnamefont {{Peterson}}}, \bibinfo {author} {\bibfnamefont
  {P.}~\bibnamefont {{G{\'e}rard-Marchant}}}, \bibinfo {author} {\bibfnamefont
  {K.}~\bibnamefont {{Sheppard}}}, \bibinfo {author} {\bibfnamefont
  {T.}~\bibnamefont {{Reddy}}}, \bibinfo {author} {\bibfnamefont
  {W.}~\bibnamefont {{Weckesser}}}, \bibinfo {author} {\bibfnamefont
  {H.}~\bibnamefont {{Abbasi}}}, \bibinfo {author} {\bibfnamefont
  {C.}~\bibnamefont {{Gohlke}}}, \ and\ \bibinfo {author} {\bibfnamefont
  {T.~E.}\ \bibnamefont {{Oliphant}}},\ }\href {\doibase
  10.1038/s41586-020-2649-2} {\bibfield  {journal} {\bibinfo  {journal} {\nat}\
  }\textbf {\bibinfo {volume} {585}},\ \bibinfo {pages} {357} (\bibinfo {year}
  {2020})},\ \Eprint {http://arxiv.org/abs/2006.10256} {arXiv:2006.10256
  [cs.MS]} \BibitemShut {NoStop}%
\bibitem [{\citenamefont {Hoy}\ and\ \citenamefont
  {Raymond}(2021)}]{Hoy:2020vys}%
  \BibitemOpen
  \bibfield  {author} {\bibinfo {author} {\bibfnamefont {C.}~\bibnamefont
  {Hoy}}\ and\ \bibinfo {author} {\bibfnamefont {V.}~\bibnamefont {Raymond}},\
  }\href {\doibase 10.1016/j.softx.2021.100765} {\bibfield  {journal} {\bibinfo
   {journal} {SoftwareX}\ }\textbf {\bibinfo {volume} {15}},\ \bibinfo {pages}
  {100765} (\bibinfo {year} {2021})},\ \Eprint
  {http://arxiv.org/abs/2006.06639} {arXiv:2006.06639 [astro-ph.IM]}
  \BibitemShut {NoStop}%
\bibitem [{\citenamefont {{Hunter}}(2007)}]{2007CSE.....9...90H}%
  \BibitemOpen
  \bibfield  {author} {\bibinfo {author} {\bibfnamefont {J.~D.}\ \bibnamefont
  {{Hunter}}},\ }\href {\doibase 10.1109/MCSE.2007.55} {\bibfield  {journal}
  {\bibinfo  {journal} {Computing in Science and Engineering}\ }\textbf
  {\bibinfo {volume} {9}},\ \bibinfo {pages} {90} (\bibinfo {year}
  {2007})}\BibitemShut {NoStop}%
\bibitem [{\citenamefont {Waskom}(2021)}]{Waskom2021}%
  \BibitemOpen
  \bibfield  {author} {\bibinfo {author} {\bibfnamefont {M.~L.}\ \bibnamefont
  {Waskom}},\ }\href {\doibase 10.21105/joss.03021} {\bibfield  {journal}
  {\bibinfo  {journal} {Journal of Open Source Software}\ }\textbf {\bibinfo
  {volume} {6}},\ \bibinfo {pages} {3021} (\bibinfo {year} {2021})}\BibitemShut
  {NoStop}%
\bibitem [{\citenamefont {Kluyver}\ \emph {et~al.}(2016)\citenamefont
  {Kluyver}, \citenamefont {Ragan-Kelley}, \citenamefont {P{\'e}rez},
  \citenamefont {Granger}, \citenamefont {Bussonnier}, \citenamefont
  {Frederic}, \citenamefont {Kelley}, \citenamefont {Hamrick}, \citenamefont
  {Grout}, \citenamefont {Corlay}, \citenamefont {Ivanov}, \citenamefont
  {Avila}, \citenamefont {Abdalla}, \citenamefont {Willing},\ and\
  \citenamefont {development team}}]{soton403913}%
  \BibitemOpen
  \bibfield  {author} {\bibinfo {author} {\bibfnamefont {T.}~\bibnamefont
  {Kluyver}}, \bibinfo {author} {\bibfnamefont {B.}~\bibnamefont
  {Ragan-Kelley}}, \bibinfo {author} {\bibfnamefont {F.}~\bibnamefont
  {P{\'e}rez}}, \bibinfo {author} {\bibfnamefont {B.}~\bibnamefont {Granger}},
  \bibinfo {author} {\bibfnamefont {M.}~\bibnamefont {Bussonnier}}, \bibinfo
  {author} {\bibfnamefont {J.}~\bibnamefont {Frederic}}, \bibinfo {author}
  {\bibfnamefont {K.}~\bibnamefont {Kelley}}, \bibinfo {author} {\bibfnamefont
  {J.}~\bibnamefont {Hamrick}}, \bibinfo {author} {\bibfnamefont
  {J.}~\bibnamefont {Grout}}, \bibinfo {author} {\bibfnamefont
  {S.}~\bibnamefont {Corlay}}, \bibinfo {author} {\bibfnamefont
  {P.}~\bibnamefont {Ivanov}}, \bibinfo {author} {\bibfnamefont
  {D.}~\bibnamefont {Avila}}, \bibinfo {author} {\bibfnamefont
  {S.}~\bibnamefont {Abdalla}}, \bibinfo {author} {\bibfnamefont
  {C.}~\bibnamefont {Willing}}, \ and\ \bibinfo {author} {\bibfnamefont
  {J.}~\bibnamefont {development team}},\ }in\ \href@noop {} {\emph {\bibinfo
  {booktitle} {Positioning and Power in Academic Publishing: Players, Agents
  and Agendas}}},\ \bibinfo {editor} {edited by\ \bibinfo {editor}
  {\bibfnamefont {F.}~\bibnamefont {Loizides}}\ and\ \bibinfo {editor}
  {\bibfnamefont {B.}~\bibnamefont {Scmidt}}}\ (\bibinfo  {publisher} {IOS
  Press},\ \bibinfo {year} {2016})\ pp.\ \bibinfo {pages} {87--90}\BibitemShut
  {NoStop}%
\bibitem [{\citenamefont {Foreman-Mackey}(2016)}]{corner}%
  \BibitemOpen
  \bibfield  {author} {\bibinfo {author} {\bibfnamefont {D.}~\bibnamefont
  {Foreman-Mackey}},\ }\href {\doibase 10.21105/joss.00024} {\bibfield
  {journal} {\bibinfo  {journal} {The Journal of Open Source Software}\
  }\textbf {\bibinfo {volume} {1}},\ \bibinfo {pages} {24} (\bibinfo {year}
  {2016})}\BibitemShut {NoStop}%
\end{thebibliography}%

\end{document}